\documentclass[12pt]{article}
\usepackage[utf8]{inputenc}
\usepackage{amsmath}
\usepackage{amssymb}
\usepackage[mathlines,displaymath]{lineno}
\DeclareMathAlphabet{\mathbold}{OML}{txr}{b}{it}
\usepackage[font={small}]{caption}  

\usepackage{multirow}
\usepackage{dsfont}
\usepackage{graphicx}
\usepackage{rotating}
\usepackage{paralist} 
\graphicspath{{figures/}{./}}

\usepackage{xspace}
\usepackage{acronym}
\usepackage{dcolumn}
\newcolumntype{L}{D{.}{.}{2,5}}
\usepackage{booktabs} 
\usepackage{cite} 

\usepackage{hyperref} 
\hypersetup{colorlinks=true, urlcolor=blue}
\usepackage{cite} 
\hypersetup{
  colorlinks,
  citecolor=blue,
  linkcolor=red,
  urlcolor=blue
  }
  
\newlength{\dinwidth}
\newlength{\dinmargin}
\usepackage{parskip}
\newcommand*{\rom}[1]{\expandafter\@slowromancap\romannumeral #1@}

\newcommand{\Qsq}{\ensuremath{Q^2}\xspace}

\newcommand{\MeV}{\ensuremath{\mathrm{MeV}}\xspace}
\newcommand{\GeV}{\ensuremath{\mathrm{GeV}}\xspace}
\newcommand{\TeV}{\ensuremath{\mathrm{TeV}}\xspace}
\newcommand{\GeVsq}{\ensuremath{\mathrm{GeV}^2}\xspace}

\newcommand{\sw}{\ensuremath{\sin^2\hspace*{-0.15em}\theta_W}}
\newcommand{\sweff}{\ensuremath{\sin^2\hspace*{-0.15em}\theta_{\textrm{W},f}^\textrm{eff}}} 
\newcommand{\sweffl}{\ensuremath{\sin^2\hspace*{-0.15em}\theta_{\textrm{W},\ell}^\textrm{eff}}} 
\newcommand{\dr}{\ensuremath{\Delta r}}
\newcommand{\gf}{\ensuremath{G_{\rm F}}}

\newcommand{\rhopW}[2][]{\ensuremath{\rho^{\prime}_{\text{CC}#2}}}
\newcommand{\rhop}[2][] {\ensuremath{\rho^{\prime}_{\text{NC}#2}}}
\newcommand{\kapp}[2][] {\ensuremath{\kappa^{\prime}_{#2}}}

\newcommand{\Itf}{\ensuremath{I^3_{{\rm L},f}}}

\newcommand{\ad} {\ensuremath{g_A^d}}
\newcommand{\vd} {\ensuremath{g_V^d}}
\newcommand{\au} {\ensuremath{g_A^u}}
\newcommand{\vu} {\ensuremath{g_V^u}}
\newcommand{\aq} {\ensuremath{g_A^q}}
\newcommand{\vq} {\ensuremath{g_V^q}}
\newcommand{\gae}{\ensuremath{g_A^e}}
\newcommand{\ve} {\ensuremath{g_V^e}}
\newcommand{\af} {\ensuremath{g_A^f}}
\newcommand{\vf} {\ensuremath{g_V^f}}

\newcommand{\mH} {\ensuremath{m_H}}
\newcommand{\mw}{\ensuremath{m_W}}
\newcommand{\mW}{\mw}
\newcommand{\mz}{\ensuremath{m_Z}}
\newcommand{\mZ}{\mz}
\newcommand{\mt}{\ensuremath{m_t}}

\begin{document}  
\makeatletter
\makeatother

\begin{titlepage}

\noindent
\begin{flushleft}
\end{flushleft}

\noindent
\vspace{-1.5cm}
\begin{flushright}
  MITP/20-038\\
  MPP-2020-110 
\end{flushright}


\vspace{1.5cm}

\begin{center}
\begin{Large}
{\bf 
Electroweak Physics in Inclusive Deep Inelastic 
\\[1ex]
Scattering at the LHeC}
\end{Large}
\end{center}

\vspace{1.0cm}

\begin{center}
  Daniel Britzger\,$^1$, Max Klein\,$^{2}$ and Hubert Spiesberger\,$^{3}$ \\
\vspace{0.5cm}
\small
\it
$^1$~~Max-Planck-Institut f{\"u}r Physik, F{\"o}hringer Ring 6, D-80805 M{\"u}nchen, Germany \\
$^2$~~University of Liverpool, Oxford Street, UK-L69 7ZE Liverpool, United Kingdom \\
$^3$~~PRISMA$^+$ Cluster of Excellence, Institut f{\"u}r Physik, Johannes-Gutenberg-Universit{\"a}t, 
Staudinger Weg 7, D-55099 Mainz, Germany 
\end{center}

\vspace{1.5cm}

\begin{abstract}
\noindent
The proposed electron-proton collider LHeC is a unique facility 
where electroweak interactions can be studied with a very high 
precision in a largely unexplored kinematic regime of 
spacelike momentum transfer. 
We have simulated inclusive neutral- and charged-current 
deep-inelastic lepton proton scattering cross section data at 
center-of-mass energies of 1.2 and 1.3\,\TeV including their systematic uncertainties.
Based on simultaneous fits of electroweak physics parameters and 
parton distribution functions, we estimate the uncertainties of 
Standard Model parameters as well as a number of parameters 
describing physics beyond the Standard Model, for instance 
the oblique parameters $S$, $T$, and $U$. 
An unprecedented precision at the sub-percent level is expected 
for the measurement of the weak neutral-current couplings of the 
light-quarks to the $Z$ boson, $g_{A/V}^{u/d}$, improving their 
present precision by more than an order of magnitude.
The weak mixing angle can be determined with a precision of 
about $\Delta \sw = \pm 0.00015$, and its scale dependence
can be studied in the range between about $25$ and 
$700\,\GeV$.
An indirect determination of the $W$-boson mass in the on-shell 
scheme is possible with an experimental uncertainty down to 
$\Delta\mW=\pm6\,\MeV$. 
We discuss how the uncertainties of such measurements in 
deep-inelastic scattering compare with those from measurements 
in the timelike domain, e.g.\ at the $Z$-pole, and which aspects 
of the electroweak interaction are unique to measurements at the 
LHeC, for instance electroweak parameters in charged-current 
interactions.
We conclude that the LHeC will determine electroweak 
physics parameters, in the spacelike region, with unprecedented
precision leading to thorough tests of the Standard Model
and possibly beyond.
\end{abstract}

\end{titlepage}
\sloppy

\clearpage


\section{Introduction}

With the discovery of the Standard Model (SM) Higgs boson at 
the CERN Large Hadron Collider (LHC) experiments and subsequent 
measurements of its parameters, the fundamental parameters of the 
SM have been measured directly and with remarkable precision.
To further map out the validity of the theory of electroweak
interactions, more and higher precision electroweak measurements 
have to be performed. Such high-precision measurements can also  
be considered as a portal to new physics, since non-SM contributions 
may lead to significant deviations for some precisely measurable 
and calculable observables. 

The Large Hadron-electron Collider
(LHeC)~\cite{AbelleiraFernandez:2012cc,Bruning:2019scy,%
Bruning:2706220}, planned at the LHC, may complement the proton ring 
with an electron beam, allowing to perform deep inelastic scattering 
(DIS) with electrons and protons at TeV energies. Its electron beam
energy may be chosen to be 60 or 50\,GeV.
Considerations in this choice, as for example cost reasons,
are discussed in the forthcoming thorough update of the physics and conceptual 
accelerator and detector design report ~\cite{Bruning:2706220}. 
In both cases its kinematic reach extends to much 
higher scales in comparison to HERA, which together with the huge 
increase of the expected 
integrated luminosity will allow to perform 
high-precision electroweak measurements at high scales in DIS
for the first time.  

Since the discovery of weak neutral currents in 
1973~\cite{Hasert:1973ff,Haidt:2015bgg} and the formulation 
of the Glashow-Weinberg-Salam
model~\cite{Glashow:1961tr,Weinberg:1967tq,Weinberg:1971fb, 
Weinberg:1972tu,Salam:1964ry,Higgs:1964ia,Higgs:1964pj, 
Englert:1964et}, deep-inelastic lepton nucleon scattering has 
played an important role in testing the Standard Model. One of
the first measurements of the electroweak mixing angle, $\sin^2 \theta_W$,  
was obtained from polarized electron-deuteron scattering at SLAC 
\cite{Prescott:1978tm,Prescott:1979dh}. With the advent of HERA, 
the first electron-proton collider, a much larger range of 
momentum transfers squared, $Q^2$, became accessible -- an important 
prerequisite for probing electroweak interactions in DIS. 
First measurements of electroweak effects at HERA were 
undertaken in Refs.~\cite{Ahmed:1994fa,Aktas:2005iv}, and more 
thorough electroweak analyses have been performed recently,
for example with the complete set of H1 data in 
Ref.~\cite{Spiesberger:2018vki}.

\begin{figure}[t!hb]
  \centering
  \includegraphics[width=0.56\textwidth]{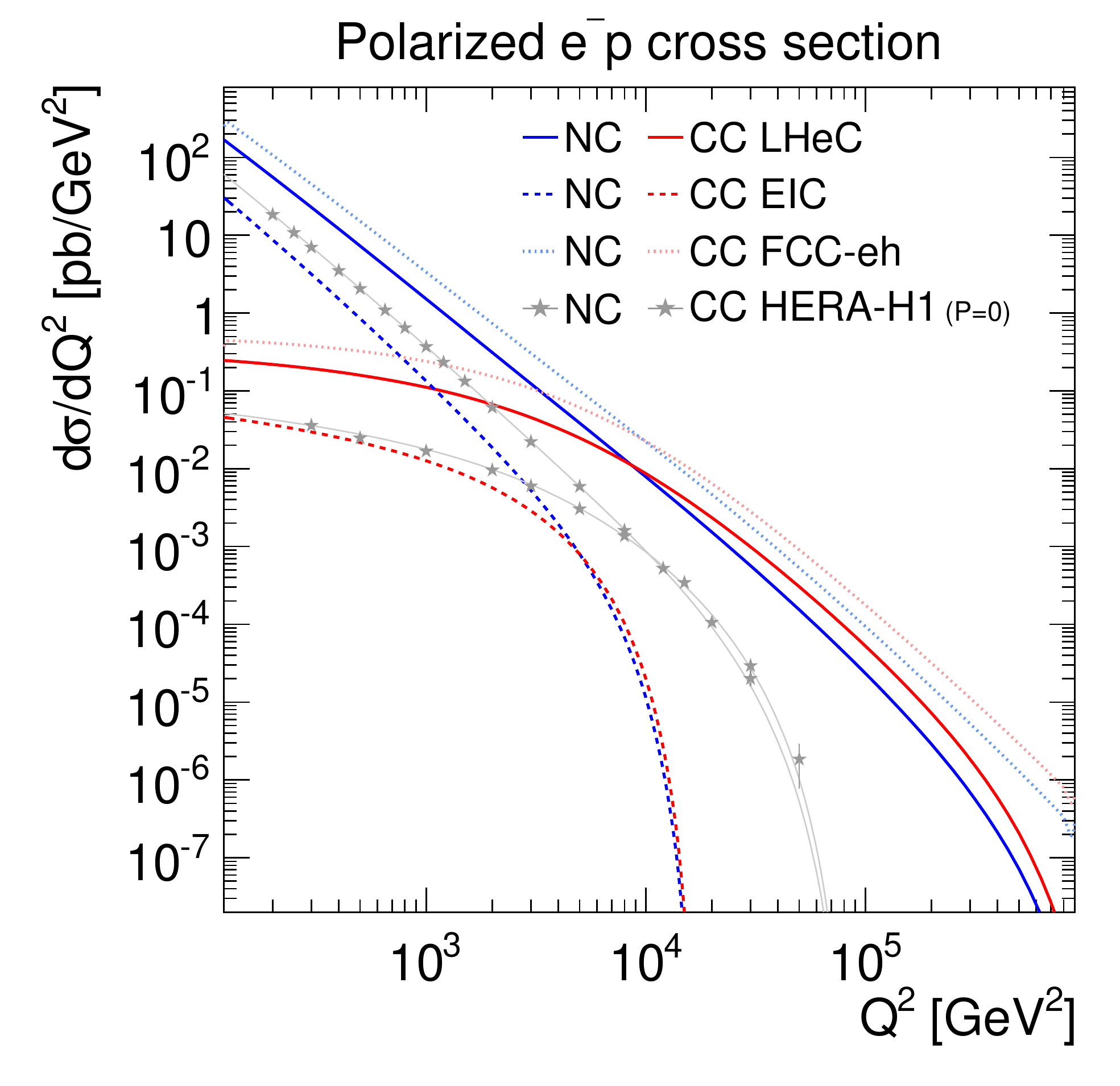}
  \caption{
    Single differential inclusive DIS cross sections for
    neutral- and charged-current $e^-p$ DIS with longitudinally 
    polarized electrons ($P_e=-0.8$) at LHeC, EIC, FCC-$eh$, and 
    HERA. For HERA, unpolarized cross sections are displayed 
    together with data from the H1 experiment.
  }
  \label{fig:dSigmaFacilities}
\end{figure}

Apart from the LHeC, other options for electron-hadron colliders 
are presently considered. A DIS option is studied as part 
of the possible Future Circular Collider (FCC) at CERN, the 
FCC-$eh$~\cite{Abada:2019lih}, and will reach center-of-mass 
energies still higher than at the LHeC. At Brookhaven, the  
Electron Ion Collider 
(EIC)~\cite{Accardi:2012qut,Aschenauer:2017jsk,EICpreCDR}
is under development to perform DIS measurements at lower energies but with higher luminosities than were 
achieved at HERA. 
For comparison, in Fig.~\ref{fig:dSigmaFacilities} 
we show single-differential neutral- and charged-current (NC and 
CC) inclusive DIS cross sections for polarized electron-proton 
scattering as a function of \Qsq\ comparing the future facilities 
LHeC, FCC-$eh$, and EIC, with H1 data from the past HERA
collider~\cite{Aaron:2012qi}. For studies of electroweak 
effects, data at higher values of \Qsq\ will be particularly 
suitable. However, it is 
  expected, that also the EIC may contribute to electroweak
  physics~\cite{Zhao:2016rfu}.
The LHeC might be realized during the lifetime of the LHC and 
could start taking data in the 2030s, and it has recently been 
described as a realistic option in the EPPSU deliberation 
document~\cite{European:2720131}.
The newly proposed energy-recovery 
linac (ERL) for a high-quality electron beam, together with the 
high-luminosity upgrade of the LHC (HL-LHC), are expected to 
provide more than an order of magnitude increase in the reach towards 
higher \Qsq\ compared to HERA and furthermore an extraordinary increase of the integrated luminosity compared to what was assumed in all previous studies. 
This motivates us to perform a novel exploratory study for the LHeC 
investigating new possibilities for the measurement of 
electroweak physics effects. Previously, studies of electroweak effects for 
similar energies have been performed for the 
LHeC~\cite{AbelleiraFernandez:2012cc} and earlier, to some extent, 
for the LEP$\otimes$LHC proposal \cite{Jarlskog:1990bk}. 

We will put the focus on the measurement prospects 
of inclusive NC and CC 
cross sections at the LHeC with the aim to determine parameters 
of the electroweak interaction by analysing pseudo-data 
which we simulated with different assumptions on the experimental 
uncertainties or the center-of-mass energy. 
Measurements in the 
regime of space-like momentum transfer, where the interaction is 
mediated by gauge boson exchange in the $t$-channel, are essentially 
complementary to other experiments, such as proton-proton collisions 
or electron-positron annihilation, or experiments at lower energies, 
like neutrino or muon scattering. The potential of experiments 
at the LHeC with exclusive final states, for example $W$- or 
$Z$-boson production, or production of the Higgs boson, has 
been studied 
elsewhere~\cite{Baur:1991pp,Baur:1989gh,Blumlein:1992eh,Li:2017kfk};
the possible improvement in our knowledge of parton distribution 
functions due to LHeC experiments was described in 
Refs.~\cite{Klein:1564929,AbdulKhalek:2019mps}
(see also 
Refs.~\cite{AbelleiraFernandez:2012cc,Abada:2019lih,Bruning:2706220} 
and references therein).

Our goal is to study tests of the electroweak SM. We therefore 
start with laying out the theoretical framework and summarize the 
SM predictions for NC and CC DIS cross sections, including 
higher-order electroweak corrections in the following 
Sec.~\ref{sec:theo}.
In subsequent sections we describe the main 
features of the cross section predictions 
(Sec.\,\ref{sec:xsections}), the simulated data that we use 
(Sec.\,\ref{sec:data}), and the methodology for fitting these 
data to extract electroweak physics parameters  
(Sec.\,\ref{sec:fitmethods}). Then we present a first group of 
results in Sec.\,\ref{sec:mass} for the determination of mass 
parameters, i.e.\ the masses of the $W$ and $Z$ bosons and in 
Sec.\,\ref{sec:sw2eff} for the weak mixing angle. The expected high 
precision of measurements at the LHeC motivates to also 
envisage an indirect determination of the top-quark mass through 
higher-order corrections (Sec.\,\ref{sec:mt}). These studies 
will allow one to perform tests of the SM by comparing 
different determinations of the electroweak physics parameters. 

A high precision measurement of parameters of the SM is important 
in order to study the validity of the theory of electroweak interactions.
In addition, we will study a number of possible ways to 
generically parameterize new physics beyond the SM
In Sec.\,\ref{sec:stu} 
we study the well-known $STU$-parameters which describe new 
physics entering through loop insertions in the self energy 
corrections of the gauge bosons. Then we follow the wide-spread 
convention to generalize the SM gauge-boson fermion couplings 
by introducing $\rho$ and $\kappa$ parameters, both for NC 
(Sec.\,\ref{sec:rhokappa}) and for CC (Sec.\,\ref{sec:ewcc}), 
or, eventually, allowing the vector and axial-vector coupling 
constants to be independent free parameters, not obeying any 
restriction as imposed by the SM (Sec.\,\ref{sec:nccouplings}). 
We will be able to show that in particular the quark coupling 
constants, separately for up- and down-type quarks, can be 
determined with a precision at the sub-percent level. The 
large kinematic reach of the LHeC will also allow us to 
study the scale-, i.e.\ \Qsq-dependence of coupling parameters. 
This opportunity is in fact unique to the LHeC. 
Finally, we conclude and summarize the most important results 
in Sec.\,\ref{sec:conclusions}. The impact of the LHeC 
measurements on possible future global fits of the electroweak 
SM parameters is discussed in an appendix (\ref{app:globalfit}). 

A summary of our results is also part of the description of 
the electroweak physics potential within the forthcoming publication 
of the update~\cite{Bruning:2706220} of the 2012 Conceptual Design 
Report on the LHeC.

\section{Electroweak effects in inclusive NC and CC DIS}
\label{sec:theo}

In this section we lay out the general 
properties of DIS cross sections, first at leading order, 
taking into account single boson exchange diagrams at tree level. 

Inclusive NC DIS cross sections are expressed in terms of
generalized structure functions $\tilde{F}_2^\pm$, 
$x\tilde{F}_3^\pm$ and $\tilde{F}_{\rm L}^\pm$ at electroweak 
(EW) leading order (LO) as
\begin{equation}
\frac{d^2\sigma^{\rm NC}(e^\pm p)}{dxd\Qsq} 
= 
\frac{2\pi\alpha^2}{xQ^4} 
\left[Y_+\tilde{F}_2^\pm(x,\Qsq) 
     \mp Y_{-} x\tilde{F}_3^\pm(x,\Qsq) 
     - y^2 \tilde{F}_{\rm L}^\pm(x,\Qsq) 
\right]~,
\label{eq:cs}
\end{equation}
where $\alpha$ denotes the fine structure constant, $x$ is the
Bjorken scaling variable, and $y$ the inelasticity. The factors 
$Y_\pm = 1\pm(1-y)^2$ encode the helicity dependence of the 
underlying lepton quark hard-scattering process. The generalized 
structure functions can be separated into contributions from 
pure $\gamma$- and $Z$-exchange, and their 
interference~\cite{Klein:1983vs}:
\begin{eqnarray}
  \tilde{F}_2^\pm
  &=& F_2
  -(\ve\pm P_e\gae)\varkappa_ZF_2^{\gamma Z}
  +\left[(\ve\ve+\gae\gae)\pm2P_e\ve\gae\right]\varkappa_Z^2F_2^Z~,
\label{eq:strfun1}
  \\
  \tilde{F}_3^\pm
  &=& ~~~~
  -(\gae\pm P_e\ve)\varkappa_ZF_3^{\gamma Z}
  +\left[2\ve\gae\pm P_e(\ve\ve+\gae\gae)\right]\varkappa_Z^2F_3^Z~,
\label{eq:strfun2}
\end{eqnarray}
where $P_e$ is the degree of longitudinal polarization ($P_e = -1$ 
for a purely left-handed polarized electron beam). 
A similar decomposition exists for $\tilde{F}_L$.
The naive quark-parton model corresponds to the LO approximation 
of Quantum Chromodynamics (QCD). In this approximation the 
structure functions are calculated from quark and anti-quark 
parton distribution functions, $q(x)$ and 
$\bar{q}(x)$: 
\begin{eqnarray}
  \left[F_2,F_2^{\gamma Z},F_2^Z\right]
  &=& 
  x\sum_q\left[Q_q^2,2Q_q\vq,\vq\vq+\aq\aq \right]\{q+\bar{q}\}~, 
\label{eq:last1}
  \\
  x\left[F_3^{\gamma Z},F_3^Z\right]
  &=& 
  x\sum_q\left[2Q_q\aq,2\vq\aq\right]\{q-\bar{q}\}~. 
\label{eq:last2}
\end{eqnarray}
In Eqs.~\eqref{eq:strfun1} and~\eqref{eq:strfun2}, the 
coefficient $\varkappa_Z$ accounts for the $Z$-boson 
propagator and the normalization of the weak, relative to the 
electromagnetic, interaction. It is calculated, at LO, as 
\begin{equation}
  \varkappa_Z(\Qsq)
  = \frac{\Qsq}{\Qsq+m^2_Z}
  \frac{1}{4\sw \cos^2\theta_W}
  = \frac{\Qsq}{\Qsq+m^2_Z}
  \frac{\gf m_Z^2}{2\sqrt{2}\pi\alpha}~. 
\label{eq:kappaZ-LO}
\end{equation}
Thus, depending on the choice of independent theory parameters, 
the normalization of $\varkappa_Z$ is fixed by an input value 
for $\sw$, or, alternatively, using the Fermi coupling constant 
$\gf$. The first option where $\sw = 1 - \cos^2\theta_W = 
1 - m_W^2/m_Z^2$ is fixed, is called the {\em on-shell scheme}, 
while the second option with $\gf$ as input parameter is known 
as the {\em modified on-shell scheme}. 

The vector and axial-vector coupling constants of the lepton 
or quark to the $Z$-boson, $g_V^f$ and $g_A^f$ (with $f = e$, 
$q$ and $q = u$, $d$) in Eqs.~\eqref{eq:strfun1} 
and~\eqref{eq:strfun2}, are given by the SM electroweak theory. 
They depend on the electric charge, $Q_f$, in units of the 
positron charge, and on the third component of the weak-isospin 
of the fermion, $I^3_{{\rm L},f}$. They are given, at LO, by
\begin{eqnarray}
  g_A^{f} 
  &=& I^3_{{\rm L},f}
\label{eq:gA-LO} \,, \\
  g_V^{f} 
  &=& I^3_{{\rm L},f} - 2 Q_{f} \sw \, . 
\label{eq:gV-LO} 
\end{eqnarray} 

The CC DIS cross section is written, in the LO approximation, as 
\begin{eqnarray}
  \frac{d^2\sigma^{\rm CC}(e^\pm p)}{dxd\Qsq}
  &=& 
  \frac{1 \pm P_e}{2}
  \frac{\pi \alpha^2}{4\sin^4\theta_W} 
  \frac{1}{x}
  \left[\frac{1}{\Qsq+m_W^2}\right]^2 \times
  \nonumber \\[1ex] 
  &&
  \left(Y_+ W_2^\pm(x,\Qsq) \mp Y_{-} xW_3^\pm(x,\Qsq)
  - y^2 W_{\rm L}^\pm(x,\Qsq)\right)~.
\label{eq:cc-cs}
\end{eqnarray}
Here, an incoming electron can scatter only with positively 
charged quarks. Therefore, in the naive quark-parton model the 
structure functions $W_2^\pm$ and $xW_3^\pm$ are obtained 
from parton distribution functions for up-type quarks and 
down-type anti-quarks as 
\begin{equation}
  W_2^- =
  x \left( U + \overline{D} \right)
  \, ,
  \quad xW_3^- =
  x \left( U - \overline{D} \right)
  \, ,
\label{eq:w23el-LO}
\end{equation}
where $U = u+c$ and $\overline{D} = \bar{d} + \bar{s}$. For 
positron scattering, the combinations $\overline{U} = \bar{u} 
+ \bar{c}$ and $D = d+s$ are needed and one has 
\begin{equation}
  W_2^+ =
  x \left( \overline{U} + D \right)
  \, , 
  \quad xW_3^+ =
  x \left( D - \overline{U} \right)
  \, . 
\label{eq:w23po-LO}
\end{equation}
At LO of QCD, one has for the longitudinal structure function 
$W_{\rm L}^\pm = 0$. 

Higher-order perturbative corrections of QCD are included in 
the $\overline{\rm MS}$ scheme by using $Q^2$-dependent parton 
distribution functions, $q(x, Q^2)$ and $\bar{q}(x, Q^2)$, 
evolved according to the 
Dokshitzer-Gribov-Lipatov-Altarelli-Parisi equations. 
In addition, there are corrections of order $O(\alpha_s)$ to 
the relations (\ref{eq:last1}, \ref{eq:last2}) and 
(\ref{eq:w23el-LO}, \ref{eq:w23po-LO}) between PDFs and 
structure functions, and the longitudinal structure functions 
for NC and CC are predictions of perturbative QCD. 

We will see below that the precision of LHeC measurements 
is expected to be at a level which makes the inclusion of 
higher-order electroweak corrections indispensable. In particular, 
QED radiative corrections (bremsstrahlung) have to be taken into 
account. We assume that these corrections will be removed from 
the data at the required level of precision. 
One-loop EW corrections have been calculated in 
Refs.~\cite{Bohm:1986na,Bardin:1988by,Hollik:1992bz}
for NC and in Refs.~\cite{Bohm:1987cg,Bardin:1989vz} for CC
scattering (see also ref.~\cite{Heinemann:1998kk} for a study 
of numerical results). We have adapted the implementation in 
the program EPRC~\cite{Spiesberger:1995pr} for our present 
study. 

The dominating universal higher-order EW corrections can be 
described by a modification of the fermion gauge-boson couplings. 
For NC scattering, vacuum polarization leads to the running of 
the fine structure constant. The NC couplings are affected by 
$\gamma Z$ mixing and $Z$ self energy corrections. These 
corrections are taken into account by replacing 
Eqs.~(\ref{eq:gA-LO}, \ref{eq:gV-LO}) with corrected couplings 
\begin{eqnarray}
  g_A^{f} 
  &=& \sqrt{\rho_{\text{NC}, f}} \, I^3_{{\rm L},f}
\label{eq:gA-NLO} \,, \\
  g_V^{f} 
  &=& \sqrt{\rho_{\text{NC}, f}} \, \left(I^3_{{\rm L},f} - 2
  Q_{f} \, \kappa_{f} \, \sw \right)
\label{eq:gV-NLO} \,.
\end{eqnarray} 
At LO, the coefficients $\rho_{\text{NC}, f}$ and 
$\kappa_{f}$ are unity, but at NLO they are promoted 
to form factors which are flavor and scale dependent. Since they 
depend on \Qsq, they render the coupling constants `effective' 
running couplings. The coefficient $\kappa_{f}$ can 
be combined with \sw\ to define an effective, flavor and 
scale-dependent ($\mu^2$)
weak mixing angle, 
\begin{equation}
  \sin^2 \theta_{W,f}^{\rm eff} (\mu^2) = 
  \kappa_{f}(\mu^2) \sw \, . 
\label{eq:sin2w-eff}
\end{equation} 
The leptonic weak mixing angle, $\sweffl(m_Z^2)$,
has been used to describe LEP/SLD observables at the 
$Z$-pole (see e.g.~\cite{PDG2020}). We emphasize that the $\mu^2$ 
dependence of the effective weak mixing angle is not negligible 
for LHeC physics ($\mu^2 = - Q^2$), while only its value at and close 
to $\mu^2 = +m_Z^2$ was relevant for $Z$-pole observables. 

For CC scattering, a corresponding correction factor 
$\rho_{\text{CC}, eq}$ is introduced for $e^- q$ and 
$e^+ \bar{q}$ scattering, and $\rho_{\text{CC}, e\bar{q}}$ 
for $e^- \bar{q}$ and $e^+ q$ scattering, by the replacement 
of Eqs.~(\ref{eq:w23el-LO}, \ref{eq:w23po-LO}) with 
\begin{equation}
  W_2^- =
  x \left( \rho_{\text{CC}, eq}^2 U 
  + \rho_{\text{CC},e\bar{q}}^2 \overline{D} \right)
  \, ,
  \quad xW_3^- =
  x \left( \rho_{\text{CC}, eq}^2 U 
  - \rho_{\text{CC},e\bar{q}}^2 \overline{D} \right)
  \, ,
\label{eq:w23el-NLO}
\end{equation}
and 
\begin{equation}
  W_2^+ =
  x \left( \rho_{\text{CC},e\bar{q}}^2 \overline{U} 
  + \rho_{\text{CC}, eq}^2 D \right)
  \, , 
  \quad xW_3^+ =
  x \left( \rho_{\text{CC},e\bar{q}}^2 D 
  - \overline{U} \rho_{\text{CC}, eq}^2 \right)
  \, . 
\label{eq:w23po-NLO}
\end{equation}
In addition, box graph corrections, which are $Q^2$- and 
energy-dependent, are added as separate correction terms to the 
NC and CC cross sections. Higher-order EW corrections are defined 
in the on-shell scheme~\cite{Sirlin:1980nh,Sirlin:1983ys}, 
using \mz\ and \mw\ as independent parameters (see also 
Refs.~\cite{Bohm:1986rj,Hollik:1988ii}).

In order to calculate predictions in the SM electroweak theory 
at LO, only two independent parameters are needed in addition 
to $\alpha$. At higher orders, loop corrections involve a 
non-negligible dependence on the complete set of SM parameters, 
where the most important ones are the top-quark mass, $m_t$, 
and the Higgs-boson mass, $m_H$. In addition, hadronic 
contributions to the running of the effective couplings 
have to be provided as independent 
input~\cite{Jegerlehner:1985gq,Jegerlehner:2011mw}, since the 
corresponding higher-order corrections can not be calculated 
in perturbation theory. 

In the on-shell scheme, the masses of all particles are 
taken as independent input parameters. The weak mixing angle 
is defined by the masses of the $W$ and $Z$ bosons, 
$\sw = 1-m_W^2/m_Z^2$, also at NLO. Since the Fermi constant 
\gf\ has been measured with a very high precision
in muon-decay experiments~\cite{Tishchenko:2012ie} it is 
often preferred to calculate the less well-known $W$-boson 
mass from the relation 
\begin{equation}
  \gf = \frac{\pi \alpha}{\sqrt{2} m_W^2} \frac{1}{\sw} 
  \frac{1}{1 - \Delta r} \, , 
\label{eq:deltar-NLO}
\end{equation}
where higher-order corrections enter through the
quantity $\dr =$ $\dr(\alpha,$ $m_W$, $m_Z$, $m_t$, $m_H, \ldots)$
\cite{Sirlin:1980nh}, which depends on all mass parameters 
of the EW SM. The correction \dr\ has also to be taken into 
account when the propagator factor $\varkappa_Z(\Qsq)$ (see 
Eq.~(\ref{eq:kappaZ-LO})) is calculated, using either 
$\alpha$, $m_W$ and $m_Z$ (the naive on-shell scheme), or 
$\alpha$, \gf\ and $m_Z$ (the modified on-shell scheme) to 
fix input parameters. The choice of a scheme for input 
parameters is important since it leads to very different 
sensitivities to parameter variations.


\section{Inclusive DIS cross sections at the LHeC}
\label{sec:xsections}

\begin{figure}[t!b]
  \centering
  \includegraphics[width=0.56\textwidth,trim=0 10 0 40,clip]{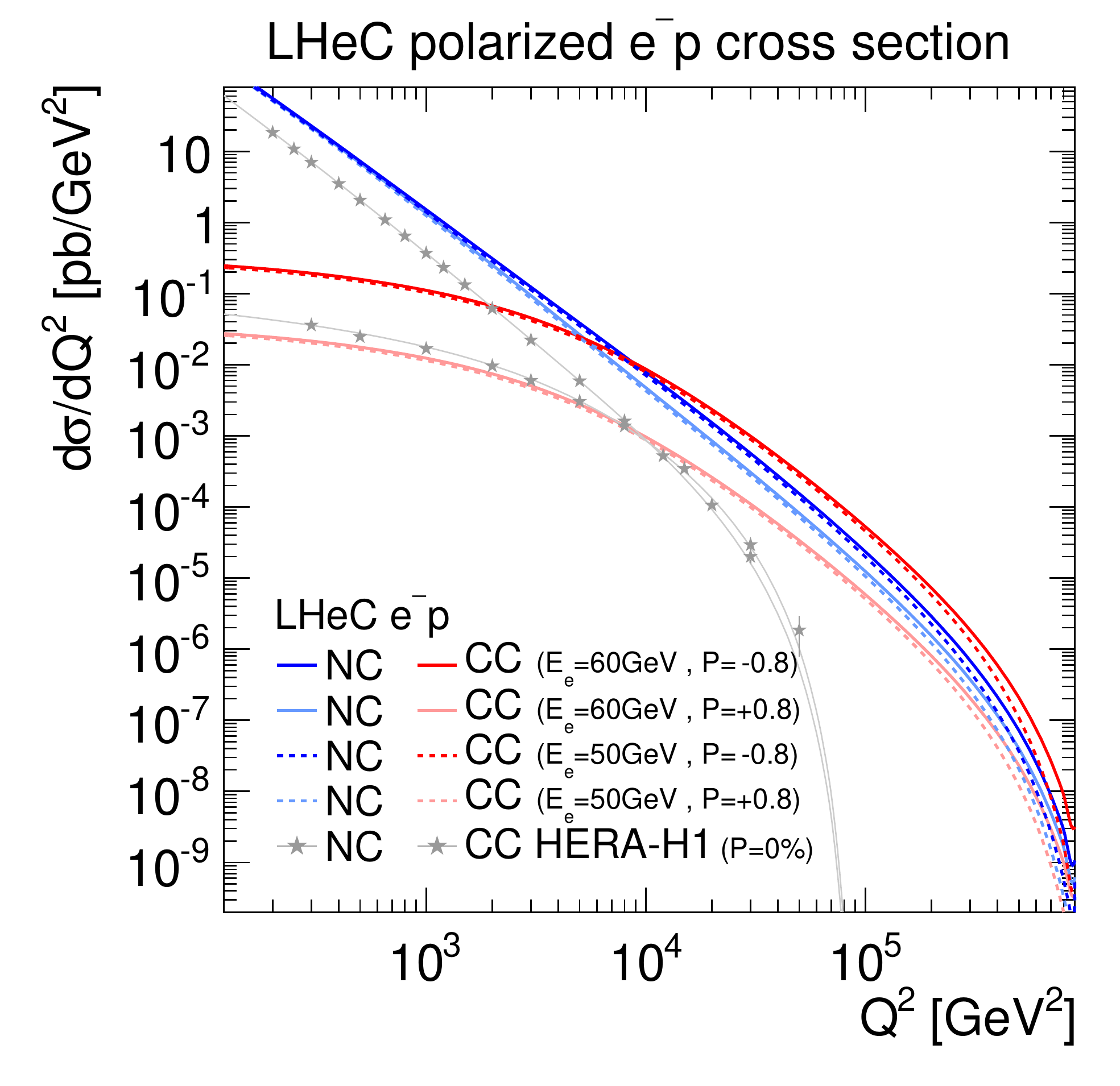}
  \caption{
      Single differential inclusive DIS cross sections for
      polarized $e^-p$ NC and CC DIS at the LHeC for two different 
      electron beam energies ($E_e=50$ and 60~\GeV). Cross 
      sections for longitudinal electron beam polarizations of 
      $P_e=-0.8$ and $+0.8$ are displayed. For comparison also 
      data measured by H1 at HERA~\cite{Aaron:2012qi}
      at center-of-mass energies of $\sqrt{s}=920~\GeV$ with 
      unpolarized ($P=0$) electron beams are displayed. 
      }
\label{fig:dSigma}
\end{figure}

The contribution of the weak interaction to inclusive NC and CC
DIS cross sections becomes large at high momentum transfers squared
and competes with the purely electromagnetic neutral current interaction. This 
is most clearly illustrated in Fig.~\ref{fig:dSigma} where we 
show predictions for the single-differential cross sections for 
polarized $e^-p$ scattering as a function of \Qsq. Here, LHeC 
electron beam energies of $E_e=50~\GeV$ and 60~\GeV, and a 
proton beam energy of $E_p=7000~\GeV$ are chosen. The LHeC 
predictions are compared to data for unpolarized scattering 
measured at HERA, where the electron and proton beam energies 
had been $E_e = 27.6$~GeV and $E_p = 920$~GeV, respectively.  

At lower values of \Qsq, the NC cross section is dominated by
the photon-exchange contribution, determined by the structure 
function $F_2$ (cf.\ Eqs.~(\ref{eq:strfun1}, \ref{eq:strfun2})), 
and much larger than the cross section for CC scattering. 
At values of \Qsq below the mass of the $W$ 
boson, $\Qsq\ll\mW^2$, the propagator term in the CC cross 
section becomes $\mW^2 / (\mW^2+\Qsq) \simeq 1$ and, 
therefore, the CC cross section depends only little on \Qsq.

Weak contributions to the NC cross section become important 
at \Qsq\ values around the electroweak scale, $\Qsq\approx\mZ^2$. 
As a consequence, the dependence of the NC cross section on 
the longitudinal beam polarization, $P_e$, becomes strong, and 
the cross sections for positive and negative helicities differ 
significantly. Since CC scattering is purely left-handed, the 
dependence on the longitudinal beam polarization is strongest 
in this case: the CC cross section scales linearly with the 
fraction of left-handed electrons in the beam, i.e.\ with 
$1 - P_e$ (cf.\, Eq.~\eqref{eq:cc-cs}). Note that, since DIS 
is mediated by gauge boson exchange with spacelike momentum 
transfer, $\mu^2=-\Qsq$, no resonance of a weak boson is present in the 
\Qsq-dependent cross section. 

The cross sections increase slowly with the center-of-mass 
energy, mainly because the reach towards smaller values of 
the Bjorken variable $x$ gets larger. For an electron beam 
energy of $E_e=60~\GeV$, the cross sections for NC or CC 
scattering in the typical range of \Qsq in $10\,000 < \Qsq 
< 100\,000~\GeVsq$ are larger by about 10 to 15~\%, compared 
to the case of $E_e=50~\GeV$. The difference of cross sections 
between $E_e=50$ and 60~\GeV increases with \Qsq.


\section{LHeC pseudo data}
\label{sec:data}

In this section, details of the simulation of LHeC pseudo-data\,\footnote{In 
   the following, the simulated \emph{pseudo-data} is simply 
   denoted as \emph{data} in order to facilitate reading.}
used subsequently for an extraction of electroweak parameters are 
described.

In the present analysis simulated double-differential inclusive 
NC and CC DIS cross section data are exploited.
The data have been simulated based on a numerical
procedure~\cite{Blumlein:1990dj} for the purpose of the LHeC CDR
update~\cite{Bruning:2706220}.
The data are briefly described in the following.

The data sets include electron and positron scattering, 
different lepton beam polarization settings, and different 
proton beam energies. Since a decision about the actual layout 
of the LHeC energy-recovery linac for the lepton beam has not 
yet been taken, we will study scenarios for two lepton beam
energies, i.e.\ $E_e = 50$ and 60\,\GeV. Most of the 
data were generated with the nominal LHC proton beam energy 
of $E_p = 7000\,\GeV$, but in addition, a small sample with 
reduced proton energy of $E_p = 1000$\,GeV is also considered.
A summary of the data sets is given in Tab.~\ref{tab:datasets}.

\begin{table}[tbh!]
  \centering
  \small
  \begin{tabular}{lccccccc}
    \toprule
    Processes & $E_p$ & $Q_e$ & $P_e$ &
    $\mathcal{L}$ & \Qsq\ range & 
    \multicolumn{2}{c}{No. of data points (NC, CC)} \\
    \cmidrule(lr){7-8}
    &  [TeV] & & &  [fb$^{-1}$]&  [GeV$^2$] & LHeC-60 & LHeC-50  \\
    \midrule
    NC, CC &  7  & $-1$ &  $-0.8$ & 1000 & 5 -- $10^6$ 
       & 150, 114 & 150, 123
    \\ 
    NC, CC &  7  & $-1$ &  $+0.8$ & 10   & 5 -- $10^6$  
       & 150, 113 & 146, 117
    \\ 
    NC, CC &  7  & $+1$ &  $ 0  $ & 10   & 5 -- $5\cdot10^5$  
       & 148, 109 & 145, 111
    \\ 
    NC, CC &  1  & $-1$ &  $ 0  $ & 1    & 5 -- $10^5$  
       & 128, ~~93 & 120, ~~92
    \\ 
    \bottomrule
  \end{tabular}
  \caption{
    Summary of data sets used in our analysis. Each set is 
    simulated for the two electron beam energies $E_e=50~\GeV$ 
    and 60~\GeV. 
  }
  \label{tab:datasets}
\end{table}

The majority of the data will be collected with an electron
beam ($Q_e=-1$) and with a longitudinal beam polarization
of $P_e=-0.8$, expected to reach an integrated luminosity of about
$\mathcal{L}\simeq1000~\textrm{fb}^{-1}$. This will allow 
us to consider measurements of NC and CC DIS cross sections 
up to values 
of $\Qsq\simeq1\,000\,000\,\GeVsq$. A considerably smaller 
data sample will be collected with a positive electron beam 
polarization of $P_e=+0.8$, i.e.\ with right-handed electrons. 
For this sample, an integrated luminosity of 10\,fb$^{-1}$ 
was assumed.
Another data sample may be collected with a 
positron beam, where it is assumed that polarization will not 
be available. Technical limitations for the positron source 
put constraints on the achievable beam current and thus on 
the instantaneous luminosity. Therefore, an integrated luminosity 
of 10\,fb$^{-1}$ is assumed for this sample\footnote{This luminosity value may eventually be smaller
 due to difficulties to generate intense positron beams.}. 
Such reduced luminosity values  still allow
to consider measurements with 
positrons reaching up to \Qsq\ values of 500\,000\,\GeVsq. Finally, 
another data sample will be collected with a reduced proton beam
energy. This will be important for a determination of $F_L$ 
and to access higher values of $x$ at fixed medium $Q^2$. For this low-energy sample 
an integrated luminosity of 1\,fb$^{-1}$ was assumed. 

The analysis of all data sets is restricted to $\Qsq\geq5\,\GeVsq$ in order 
to avoid regions where non-perturbative QCD effects are important, 
which could deteriorate the determinations of parton distribution 
functions. For our purpose, the low-\Qsq region is anyway of 
less interest since it does not contribute much to the 
sensitivity to EW parameters. CC DIS data are simulated only for 
$\Qsq\geq100\,\GeVsq$, since CC scattering events with
significantly lower \Qsq 
may be difficult to measure due to limitations of the trigger 
system.

The data simulation accounts for the acceptance of the LHeC
detector, the kinematic reconstruction, and trigger restrictions. 
The resulting coverage of the kinematic plane can be found, for
instance, in Ref.~\cite{AbdulKhalek:2019mps}. 

\begin{table}[tbh!]
  \centering
  \small
  \begin{tabular}{lccc}
    \toprule
    Source of uncertainty & Size of uncertainty & \multicolumn{2}{c}{Uncertainty on cross section} \\
    \cmidrule(lr){3-4}
    &  & $\Delta\sigma_\textrm{NC}$ & $\Delta\sigma_\textrm{CC}$ \\
    \midrule
    Scattered electron energy scale $\Delta E_e' /E_e'$ & 0.1 \%  & 0.1 -- 1.7\,\% & --  \\
    Scattered electron polar angle  & 0.1\,mrad  &   0.1 -- 0.7\,\% &  --  \\
    Hadronic energy scale $\Delta E_h /E_h$ & 0.5\,\%  & 0.1 -- 4\,\% &  1.0 -- 8.6\,\%  \\
    Calorimeter noise (only $y < 0.01$) &   & 0.0 -- 1.1\,\% &  included above  \\
    \addlinespace    
    Radiative corrections &  &  0.3\,\%  & -- \\
    Photoproduction background ($y > 0.5$) & 1\,\%  & 0.0 or 1.0\,\% &  --  \\
    Uncorrelated uncertainty (efficiency) &  &  0.5\,\% &   0.5\,\%  \\ 
    Luminosity uncertainty (normalization) & & 1.0\,\%   &  1.0\,\%    \\
    \bottomrule
 \end{tabular}
\caption{
  Summary of the assumptions for uncertainties from various 
  sources used in the simulation of the NC and CC cross sections.
  The first three items are calibration uncertainties and 
  affect the event reconstruction. The last four items are 
  uncertainties which can be assigned directly to the cross 
  section. 
}
\label{tab:uncert}
\end{table}

The data include a full set of systematic uncertainties and
the individual sources are summarized in Tab.~\ref{tab:uncert}. 
For the bulk of the phase space, the `electron' 
reconstruction method is used where the kinematic variables $x$ and 
$Q^2$ are determined from the energy and polar angle of the 
scattered electron. Important uncertainties originate from the 
electron energy scale and polar angle measurement, and
uncertainties of $\Delta E_e^\prime / E_e^\prime = 0.1~\%$ and
$\Delta\theta^\prime_e = 0.1$~mrad are assumed. However, at lower 
values of $y$ the electron method leads to a deterioration of 
the measurement resolution $\propto 1/y$. Thus one has to exploit
the hadronic final state in the determination of the inelasticity.
The present simulation accounts for this by using a simple
`mixed' (i.e. $Q^2_e,~y_h$) reconstruction
method~\cite{Blumlein:1990dj} to
determine  $x=\Qsq/(sy)$. 
For the measurement of the hadronic final state, an uncertainty 
on the hadronic energy scale of $\Delta E_h/E_h = 0.5~\%$
is imposed.
As discussed in Sec.~\ref{sec:theo}, for the analysis of inclusive DIS
data, we expect that these data are corrected for QED radiative
corrections, including QED bremsstrahlung off the lepton, photonic
lepton-vertex corrections, self-energy contributions at the external
lepton lines, and fermionic contributions to the running of the
finestructure constant.
An uncertainty of 0.3~\% on the cross sections due to these
corrections is expected and the resulting experimental uncertainties
will therefore comprise a corresponding contribution.
An uncertainty due to the background from photoproduction events of 
$1.0~\%$ in the high-$y$ region is assumed. The statistical 
uncertainty of each data point is taken to be at least 
0.1~\%. A global normalization uncertainty of 1~\% is taken into 
account, which includes the luminosity uncertainty. 

Finally, 
potential additional sources of measurement errors are combined 
in an uncorrelated uncertainty component of 0.5\,\%.
These may comprise unfolding and model uncertainties, efficiency
uncertainties, beam background related uncertainties, possible
small stochastic uncertainties related to the calibration procedure,
or uncorrelated components of any of the above sources. 
In fact, the actual size of this uncorrelated uncertainty is very 
difficult to estimate for the future LHeC, but we consider the
assumption of 0.5\,\% to be rather conservative.
In order to address the effect of the unknown size of the uncorrelated
uncertainty in some detail, we consider in the following two 
alternative scenarios, one
with an uncorrelated uncertainty of 0.5~\%, as well as one with a 
more optimistic value of 0.25\,\%.
These will be denoted in the following as the `a' or 'b' scenarios, 
respectively.
Our data samples have been simulated for simplicity
with an \emph{ad hoc} and rather coarse $x$-\Qsq
grid (see Tab.~\ref{tab:datasets}).
Yet, real data may allow a much finer 
binning, in particular at medium $x$ values or at higher \Qsq,
depending on the actual detector performance and its resolution.
In fact, the effect of a possibly finer binning may be simulated 
to a very good
approximation by changing the size of the uncorrelated uncertainty,
which would then be equivalent by comparing the `a' and `b' scenarios.  
The properties of the generated four sets of data samples are
summarized in Table\,\ref{tab:scenarios}.
\begin{table}[thb!]
  \centering
  \small
  \begin{tabular}{lccc}
    \toprule
    Scenario & $E_e$ & Uncorrelated uncertainty \\
    \midrule
    LHeC-50a  &   50\,\GeV&  0.5\,\%            \\
    LHeC-50b  &   50\,\GeV&  0.25\,\%           \\
    LHeC-60a  &   60\,\GeV&  0.5\,\%            \\
    LHeC-60b  &   60\,\GeV&  0.25\,\%            \\
    \bottomrule
 \end{tabular}
  \caption{
    Summary of the LHeC measurement scenarios. 
    The LHeC data scenarios differ by the assumption on the
    electron beam energy, $E_e$, and the assumptions made for
    the uncorrelated uncertainty (see text).
    They will be referred to by the names shown in the first column.
  }
  \label{tab:scenarios}
\end{table}

In previous similar studies (see, 
e.g.~\cite{Blumlein:1987fd,Spiesberger:1993jg}) it was often 
assumed that cross section ratios are measured.
These are for example the 
ratio of CC over NC cross sections, $R_{\rm CC/NC}$, the 
polarization asymmetry $A_{\rm LR}$, or the charge asymmetry 
$B_{\pm}$ measuring the difference between cross sections 
for electron and positron scattering.
In fact, our inclusive DIS data implicitly comprise such
a collection of cross section ratios, but while we do not construct
these ratios explicitly we instead leave it to the parameter
extraction procedure to exploit the corresponding information.
In the following, however, it is often informative to consider
these ratios for the purpose of exposing the parameter dependence 
and estimate the potential impact of data on the parameter determinations.
In such ratios, we can then expect that most of the correlated 
uncertainties, such as normalization errors, become largely 
constrained by the fit while uncorrelated uncertainties are 
reduced by taking the properly weighted average of all data. 
As a consequence, due to the large number of data points, 
in the order of a few hundred, 
uncertainties at the per mille level can be expected for the 
observables which we are going to study in the following. The 
two measurement scenarios labeled with `a' and `b' described 
above will help us to verify this estimate of expected 
parameter uncertainties and its dependence on our assumption 
for correlated measurement errors.


\section{Methodology of a combined EW+PDF fit}
\label{sec:fitmethods}

By the time when the LHeC is realized, one should expect that 
the determination of PDFs will be dominated by NC and CC DIS 
data obtained with it. The uncertainties of PDF parameterizations 
will mainly represent the propagated uncertainties of these 
inclusive LHeC data. The uncertainties of EW parameters determined 
from cross section data will therefore be correlated with PDF 
uncertainties. In order to account for these correlations, EW 
parameters have to be determined in a combined fit simultaneously 
with PDFs. This allows  the complete 
set of statistical, as well as correlated and uncorrelated 
systematic uncertainties to be taken into account. We denote such an approach in the 
following by `EW'+PDF fit, while `EW' may be replaced by the 
parameter of interest.

The $x$-dependence of the PDFs is parameterized at a starting 
scale of $\mu_0 = 1.3784~\GeV$, i.e.\ below the charm threshold. The NC and CC DIS inclusive cross sections
determine four independent combinations of quark distributions, besides the gluon. Here the following 
five PDFs are chosen as independent input at this scale: 
the $u$ and $d$ valence quark distributions ($xu$ and $xd$), 
the up-type and down-type anti-quark distributions ($x\overline{U}$ 
and $x\overline{D}$), and the gluon distribution ($xg$). The choice 
of the parameterization follows previous LHeC PDF
studies~\cite{AbelleiraFernandez:2012cc,Klein:1564929}, 
which are closely related to HERAPDF-style
PDFs~\cite{Adloff:2000qk,Abramowicz:2015mha,Andreev:2017vxu}. 
The following functional form is used: 
\begin{equation}
  xf = f_A x^{f_B} (1-x)^{f_C} (1+f_Dx+f_Ex^2) -
      f_{A^\prime}x^{f_{B^\prime}}(1-x)^{0.25}\,,
\end{equation}
where $f$ denotes any of the five input PDFs, $f = u$, $d$, 
$\overline{U}$, $\overline{D}$, $g$. The second term in this 
ansatz is taken into account only for the gluon 
distribution\footnote{
   The second term is commonly considered 
   to be of importance for PDF determinations as it introduces 
   additional freedom at lower values of $x$. This may be 
   important to describe LHeC data which probes the $x$ region 
   down to $x \simeq 5\cdot10^{-6}$. However, we find that 
   this has no significant impact on the resulting uncertainties 
   of the electroweak parameters.}, i.e.\ 
$u_{A^\prime} = d_{A^\prime} = \overline{U}_{A^\prime} = 
\overline{D}_{A^\prime} = 0$. The normalization of each PDF is 
determined by the quark number sum-rule ($u_A$, $d_A$) or 
the momentum sum-rules ($g_A$). For the anti-quark PDF, we fix 
$\overline{U}_A = \overline{D}_A(1-0.4)$. Furthermore, we use 
$\overline{D}_B = \overline{U}_B$. Altogether, 13 independent 
PDF parameters are determined in each fit ($g_B$, $g_C$, 
$g_{A^\prime}$, $g_{B^\prime}$, $u_B$, $u_C$, $u_E$, $d_B$, $d_C$, 
$\overline{U}_C$, $\overline{D}_A$, $\overline{D}_B$, 
$\overline{D}_C$). The values of the PDF parameters used for 
the generation of pseudo data are not of particular relevance 
here. They have been obtained from a private fit to HERA data, 
similar to Refs.~\cite{Aaron:2012qi,Andreev:2017vxu}. 

QCD higher-order corrections are taken into account at NNLO 
in the zero-mass variable flavor number scheme. They are 
implemented for the evolution of the PDFs and the calculation 
of the structure functions using 
QCDNUM~\cite{Botje:2010ay}. The strong coupling 
is fixed, $\alpha_s^{\overline{\text{MS}},N_f=5}(\mz) = 0.118$.  
We do not consider QED or EW corrections for the PDF 
evolution~\cite{DeRujula:1979grv,Kripfganz:1988bd,
Blumlein:1989gk,Spiesberger:1994dm}, since their impact on 
the determination of EW parameters is expected to be small 
and does not change the uncertainties estimated in the present 
study. EW effects are, however, included in the calculation 
of cross sections, as described above in Sec.~\ref{sec:theo}. 

The $\chi^2$ quantity which is subject to the minimization and
error propagation is based on normal-distributed relative
uncertainties,
\begin{equation}
  \chi^2 =\sum_{ij} \log{\frac{\varsigma_i}{\sigma_i}}V^{-1}_{ij}
  \log{\frac{\varsigma_j}{\sigma_j}}
\end{equation}
where the sum runs over all data points, $\varsigma_i$ are the 
measured cross section values and $\sigma_i$ their corresponding 
theory predictions (cf.\ Eqs.~\eqref{eq:cs} and~\eqref{eq:cc-cs}),
which incorporate the dependence on the fit parameters. The 
covariance matrix $V$ represents the relative uncertainties of 
the data points. The Minuit library is employed, and the resulting 
uncertainties of the fit parameters are calculated using the HESSE 
or MINOS algorithm~\cite{James:1975dr}. For our study, we set the 
data values equal to the predictions, i.e.\ our data represent 
an \emph{Asimov data set} \cite{Cowan:2010js} and resulting 
uncertainties refer to expected uncertainties. It is important 
to note that with the above definition of $\chi^2$ the actual 
value of the cross section at a given point does not enter the 
calculation of the uncertainties, but only the relative size of 
the uncertainties is of relevance.
A very similar methodology has previously been used by H1 for
the determination of the expected uncertainties in their 
electroweak analysis of
inclusive DIS data~\cite{Spiesberger:2018vki}.
We have validated that the 
uncertainties on the PDFs from a pure PDF fit, i.e.\ with fixed 
electroweak parameters, are in good agreement with dedicated PDF
studies based on the same LHeC data
samples~\cite{AbdulKhalek:2019mps,Bruning:2706220}.


\section{Weak boson masses}
\label{sec:mass}

First, we investigate the possibility to determine the 
fundamental parameters of the EW theory from LHeC inclusive 
DIS data. In this initial part of our study we are further
interested to understand how the uncertainty estimates 
depend on the assumptions of the simulated data. 

Since our analysis is based on theory predictions derived in the 
on-shell scheme, the free parameters at the LO are only the masses 
of the weak gauge bosons, $m_W$ and $m_Z$, and the fine structure 
constant $\alpha$. The latter is fixed in the analysis, i.e.\ it 
is considered to be known with ultimate precision. The weak mixing 
angle is defined by the ratio of the gauge boson masses
and thus not independent.
At higher orders, there is in addition 
a sensitivity to the top-quark and the Higgs-boson mass, which 
will be studied in a subsequent section. 

We determine the expected uncertainties for \mw\ in an \mw+PDF fit,
where the value of \mz\ is considered as an external input, e.g.\ 
taken from the LEP+SLD combined measurement~\cite{ALEPH:2005ab}.
For the $W$-boson mass parameter, we then find expected 
uncertainties of
\begin{eqnarray}
  \Delta\mW(\text{LHeC-60a}) 
  &=& 
  \pm8_{({\rm exp})}\pm5_{({\rm PDF})}\,\MeV  
  =\, \pm10_{\text{(tot)}}\,\MeV{\rm ~~~~and~~} 
  \\
  \Delta\mW(\text{LHeC-50a}) 
  &=& 
  \pm9_{({\rm exp})}\pm8_{({\rm PDF})}\,\MeV  
  =\, \pm12_{\text{(tot)}}\,\MeV
\nonumber
\end{eqnarray}
for the scenarios LHeC-60a and LHeC-50a (cf.\ 
section~\ref{sec:data}), and 
\begin{eqnarray}
  \Delta\mW(\text{LHeC-60b}) 
  &=& 
  \pm5_{({\rm exp})}\pm6_{({\rm PDF})}\,\MeV 
  =\, \pm6_{\text{(tot)}}\,\MeV{\rm ~~~~and~~} 
  \\
  \Delta\mW(\text{LHeC-50b}) 
  &=& 
  \pm6_{({\rm exp})}\pm6_{({\rm PDF})}\,\MeV 
  =\, \pm8_{\text{(tot)}}\,\MeV
\nonumber
\end{eqnarray}
for LHeC-60b and LHeC-50b, respectively. The breakdown of the
uncertainty into contributions due to systematic experimental and 
PDF uncertainties was obtained by repeating the fit with PDF 
parameters kept fixed, which yields the \emph{exp} 
uncertainty, while the \emph{PDF} uncertainty is then calculated as 
the quadratic difference from the total uncertainty. The size 
of the uncertainty component associated to the PDFs is found to 
be of similar size as the \emph{exp} uncertainty.

\begin{figure}[tb!]
  \centering
  \includegraphics[width=0.42\textwidth]{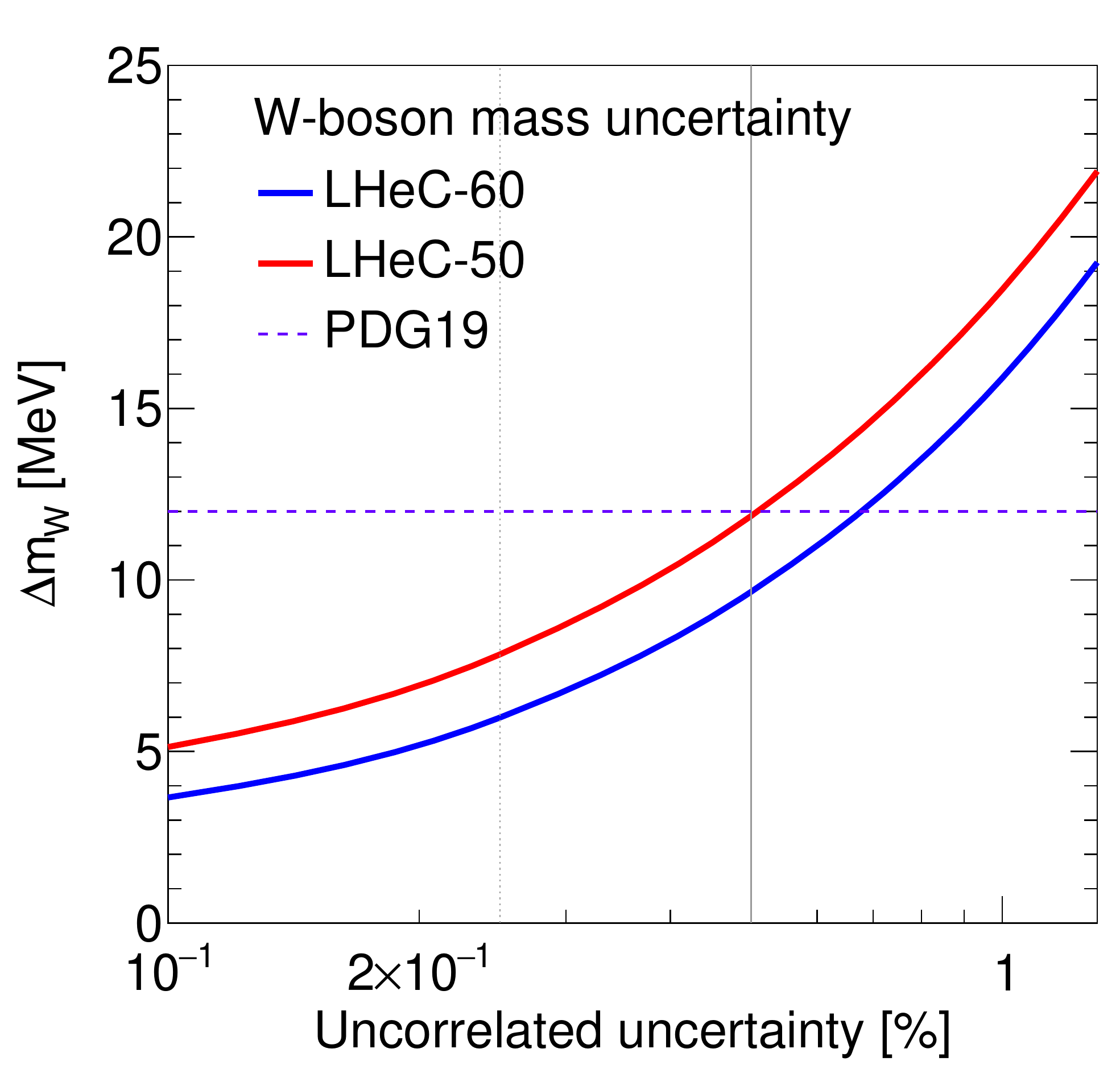}
  \hskip0.05\textwidth
  \includegraphics[width=0.42\textwidth]{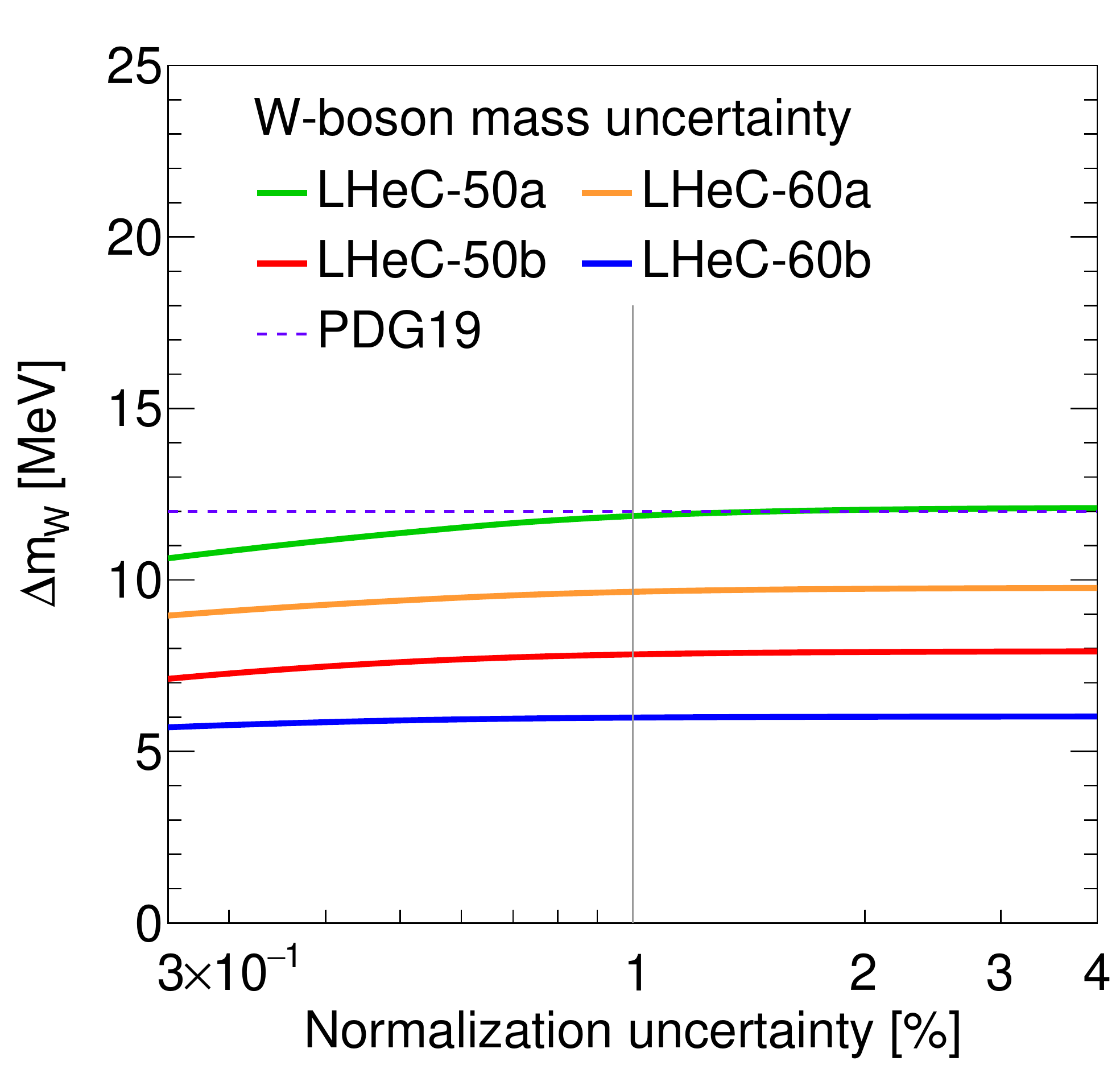}
  \caption{
    Left: The total uncertainty $\Delta m_W$ as a function of 
    the size of the uncorrelated uncertainty. The horizontal 
    line marks the uncertainty of the present world average.
    The `a' scenarios LHeC-60a and LHeC-50a (uncorrelated 
    uncertainty of 0.5~\%) and the `b' scenarios LHeC-60b and 
    LHeC-50b (0.25~\%) are indicated by vertical lines. 
    Right: The uncertainty $\Delta m_W$ as a function of the size 
    of the normalization uncertainty of the DIS cross sections.
    The nominal assumption of 1\,\% is indicated by a vertical line.
    All other systematic uncertainties are kept as listed in
    Tab.~\ref{tab:uncert}.
    }
    \label{fig:mW2}
\end{figure}

Altogether, we find a relative uncertainty for \mw\ of the order 
of $10^{-4}$, which is compatible with our rough initial estimate 
for cross section ratios\,\footnote{  
  In the previous section we have outlined that due to the large
  number of data points one expects relative uncertainties
  of a per mille for ratios of bin cross sections. Such cross 
  section ratios are determined by coefficients containing \sw\ 
  (see Eqs.~\eqref{eq:kappaZ-LO} and~\eqref{eq:cc-cs}). 
  Simple error propagation allows one to infer 
  $\Delta m_W / m_W = (\sw/2\cos^2\theta_W) (\Delta \sw /\sw)$.
  Therefore a factor of about $\sw/2\cos^2\theta_W \simeq 0.15$ 
  applies if the relative uncertainty of the cross section ratios
  is translated into an uncertainty of \mw. This results in a 
  relative uncertainty of 
  $\Delta\mw/\mw\simeq\mathcal{O}(10^{-4})$. 
}. 
The two scenarios (`a' and `b') differ by the assumption for the 
size of the single-bin uncorrelated uncertainty, but have 
otherwise the same experimental uncertainties. The dependence 
of $\Delta\mW$ on this uncertainty component is displayed 
in the left panel of Fig.~\ref{fig:mW2}.
Obviously, a good control of the uncorrelated uncertainty 
component will help to improve the precision of a potential 
$W$-boson mass determination. We re-iterate that a smaller 
uncorrelated uncertainty can be achieved through a higher 
resolution which allows one to choose finer binning of the 
data. In the right panel of Fig.~\ref{fig:mW2} we show how 
the uncertainty of $m_W$ depends on the cross section 
normalization uncertainty for the different LHeC scenarios. 
Obviously this component of the uncertainty for the cross 
section measurement cancels to a large extent, as already 
discussed in the previous section.
Other (correlated) systematic uncertainty components behave
similar as the normalization uncertainty.

\begin{figure}[tb!]
  \centering
  \includegraphics[width=0.48\textwidth,trim=0 0 0 40,clip]{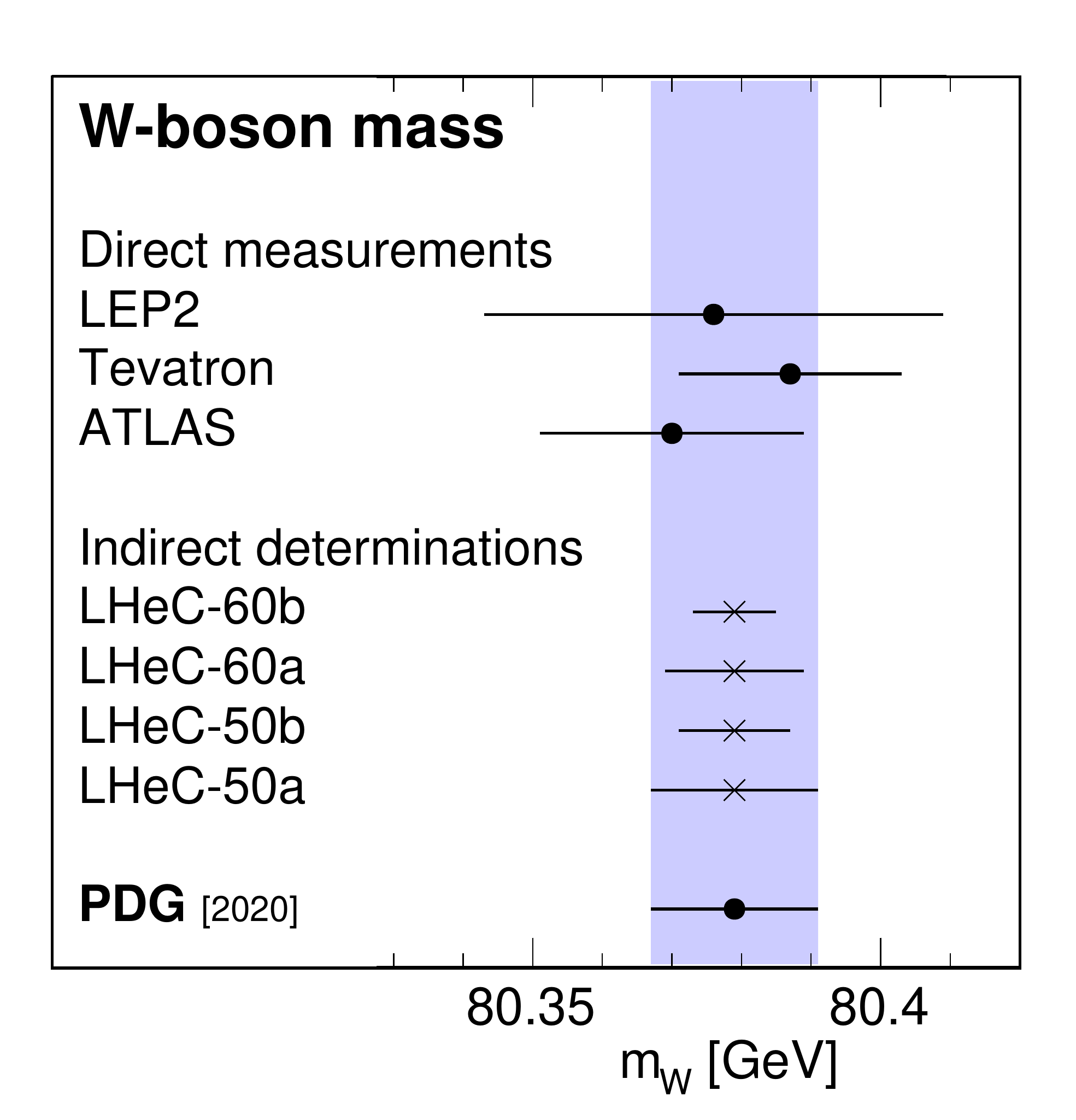}
  \caption{
    Determination of the $W$-boson mass from a combined 
    \mw+PDF fit, assuming fixed values for all other EW parameters. 
    Different LHeC scenarios with beam energies of $E_e=60~\GeV$ 
    and 50~\GeV as described in the text are considered and 
    compared with existing 
    measurements~\cite{Group:2012gb,Schael:2013ita,Aaboud:2017svj}
    and with the world average value 
    (PDG2020)~\cite{PDG2020}. 
  }
    \label{fig:mW}
\end{figure}
The expected uncertainties $\Delta\mw$ are displayed in
Fig.~\ref{fig:mW} and compared with the measurements\,\footnote{In
  Fig.~\ref{fig:mW}, the values from LEP2 and Tevatron represent
  combined results taking into account measurements from a number 
  of independent experiments. This procedure benefits from a 
  reduction of the systematic uncertainties. The same remark 
  applies to the PDG world average.} 
by LEP2~\cite{Schael:2013ita}, Tevatron~\cite{Group:2012gb}, 
ATLAS~\cite{Aaboud:2017svj} and the PDG~\cite{PDG2020}.
We conclude that the LHeC can be expected to yield a $W$-boson 
mass determination with the smallest experimental uncertainty 
from a single experiment. It will even be superior to the current 
world average. Therefore, when real data are available, a detailed 
assessment of associated theoretical uncertainties will be 
needed to determine the accurate central value of the $W$-boson 
mass. For example, a theoretical uncertainty due to the 
top-quark mass dependence entering through radiative corrections 
in \dr\ (see Eq.~(\ref{eq:deltar-NLO})) will have to be taken 
into account. Assuming $\Delta m_t = 0.5~\GeV$, one should expect 
an additional uncertainty of $\Delta\mW=2.5\,\MeV$. The estimate 
of experimental and PDF uncertainties given above, is, however, 
not sensitive by itself to higher-order corrections beyond NLO, while the actual values would be.

The high precision of the $W$-boson mass parameter requires an
in-depth discussion of its interpretation and the relation to 
other, more direct, measurements.
We find, that the sensitivity of the DIS cross sections to \mw\ arises
mainly from the weak mixing angle in the NC vector couplings $g_V^f$,
Eqs.~\eqref{eq:gV-NLO} (cf.\ also next section), 
whereas the contribution from the NC and CC normalization,
Eqs.~\eqref{eq:kappaZ-LO} and~\eqref{eq:cc-cs}, and from
the $W$-boson propagator term in CC DIS, $(1/(\Qsq+\mw^2))^2$, 
cf.\ Eq.~\eqref{eq:cc-cs},
is only small\,\footnote{A determination of \mw\ from the
  $W$-boson propagator alone yields an uncertainty of $\pm17$ or
  $\pm36\,\MeV$ for LHeC-60b or LHeC-50a, respectively.}. 
Therefore, the precise measurement of DIS 
cross sections yields primarily (only) an indirect determination 
of the SM mass parameters. In fact, the philosophy of this 
indirect parameter determination is similar to the one of the 
so-called `global EW  
fits'~\cite{Haller:2018nnx,deBlas:2016ojx,Erler:2019hds},
where a collection of observables is fitted 
to SM predictions calculated as a function of properly chosen 
free theory parameters. The `measurement' of \mw\ from inclusive 
DIS cross sections at the LHeC, therefore, provides a consistency 
check of the SM and is complementary to direct, true, mass 
measurements of the $W$-boson mass.


A determination of the $Z$-boson mass from an \mz+PDF fit yields 
expected experimental uncertainties of $\Delta\mZ=11$~MeV 
(13~MeV) for LHeC-60a (LHeC-50a), respectively. These 
uncertainties are of a similar size as those for \mW.
However, they cannot compete with the high precision 
measurements at the $Z$-pole by LEP+SLD~\cite{ALEPH:2005ab}. 
Moreover, future $e^+e^-$ colliders are expected to provide 
a substantial improvement of the precision of 
\mZ~\cite{Abada:2019lih,Abada:2019zxq,Fan:2014vta}.


\begin{figure}[tb!]
  \centering
  \includegraphics[width=0.54\textwidth,trim=0 10 0 25,clip]{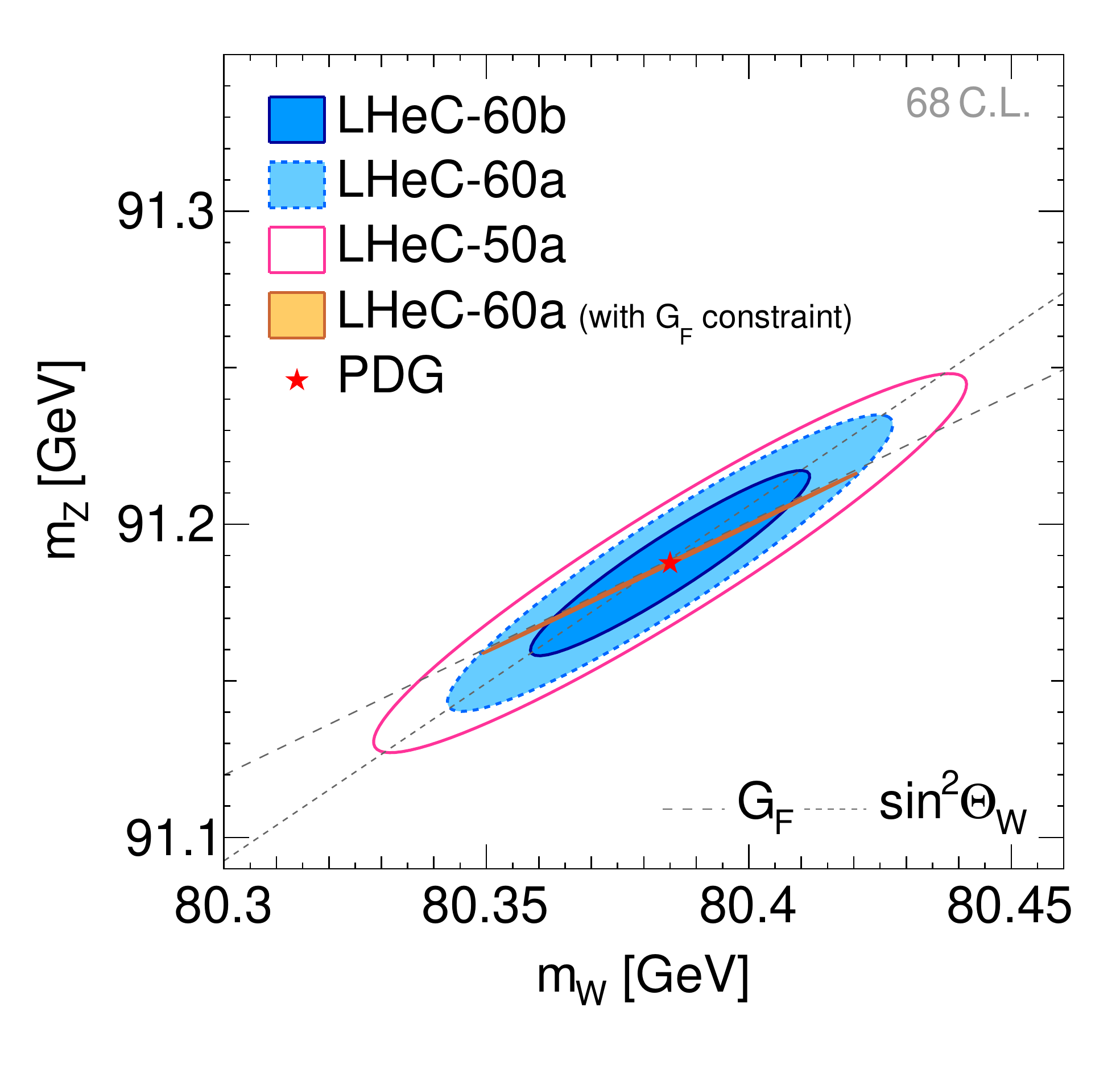}
  \caption{
    Simultaneous determination of the $Z$-boson and $W$-boson 
    masses $m_Z$ and \mw\ from LHeC-60a or LHeC-50a data.
    The additional precision measurement of \gf\ yields a
    strong constraint and its combination with the \mZ\ and \mW\ 
    determination leads to a very shallow ellipse. 
    }
\label{fig:mWmZ}
\end{figure}
Finally we investigate the possibility to perform a combined 
determination of \mW\ and \mZ. The result is shown in 
Fig.~\ref{fig:mWmZ}, where the 68\,\% confidence level
contours are displayed. The precision of \mW\ and \mZ\ if 
taken from the projections of these contours, is only moderate. 
However, the observed strong correlation provides a test of 
the high-energy behavior of the EW SM theory. Indeed, the 
68\,\% C.L.-contour is aligned along the line of a constant 
value of \sw\ (dotted line in Fig.~\ref{fig:mWmZ}). 
Imposing the additional constraint for the very precisely 
known value of $G_F$~\cite{Tishchenko:2012ie} (dashed line, 
see Refs.~\cite{Brisson:1991vj,Spiesberger:1993jg}) results in a 
very shallow ellipse (yellow). Real data would 
have to lead to a 
consistent picture of the different constraints shown in 
this figure. Their comparison provides a test for the 
consistency of high-energy data from the LHeC with low-energy 
input from $\alpha$, $G_F$ and $\sin^2\theta_W$.


\section{\boldmath The weak mixing angle \sw}
\label{sec:sw2eff}

In the SM, the weak neutral-current couplings of the fermions 
are fixed by one single parameter, i.e.\ through the weak mixing 
angle $\theta_\textrm{W}$. High-precision measurements of 
\sw\ in as many as possible different processes are therefore 
considered as a key to test and to restrict extensions of the SM. 
Therefore, we study in this section the prospects for a 
determination of \sw\ from DIS data at the LHeC, i.e.\ we 
assume the weak mixing angle in the fermion neutral-current 
couplings as a free fit parameter while all other parameters 
are fixed. This way we allow the weak neutral-current couplings 
to deviate from their SM values, however only in a correlated way, 
instead of allowing independent, flavor-dependent variations for 
vector and axial-vector couplings as we will do in a subsequent 
section. 

The highest precision on \sw\ so far has been obtained from 
interpretations of dedicated measurements in $e^+e^-$ collisions
at the $Z$ pole~\cite{ALEPH:2005ab}. The results are conventionally 
expressed in terms of a {\em leptonic} effective weak mixing angle 
which is related to the on-shell definition of \sw\
by a well-known correction factor, 
\begin{equation}
  \sweffl
  = 
  \kappa_{\ell}(\mZ^2) \sw \,. 
\label{eq:sw2eff-leptonic}
\end{equation}
A determination of \sw\ from DIS data 
can be compared with $Z$-pole measurements, provided its 
value is mapped to the definition of the leptonic weak 
mixing angle. Also in DIS one can define an effective, scale- 
and flavor-dependent weak mixing angle, 
cf.\ Eq.~(\ref{eq:sin2w-eff}), 
\begin{equation}
  \sweff(\mu^2) 
  = 
  \kappa_{f}(\mu^2) \sw \,.
\label{eq:sw2eff}
\end{equation}

We will now consider \sw\ as a free parameter which is allowed 
to vary in a \sw+PDF fit. Note that we consider in this fit only 
the \sw-dependence in the vector couplings, taken the same for 
leptons and quarks. SM higher-order corrections are taken into 
account as described in Sec.\ \ref{sec:theo} by keeping the 
\Qsq- and flavor-dependent form factors $\kappa_{f}$ 
(see Eq.~(\ref{eq:gV-NLO})). Our estimate for the uncertainties 
in the different LHeC scenarios are 
\begin{eqnarray}
  \Delta\sw\ (\text{LHeC-60a}) 
  &=& 
  \pm0.00023_{({\rm exp})}\pm0.00009_{({\rm PDF})}
  =\, \pm0.00025_{\text{(tot)}} \, , 
  \\
  \Delta\sw\ (\text{LHeC-50a}) 
  &=& 
  \pm0.00028_{({\rm exp})}\pm0.00019_{({\rm PDF})}
  =\, \pm0.00034_{\text{(tot)}} 
  \nonumber 
\end{eqnarray} 
and 
\begin{eqnarray}
  \Delta\sw\ (\text{LHeC-60b}) 
  &=& 
  \pm0.00014_{({\rm exp})}\pm0.00006_{({\rm PDF})}
  =\, \pm0.00015_{\text{(tot)}} \, ,  
  \\
  \Delta\sw\ (\text{LHeC-50b}) 
  &=& 
  \pm0.00017_{({\rm exp})}\pm0.00014_{({\rm PDF})}
  =\, \pm0.00022_{\text{(tot)}} \, .
  \nonumber
\end{eqnarray} 
\begin{figure}[tb!]
  \centering
  \includegraphics[width=0.50\textwidth]{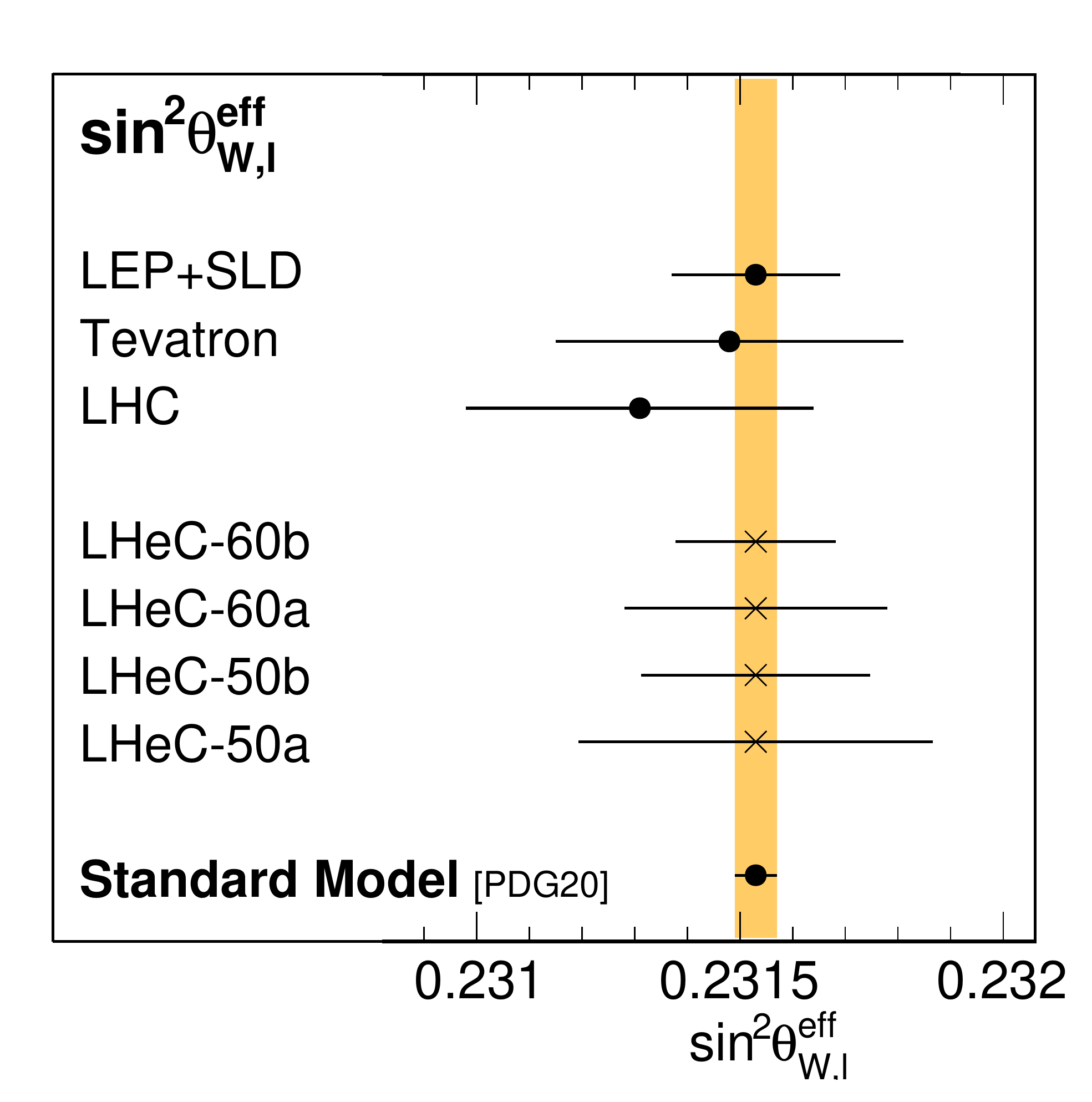}
  \caption{
    Comparison of determinations of the weak mixing angle. 
    The results from LEP+SLD~\cite{ALEPH:2005ab},  
    Tevatron~\cite{Aaltonen:2018dxj},
    LHC~\cite{Aaij:2015lka,ATLAS:2018gqq,Sirunyan:2018swq,%
    Erler:2019dcx} and the SM value refer to the leptonic 
    weak mixing angle, $\sweffl$, and include the information 
    about the $W$- and $Z$-boson masses~\cite{Erler:2019dcx}. 
    They  are all obtained from a combination of various separate 
    measurements (not shown individually) (see also 
    Ref.~\cite{Erler:2019hds} for additional information). 
    Two scenarios for the simulation of LHeC inclusive NC/CC 
    DIS data are considered. Here, the estimated uncertainties 
    refer to the fermionic effective weak mixing angle, \sweff. 
    With real data, the central values will have to be mapped to 
    each other by taking into account the proper $\kappa$-factors, 
    see Eqs.~(\ref{eq:sw2eff-leptonic}, \ref{eq:sw2eff}). 
    }
\label{fig:sw2eff}
\end{figure}
These results are collected in Fig.~\ref{fig:sw2eff} where 
we compare with presently available determinations of the 
leptonic weak mixing angle. Here we have neglected additional 
parametric uncertainties that may enter when the LHeC measurements 
are mapped to the leptonic effective weak mixing angle. The 
determination at the LHeC is superior to any current single 
measurement and of similar size as the LEP+SLD combination. Even 
measurements in a spacelike region of momentum transfers, i.e.\ 
for a non-resonant process, turns out to be competitive with 
$Z$-pole measurements, despite of the fact that the cross section 
receives large contributions from pure photon exchange at lower 
\Qsq, which is independent of the weak mixing angle. 

In the on-shell scheme, \sw\ and \mw\ are related to each other 
and a measurement of one parameter can be interpreted as a 
determination of the other. The uncertainty for \sw\ in 
scenario LHeC-60b, $\Delta\sw\ = \pm 0.00015$ would result in an 
uncertainty for the $W$-boson mass of $\Delta\mw = \pm 8$~MeV or, 
vice versa, $\Delta\mw = \pm 6$~MeV from the \mw+PDF fit (Sec.~\ref{sec:mass}) would 
correspond to $\Delta\sw\ = \pm 0.00012$. Comparing these numbers  
we can conclude that most of the sensitivity to \mw\ is due to 
the weak NC couplings; the additional \mw-dependence from the 
CC propagator mass provides little extra information for the 
determination of \mw.

\begin{figure}[tb!]
  \centering
  \includegraphics[width=0.54\textwidth]{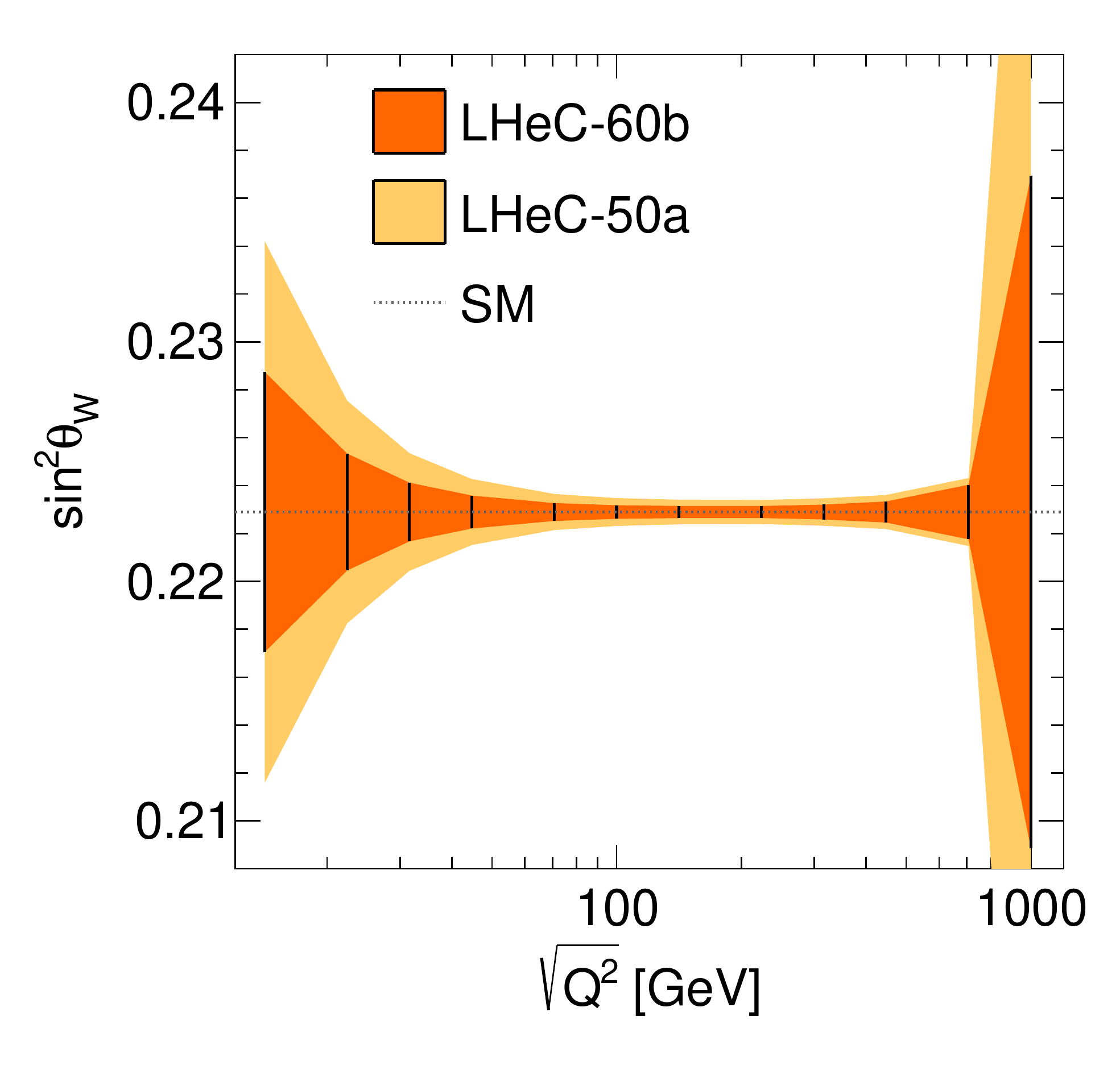}
  \caption{
    Expected uncertainties of the weak mixing angle determined 
    in sub-regions of \Qsq. Two scenarios for the simulation 
    of LHeC inclusive NC/CC DIS data are displayed, LHeC-50a and
    LHeC-60b.
    The SM expectation is displayed as a dotted line.
  }
\label{fig:sw2Q2}
\end{figure}
\begin{table}[bth!]
  \footnotesize
  \centering
  \begin{tabular}{lcllll}
    \\
    \toprule
    $\Qsq_i$ & Bin $i$ & \multicolumn{4}{c}{Expected relative 
                          uncertainty of
                          $\sin^2\theta_{\textrm{W}}$} \\
    \cmidrule(lr){3-6}      
    $[\GeVsq]$  & & LHeC-60b& LHeC-60a& LHeC-50b & LHeC-50a  \\
    \midrule
    200     &  1 & ~~$\pm0.026~$  & ~~$\pm0.049~$  & ~~$\pm0.027~$  & ~~$\pm0.051~$   \\
    500     &  2 & ~~$\pm0.011~$  & ~~$\pm0.021~$  & ~~$\pm0.011~$  & ~~$\pm0.021~$  \\
    1000    &  3 & ~~$\pm0.0055$  & ~~$\pm0.010~$  & ~~$\pm0.0061$  & ~~$\pm0.011~$  \\
    2000    &  4 & ~~$\pm0.0031$  & ~~$\pm0.0057$  & ~~$\pm0.0035$  & ~~$\pm0.0062$ \\
    5000    &  5 & ~~$\pm0.0017$  & ~~$\pm0.0030$  & ~~$\pm0.0019$  & ~~$\pm0.0034$ \\
    10000   &  6 & ~~$\pm0.0013$  & ~~$\pm0.0023$  & ~~$\pm0.0015$  & ~~$\pm0.0026$ \\
    20000   &  7 & ~~$\pm0.0011$  & ~~$\pm0.0020$  & ~~$\pm0.0014$  & ~~$\pm0.0023$ \\
    50000   &  8 & ~~$\pm0.0011$  & ~~$\pm0.0019$  & ~~$\pm0.0014$  & ~~$\pm0.0024$  \\
    100000  &  9 & ~~$\pm0.0014$  & ~~$\pm0.0022$  & ~~$\pm0.0017$  & ~~$\pm0.0026$ \\
    200000  & 10 & ~~$\pm0.0020$  & ~~$\pm0.0028$  & ~~$\pm0.0023$  & ~~$\pm0.0032$  \\
    500000  & 11 & ~~$\pm0.0051$  & ~~$\pm0.0056$  & ~~$\pm0.0057$  & ~~$\pm0.0064$ \\
    1000000 & 12 & ~~$\pm0.063~$  & ~~$\pm0.063~$  & ~~$\pm0.17~~$  & ~~$\pm0.17~~$    \\
   \bottomrule
  \end{tabular}
  \caption{
    Expected relative uncertainties for the determination of 
    \sw\ as a function of \Qsq\ for the four LHeC scenarios.
    The uncertainties are obtained in a simultaneous fit of 12 
    parameters $\sin^2\theta_{\textrm{W}}(\mu^2_i)$ with 
    $\mu_i^2 = - Q_i^2$ ($i=1,\dots,12$) together with the PDF 
    parameters. Absolute uncertainties $\Delta\sw$ can be 
    calculated by multiplication with the value of \sw. 
    }
\label{tab:sw2Q2}
\end{table}
The measurement of \sw\ can be performed in sub-regions of the 
wide kinematic range of \Qsq\ accessible at the LHeC. The results 
for twelve $Q^2$ values obtained from bin-width and bin-center 
corrected cross section data are shown in Fig.~\ref{fig:sw2Q2} 
and Tab.~\ref{tab:sw2Q2}. We find that \sw\ can be determined in 
the range of about $25 < \sqrt{\Qsq} < 700\,\GeV$ with a precision 
better than 0.1\,\% and everywhere better than 1\,\%.
We emphasize that DIS is complementary to other measurements since 
the scattering process is mediated by boson exchange with spacelike 
momenta, i.e.\ the scale is given by $\mu^2 = - Q^2$ (cf.\ 
Sec.~\ref{sec:sw2eff}). If a calculation of DIS cross sections 
including higher-order EW corrections in the $\overline{\rm MS}$ 
scheme is available, the uncertainty of this \Qsq-dependent 
\sw-determination can be translated into a test of the running 
of the weak mixing angle.

\clearpage

\section{Mass parameters through higher-order corrections}
\label{sec:mt} 

The sensitivity to the weak boson masses $m_W$ and $m_Z$ is 
high since these parameters enter at tree level through the 
cross section normalization and through the boson propagators. 
Other SM mass parameters enter only at higher orders. Cross 
sections are therefore only weakly dependent on them. However, 
measurements with a high precision may still exhibit some 
sensitivity. Their investigation is interesting since this 
provides a test of the SM at the level of quantum corrections 
which is complementary to direct determinations. 

The dominant corrections to the gauge boson self energies depend 
on the top-quark mass \mt\,\footnote{Note that
  at sufficiently high scales, e.g.\ $\Qsq \gtrsim (2\mt)^2$, 
  the top-quark contributes also to the QCD evolution of the PDFs. 
  However, these contributions to the inclusive
  DIS cross sections are very small at the LHeC. In particular, 
  their sensitivity to the actual value of the top-quark mass can 
  be neglected.
}.
In the on-shell scheme they enter in the NC coupling parameters 
$\rho_\textrm{NC}$ and $\kappa$ and in the CC correction factor 
\dr\ through the quantity $\rho_t = (3\alpha/16\pi\sw) (m_t^2/m_W^2)$.
Therefore, inclusive DIS cross sections depend quadratically 
on \mt\ and since $\rho_t$ is in the order of 1\,\%, one can 
expect to observe a sizeable effect on the DIS cross sections. 

We have determined the uncertainties of the top-quark mass \mt\ 
through DIS cross section measurements for the four scenarios in an
\mt+PDF fit. 
For LHeC-50a and LHeC-50b we find $\Delta\mt = \pm2.2\,\GeV$ 
and $\pm 1.8\,\GeV$, respectively. For the LHeC scenarios with 
$E_e=60\,\GeV$, the top-quark mass can be determined with an 
uncertainty of $\Delta\mt = \pm1.4\,\GeV$ (LHeC-60a) and 
\begin{equation}
  \Delta\mt\,\textrm{(LHeC-60b)} = \pm1.1\,\GeV\,.
\end{equation}
The size of the PDF-related uncertainty amounts to about
0.6\,\GeV\ and is already included in the values above. In these 
studies, the value of \mW\ is considered as an external, i.e.\ fixed,
parameter. However, the dominant theoretical uncertainty for an 
\mt\ determination arises in fact from the uncertainty of $\mW$. 
At present, the $W$ mass is known with an uncertainty of
$\Delta\mW = \pm 12\,\MeV$~\cite{PDG2020}. This corresponds to 
a theory uncertainty of \mt\ of about $\pm2\,\GeV$.

The size of the LHeC experimental uncertainty compares well with
uncertainties from recent LHC measurements, which are typically 
in the range between $\Delta \mt = \pm0.3$ and 
$\pm2.0$~\GeV~\cite{LHCTopWG} 
(see also Ref.~\cite{PDG2020} and references therein).
One should note, however, that the uncertainty of \mt\ at 
the LHC experiments is dominated by Monte Carlo modelling and 
theoretical uncertainties related to the proper definition of 
the top-quark mass. These theoretical uncertainties are shared 
between different LHC measurements and it is expected that they 
limit the precision of the \mt\ determination also in the future. 
In contrast, the definition of the top-quark mass entering in 
higher-order EW corrections to DIS cross sections is theoretically
very clean and free from QCD-related ambiguities. In fact, the 
definition of \mt\ corresponds to the one used in the calculation 
of observables in the SM framework, as it is also done in
the global EW fits. It will therefore be justified to include 
possible future data from the LHeC in a determination of a 
world average of \mt. We study this possibility briefly in 
appendix~\ref{app:globalfit}. Our results indicate that LHeC 
data will not improve the present uncertainty of $\Delta\mt = 
\pm2.1\,\GeV$ from global 
fits~\cite{deBlas:2016ojx,Haller:2018nnx,PDG2020} once direct 
measurements of \mt\ and \mw\ are taken into account. 

\begin{figure}[tb!]
    \centering
    \includegraphics[width=0.54\textwidth,trim=0 40 0 25,clip]{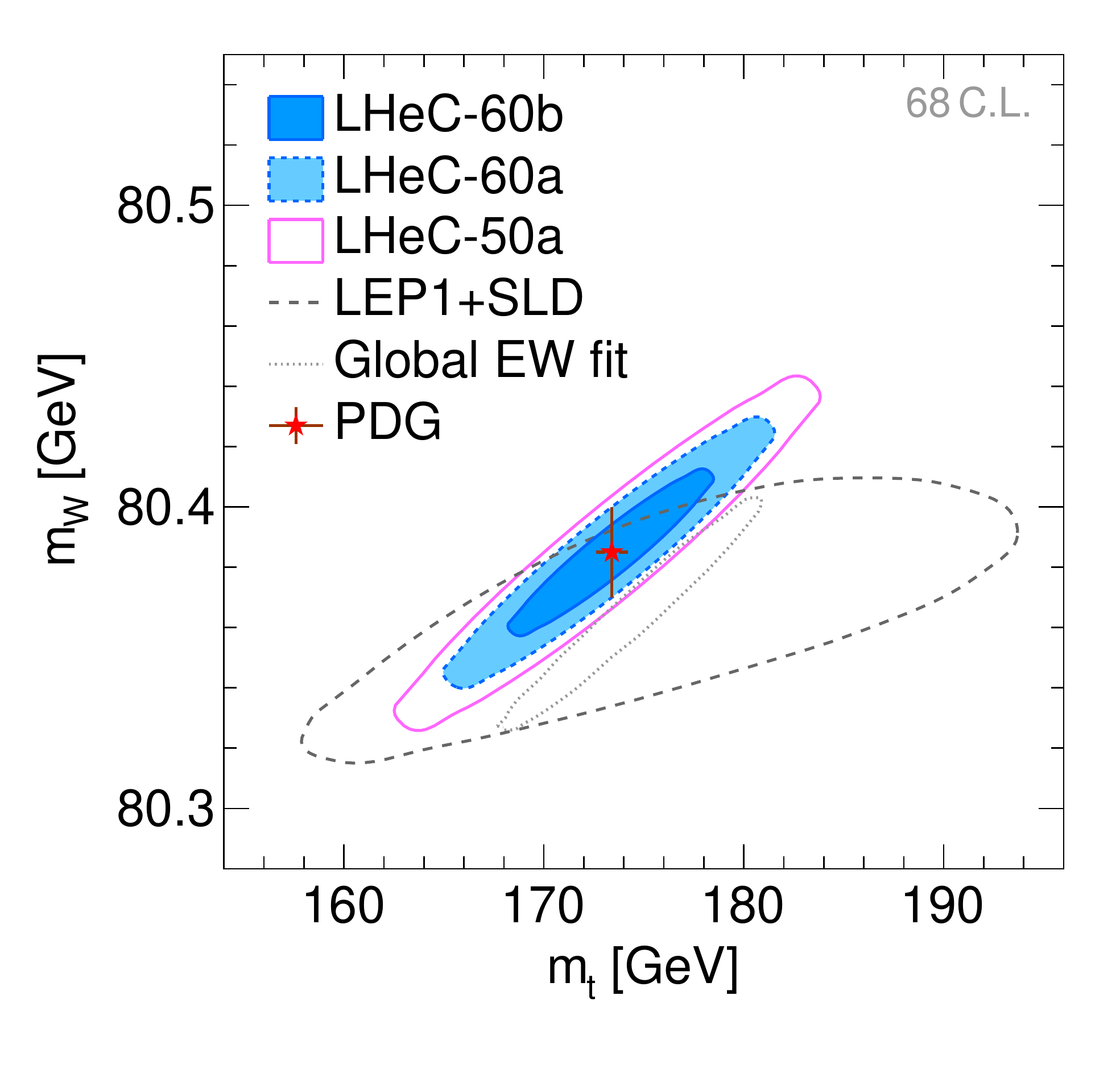}
    \caption{
       Simultaneous determination of the $W$-boson and top-quark 
       masses from LHeC-60a or LHeC-50a data. Results from the 
       $Z$-pole fit using LEP+SLD data~\cite{ALEPH:2005ab} and 
       from a global EW fit, where direct measurements of \mW\ 
       and \mt\ have been excluded~\cite{Haller:2018nnx} are 
       also shown.
    }
    \label{fig:mWmt}
\end{figure}

We also consider the possibility to determine the $W$-boson mass 
\mw\ simultaneously together with \mt. Prospects for such a 
simultaneous determination of \mt, \mw, and the PDFs are displayed 
for selected LHeC scenarios in Fig.~\ref{fig:mWmt} and compared 
with results from the LEP+SLD combination of $Z$-pole 
measurements~\cite{ALEPH:2005ab}. The figure shows also results 
from a global EW fit~\cite{Haller:2018nnx}, for which the direct 
\mt\ and \mw\ measurements have been excluded. We find that the 
uncertainties of the LHeC are better than those obtained from 
the LEP+SLD combined data. For the scenario LHeC-60b, the 
uncertainty contour is very similar in size as the global EW fit.
It is not surprising that both the global EW fit and the LHeC 
fit exhibit the same type of correlation since they exploit the 
same $\mt^2/\mw^2$-dependent terms of the radiative corrections. 

One may also attempt to determine the Higgs-boson mass \mH\ 
from inclusive DIS data. \mH\ also enters through the self-energy 
corrections in the SM, however, the \mH\ dependence is only 
logarithmic, $\propto\log(\mH^2/\mw^2)$, i.e.\ very weak. 
An \mH+PDF fit leads to an uncertainty of $\Delta m_H=^{+28}_{-23}$ 
and $^{+14}_{-13}~\GeV$ for the scenarios LHeC-50a and LHeC-60b, 
respectively. This compares well with the precision found for 
the indirect \mH\ determinations from LEP+SLD combined 
data~\cite{ALEPH:2005ab,deBlas:2016ojx,Haller:2018nnx}, but 
is, of course, much less precise than the direct determination 
from the LHC experiments 
nowadays~\cite{Sirunyan:2020xwk,ATLAS-CONF-2020-005}. 
From Higgs boson production and its decays into fermion pairs,
the LHeC has a direct Higgs mass measurement potential as
well, which surely is much better than the indirect one but
unlikely competitive to that at the LHC through the 4-lepton 
and 2-photon decays.


\section{\boldmath Oblique parameters $S$, $T$, and $U$} 
\label{sec:stu}

Many theories beyond the SM predict additional heavy particles. 
While these may be too heavy for a direct detection in present 
or future experiments, they may contribute through effective 
low-energy operators or through higher-order loop corrections 
to observables. High-precision measurements provide an opportunity 
to observe in an indirect way their presence. 

Loop insertions with particle-antiparticle pairs in gauge boson 
self energies, $\Sigma^{ij}(q^2)$, are particularly important 
since they are universal. If the masses of the non-SM particles 
are large, a low-$q^2$ expansion of the self-energy corrections, 
\begin{equation}
  \Sigma^{ij}(q^2) = \Sigma^{ij}(0) + q^2 F^{ij}(q^2) \, ,  
  \quad \quad 
  (ij) = (\gamma\gamma),~(\gamma Z),~(ZZ),~(WW) \, , 
\label{eq:def-stu}
\end{equation} 
and neglecting the $q^2$-dependence of $F^{ij}(q^2)$, can provide 
a sufficiently precise approximation by constant parameters. 
Taking into account that the electromagnetic $U(1)$ gauge 
symmetry has to stay intact and that some of these constants 
can be absorbed into renormalization constants, there are three 
free parameters, usually called $S$, $T$ and 
$U$~\cite{Peskin:1991sw}. A suitable definition is described 
in Ref.~\cite{PDG2020} which we adopt in the following, while 
their relation to alternative 
definitions~\cite{Kennedy:1988sn,Altarelli:1990zd}
is described in Ref.~\cite{Spiesberger:1993jg}.

\begin{figure}[tb!]
  \centering
  \includegraphics[width=0.32\textwidth,trim=20 48 20 25,clip]{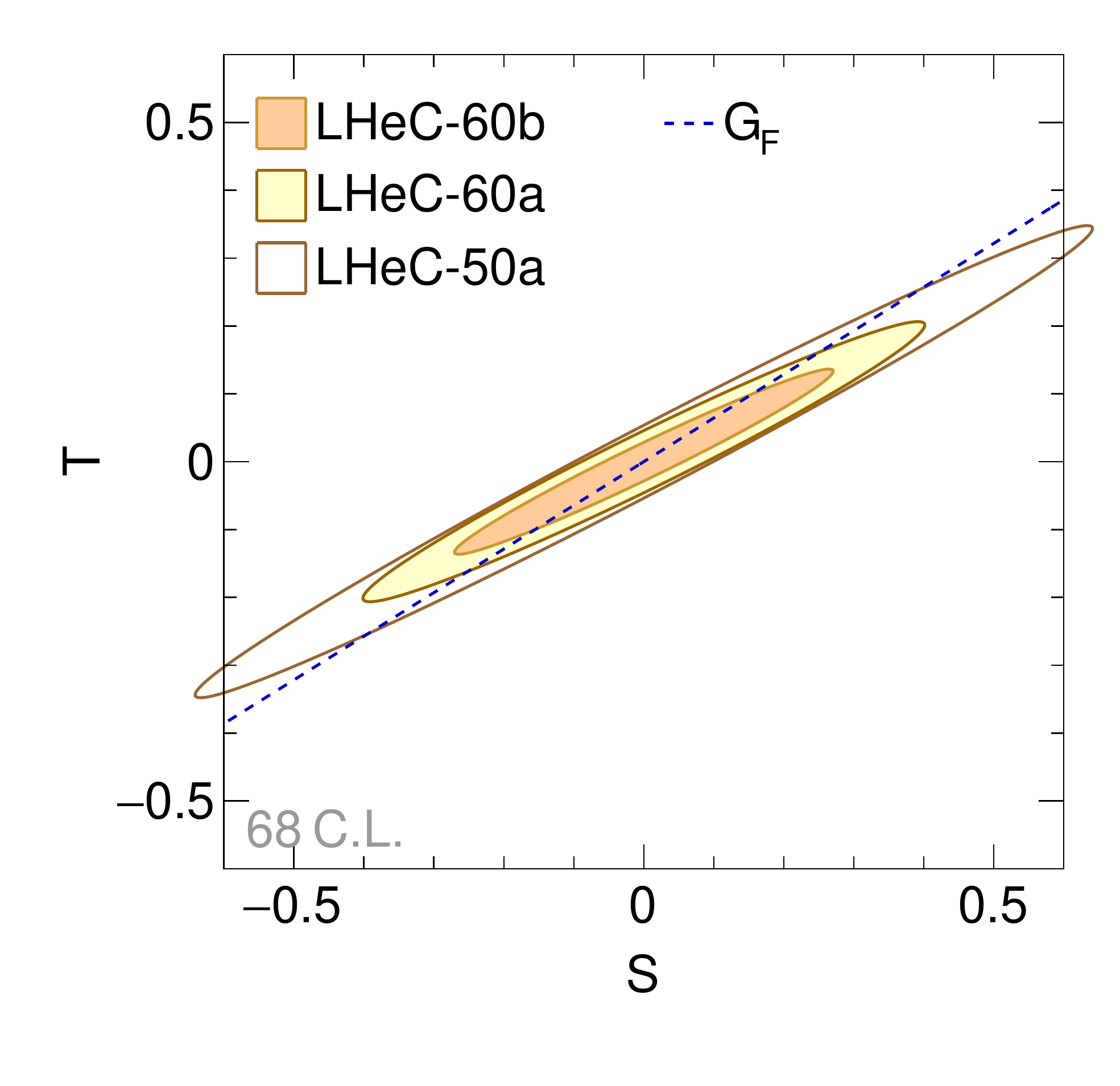}
  \includegraphics[width=0.32\textwidth,trim=20 48 20 25,clip]{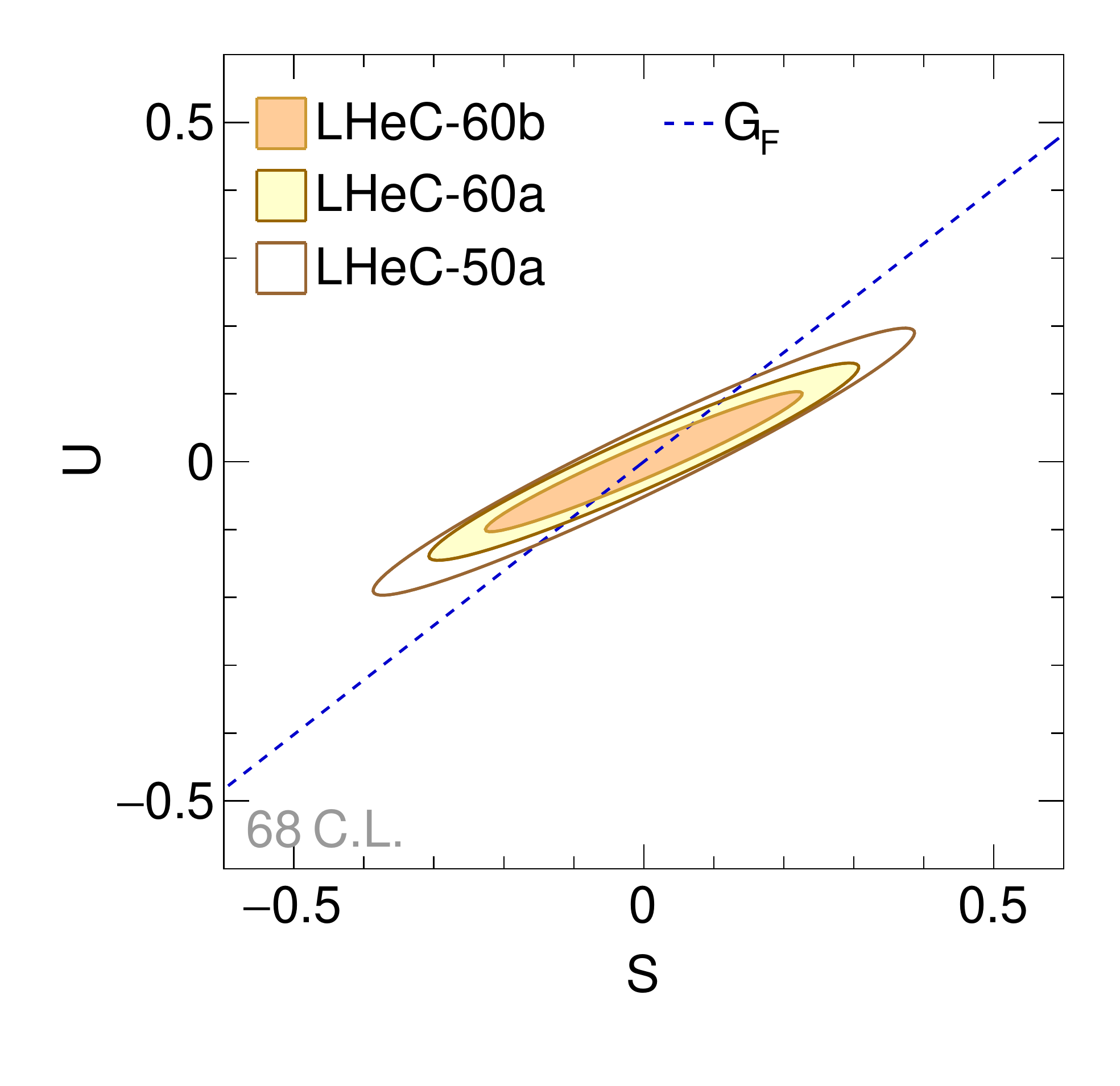}
  \includegraphics[width=0.32\textwidth,trim=20 48 20 25,clip]{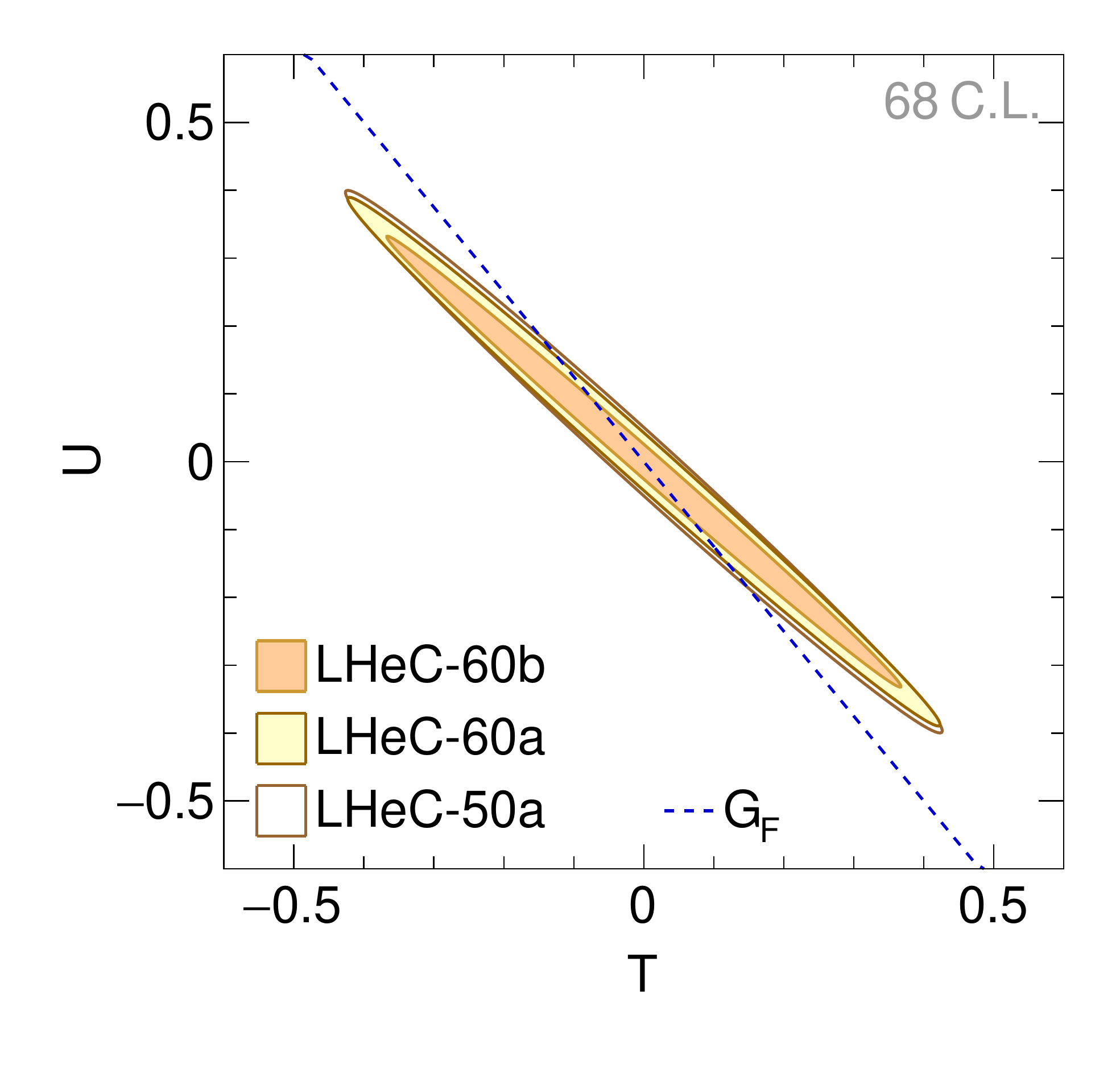}
  \caption{
     Results of 2-parameter fits to pairs of $S$, $T$, and $U$ 
     where $m_Z$ and $m_W$ are fixed SM input parameters. For 
     each choice of two of the three parameters $S$, $T$, or $U$, 
     the third oblique parameter is kept equal to zero. 
     $1\sigma$ contours are shown for three LHeC scenarios. 
     The relation how a direct measurement of \gf\ would constrain 
     the parameters is indicated in addition.
  }
  \label{fig:STU_OS}
\end{figure}

Results for various $STU$+PDF fits are presented in 
Figs.~\ref{fig:STU_OS} and Tab.~\ref{tab:STU_OS}.
These fits are performed in the on-shell scheme and the 
SM masses are fixed at their PDG values, in particular the 
values of $m_Z$ and $m_W$. Single-parameter fits of $S$, $T$ 
or $U$ can provide uncertainties that are better by a factor 
of 2 to 5 compared to the present PDG values~\cite{PDG2020}.
In 2- and 3-parameter fits we observe a very strong correlation 
of the parameters. This can be traced back to the fact that only 
certain linear combinations of $S$, $T$ and $U$ contribute to 
the NC and CC scattering cross sections and the $\gamma Z$ 
interference contribution. For instance, the values of $T$ 
and $U$ can be disentangled only if their contributions to NC 
and CC DIS are combined, but not from NC DIS alone.
By implication, however, these linear combinations can be 
determined with very high precision -- a fact which makes the DIS 
measurement particularly useful since it is complementary 
to determinations of $S$, $T$ and $U$ from $Z$-pole data 
(see, for example, 
Refs.~\cite{deBlas:2016ojx,Haller:2018nnx,PDG2020}). 

\begin{table}[bht!]
  \footnotesize
  \centering
  \begin{tabular}{l@{\hspace*{0.5em}}cc@{\hspace*{0.5em}}c@{\hspace*{0.5em}}c@{\hspace*{0.5em}}cccc}
    \toprule
    Fit parameters & Parameter & \multicolumn{4}{c}{Expected uncertainty} & \multicolumn{3}{c}{Correlation (LHeC-60b)} \\
    \cmidrule(lr){3-6}    \cmidrule(lr){7-9}
    & & LHeC-60b& LHeC-60a& LHeC-50b & LHeC-50a & $S$ & $T$ & $U$ \\
    \midrule
    $S$+PDF      &  $S$  & $\pm0.04$ & $\pm0.06$ & $\pm0.04$ & $\pm0.07$ \\
    $T$+PDF      &  $T$  & $\pm0.02$ & $\pm0.03$ & $\pm0.02$ & $\pm0.04$ \\
    $U$+PDF      &  $U$  & $\pm0.02$ & $\pm0.03$ & $\pm0.04$ & $\pm0.03$ \\
    \addlinespace
    $S$+$T$+PDF  &  $S$  & $\pm0.18$ & $\pm0.26$ & $\pm0.35$ & $\pm0.42$ & $1.00$ & $0.98$  \\
                 &  $T$  & $\pm0.09$ & $\pm0.14$ & $\pm0.19$ & $\pm0.23$ &        & $1.00$  \\
    \addlinespace
    $S$+$U$+PDF  &  $S$  & $\pm0.15$ & $\pm0.20$ & $\pm0.22$ & $\pm0.26$ & $1.00$ & & $0.97$  \\
                 &  $U$  & $\pm0.07$ & $\pm0.09$ & $\pm0.11$ & $\pm0.13$ &        & & $1.00$  \\
    \addlinespace
    $T$+$U$+PDF  &  $T$  & $\pm0.24$ & $\pm0.28$ & $\pm0.24$ & $\pm0.28$ & & $1.00$ & $-0.99$  \\
                 &  $U$  & $\pm0.22$ & $\pm0.26$ & $\pm0.22$ & $\pm0.26$ & &        & \phantom{--}$1.00$  \\
    \addlinespace
    $S$+$T$+$U$+PDF  &  $S$  & $\pm0.20$ & $\pm0.31$ & $\pm0.46$ & $\pm0.58$ & 1.00 & 0.65 & $-0.41$ \\
                 &  $T$  & $\pm0.32$ & $\pm0.42$ & $\pm0.52$ & $\pm0.64$ &      & 1.00 & $-0.96$\\
                 &  $U$  & $\pm0.24$ & $\pm0.30$ & $\pm0.30$ & $\pm0.36$ &      &      & \phantom{--}1.00 \\
    \bottomrule
  \end{tabular}
  \caption{
    Results of the $STU$+PDF fits with fixed SM gauge boson masses. 
    From top to bottom we show the expected uncertainties for 
    1-, 2- and 3-parameter fits as indicated in the first column 
    for all four LHeC scenarios. In the case of 2- and 3-parameter 
    fits, the last columns show the correlation matrices.     
  }
  \label{tab:STU_OS}
\end{table}

The $STU$+PDF fit to LHeC DIS data in the on-shell scheme 
can be combined with the constraint from the $G_F$ measurement, 
cf.\ Fig.~\ref{fig:STU_OS}. 
Since $G_F$ is known with very high precision, this constraint 
amounts essentially to fixing one linear combination of the 
$STU$ parameters, the one that enters in $\Delta r$ (see 
Eq.~(\ref{eq:deltar-NLO})).

\begin{figure}[tb!]
  \centering
  \includegraphics[width=0.32\textwidth,trim=20 20 20 25,clip]{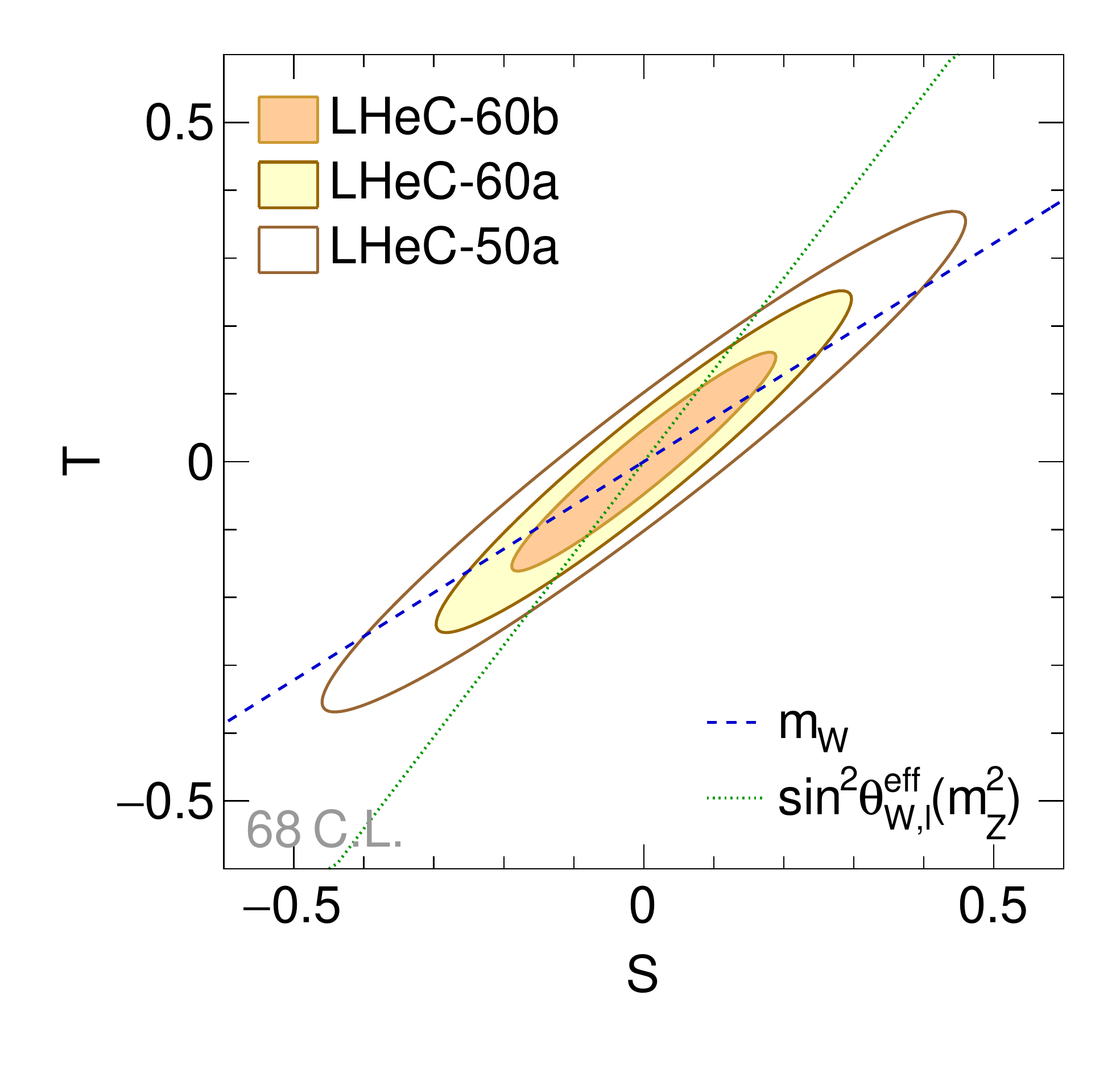}
  \includegraphics[width=0.32\textwidth,trim=20 20 20 25,clip]{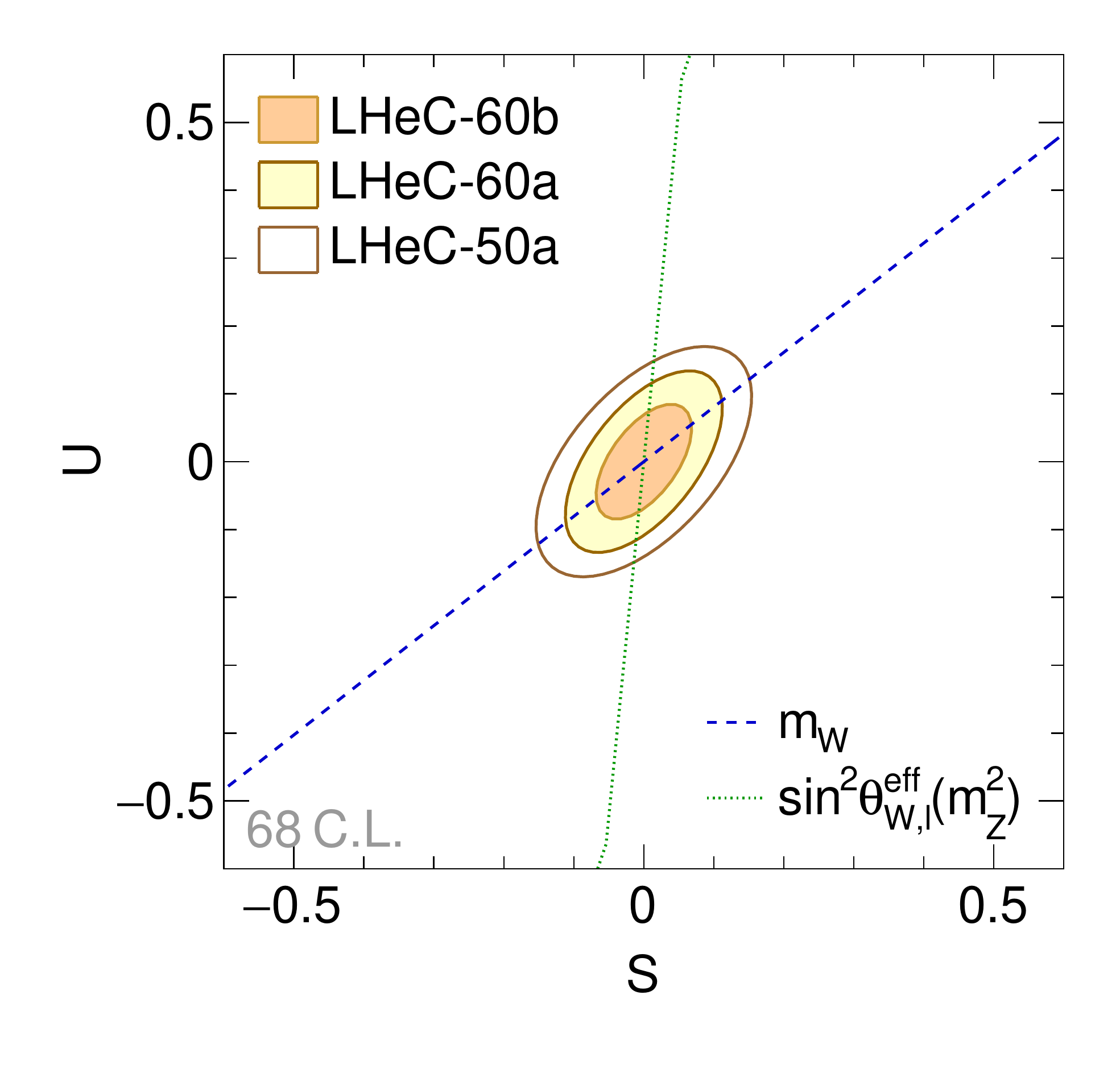}
  \includegraphics[width=0.32\textwidth,trim=20 20 20 25,clip]{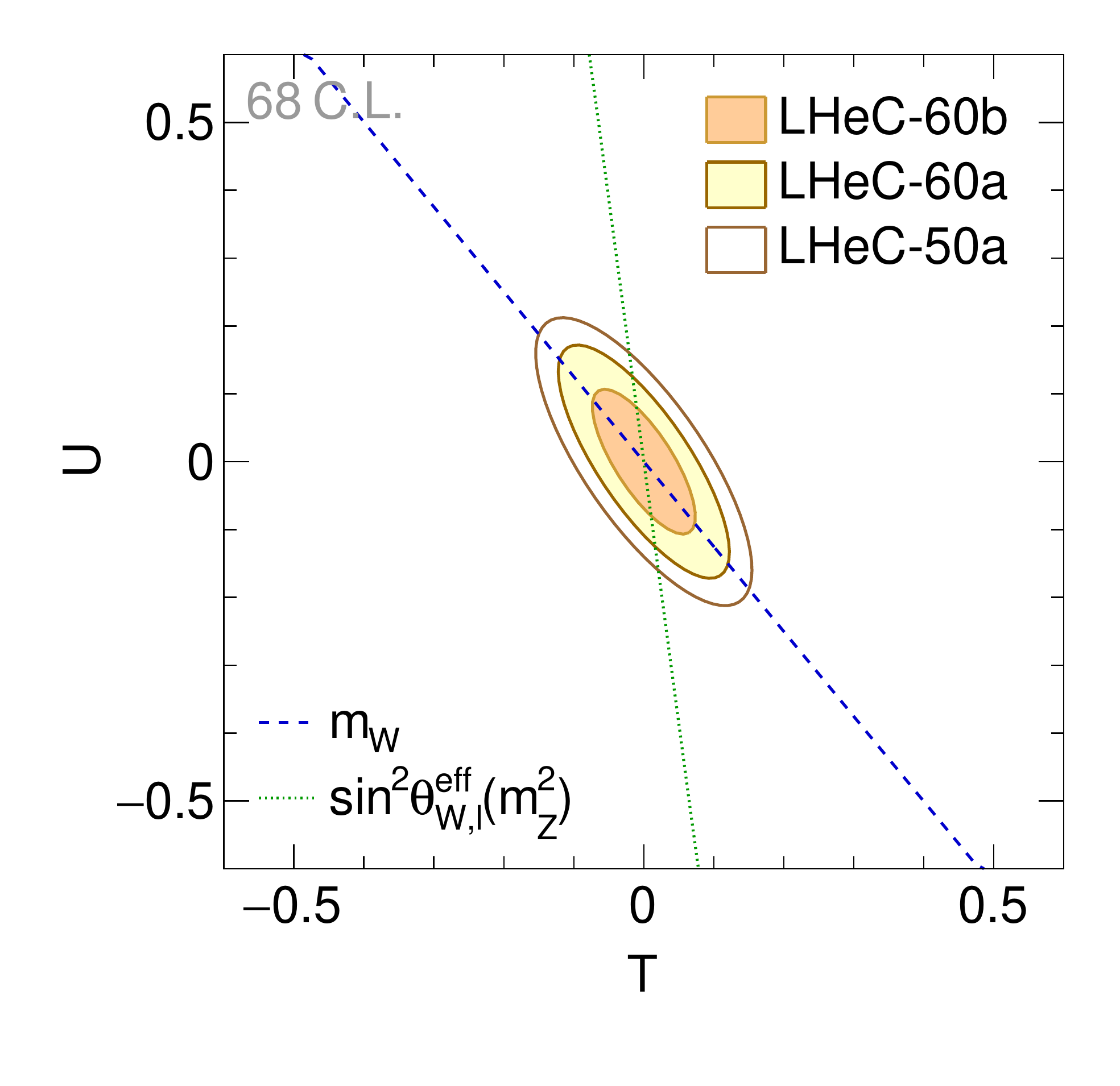}
  \caption{
    Same as Fig.~\ref{fig:STU_OS}, but the calculations are 
    performed in the modified on-shell scheme, i.e.\ 
    the value for $m_Z$ is fixed, but the $W$ boson mass is 
    calculated from its relation to the Fermi constant \gf.
    The relations how direct measurements of \mw\
    or \sweffl\ at the $Z$-pole would constrain the two oblique
    parameters are additionally indicated by dashed and dotted 
    lines, respectively.
    In the modified on-shell scheme, the measurement of \mw\ would
    constrain the same relation as a measurement of $1-\mw^2/\mz^2$
    (dashed line).
  }
  \label{fig:STU_MOMS}
\end{figure}
\begin{table}[bht!]
  \footnotesize
  \centering
  \begin{tabular}{l@{\hspace*{0.5em}}cc@{\hspace*{0.5em}}c@{\hspace*{0.5em}}c@{\hspace*{0.5em}}cccc}
    \toprule
    Fit parameters & Parameter& \multicolumn{4}{c}{Expected uncertainty} & \multicolumn{3}{c}{Correlation (LHeC-60b)} \\
    \cmidrule(lr){3-6}    \cmidrule(lr){7-9}
    & & LHeC-60b& LHeC-60a& LHeC-50b & LHeC-50a & S & T & U \\
    \midrule
    $S$+PDF      &  $S$  & $\pm0.04$ & $\pm0.06$ & $\pm0.05$ & $\pm0.08$ \\
    $T$+PDF      &  $T$  & $\pm0.03$ & $\pm0.05$ & $\pm0.04$ & $\pm0.07$ \\
    $U$+PDF      &  $U$  & $\pm0.04$ & $\pm0.07$ & $\pm0.06$ & $\pm0.09$ \\
    \addlinespace
    $S$+$T$+PDF  &  $S$  & $\pm0.12$ & $\pm0.20$ & $\pm0.21$ & $\pm0.30$ & $1.00$ & $0.95$  \\
                 &  $T$  & $\pm0.11$ & $\pm0.17$ & $\pm0.17$ & $\pm0.24$ &        & $1.00$  \\
    \addlinespace
    $S$+$U$+PDF  &  $S$  & $\pm0.05$ & $\pm0.07$ & $\pm0.06$ & $\pm0.10$ & $1.00$ & & $0.57$  \\
                 &  $U$  & $\pm0.06$ & $\pm0.09$ & $\pm0.07$ & $\pm0.11$ &        & & $1.00$  \\
    \addlinespace
    $T$+$U$+PDF  &  $T$  & $\pm0.05$ & $\pm0.08$ & $\pm0.06$ & $\pm0.10$ & & $1.00$ & $-0.76$  \\
                 &  $U$  & $\pm0.07$ & $\pm0.11$ & $\pm0.09$ & $\pm0.14$ & &        & \phantom{--}$1.00$  \\
     \addlinespace
    $S$+$T$+$U$+PDF  & $S$  & $\pm0.20$ & $\pm0.32$ & $\pm0.47$ & $\pm0.60$ & 1.00 & 0.97 & $-0.79$ \\
                 & $T$  & $\pm0.22$ & $\pm0.35$ & $\pm0.46$ & $\pm0.60$ &      & 1.00 & $-0.87$\\
                 & $U$  & $\pm0.11$ & $\pm0.19$ & $\pm0.20$ & $\pm0.28$ &      &      & \phantom{--}1.00 \\
    \bottomrule
  \end{tabular}
  \caption{
    Same as Tab.~\ref{tab:STU_OS}, but in the modified on-shell 
    scheme with $m_Z$ and $G_F$ as fixed input parameters, i.e.\ 
    $m_W$ is calculated. 
  }
  \label{tab:STU_MOMS}
\end{table}
New physics parameterized with the help of $S$, $T$ and $U$ 
will also affect the \gf-\mw\ relation, Eq.~(\ref{eq:deltar-NLO}), 
through the $W$-boson self energy correction to the muon decay. 
In the modified on-shell scheme~\cite{Marciano:1980pb}, 
where $m_W$ is calculated from $G_F$, new physics will 
therefore not only contribute by corrections to the measured 
cross sections, but also through a modification of the input 
parameters. As a consequence, the sensitivity to $S$, $T$ and 
$U$ is modified. Results of a $STU$+PDF fit in the modified 
on-shell scheme are collected in Figs.~\ref{fig:STU_MOMS} and 
Tab.~\ref{tab:STU_MOMS}. The uncertainties determined from 
single-parameter fits are slightly less favorable in this 
case. However, the 2- and 3-parameter fits exhibit weaker 
correlations leading to smaller uncertainties for their 
corresponding 1-parameter projections. In the modified on-shell 
scheme, additional constraints on the $STU$ parameters may be 
obtained by adding further direct measurements of \mw\ or \sweffl,
e.g.\ from measurements in $e^+e^-$ or hadron-hadron collisions.
The parameter relations of such measurements are also indicated 
in Fig.~\ref{fig:STU_MOMS} and in particular external 
measurements sensitive to \sweffl\ would be useful for further 
improvements.

\clearpage

\section{\boldmath Weak neutral-current couplings beyond the SM: 
$\rho_{\text{NC}}$ and $\kappa$}
\label{sec:rhokappa}

In the following we consider the option that modifications of 
the EW interaction by new physics can be parameterized directly 
with the help of the NC weak coupling constants. A systematic 
approach is based on using anomalous parameters \rhop\ and 
\kapp{}. The first, \rhop{}, affects the SU(2) component of NC 
couplings, while the second, \kapp{}, represents a modification 
of the weak mixing with the U(1) gauge field. These 
parameters can be chosen flavor-specific and are introduced 
by writing~\cite{Spiesberger:2018vki} 
\begin{eqnarray}
  g_A^f &=& 
  \sqrt{\rhop{,f}\rho_{\text{NC}, f}} \, \Itf \, , 
\label{eq:gA} 
  \\
  g_V^f &=& 
  \sqrt{\rhop{,f}\rho_{\text{NC}, f}} \, 
  \left( \Itf - 2 Q_f \kapp{f}\kappa_{f}\sw \right) \, . 
\label{eq:gV} 
\end{eqnarray}
Here, the un-primed form factors $\rho_\text{NC}$ and 
$\kappa_f$ take higher-order SM corrections into account, 
as described in Sec.~\ref{sec:theo}. In the SM, the anomalous 
parameters \rhop\ and \kapp{} are unity. In the presence of 
physics beyond the SM, they can deviate from unity and be 
\Qsq-dependent. In particular, a value of $\rhop\ \neq 1$ 
corresponds to a modification of the ratio of the strengths of 
NC and CC weak interactions. A similar study of a generalization 
of the CC form factor $\rho_\text{CC}$ will be discussed below 
in Sec.~\ref{sec:ewcc}. The parameter \kapp{} can also be 
interpreted as a modification of the weak mixing angle \sw\ (see 
Sec.~\ref{sec:sw2eff}), i.e.\ the definition of the effective 
weak mixing angle, Eq.~(\ref{eq:sw2eff}), is replaced by 
\begin{equation}
  \sweff(\mu^2) 
  = 
  \kapp{f}(\mu^2) \kappa_{f}(\mu^2) \sw \,.
\label{eq:sw2eff-chapter10}
\end{equation} 

\begin{figure}[th!]
  \centering
  \includegraphics[width=0.54\textwidth]{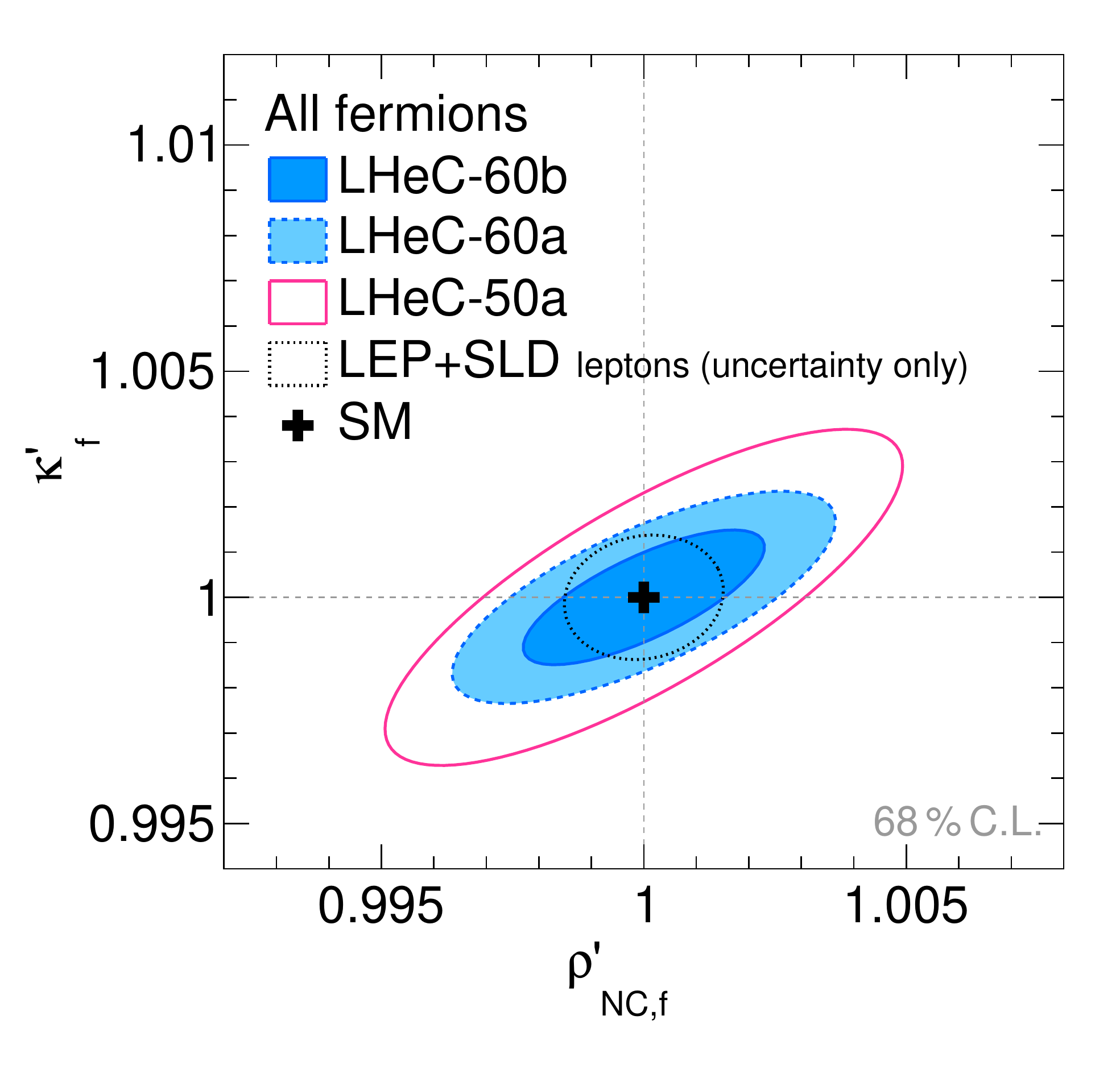}
  \caption{
    Expectation for a determination of $\rhop{,f}$ and 
    $\kapp{f}$ at the 68\,\% confidence level, assuming 
    one common anomalous factor for each fermion type. The 
    results for three different LHeC scenarios are compared 
    with the relative uncertainties obtained from an analysis 
    of LEP+SLD combined data~\cite{ALEPH:2005ab} for leptonic 
    couplings. 
  }
\label{fig:rhokappa:f}
\end{figure}
We determine the uncertainties of the anomalous form factors
 $\rho_\text{NC}^\prime$ and $\kappa^\prime$ 
in a simultaneous fit together with the PDFs, using the simulated 
LHeC inclusive NC and CC DIS data. First, we consider 
universal, i.e.\ flavor-independent, $\rho_\text{NC}^\prime$ 
and $\kappa^\prime$ parameters for both the 
quark and electron couplings. The results are displayed in 
Fig.~\ref{fig:rhokappa:f}. In this figure, we compare the 
expected LHeC uncertainties with corresponding results that 
have been obtained from combined LEP+SLD data for leptonic 
couplings~\footnote{From the combined measurements of 
  LEP+SLD, the leptonic parameters $\rho_{\textrm{NC},\ell}$ 
  and $\kappa_{\ell}$ have been determined~\cite{ALEPH:2005ab}. 
  For our comparison, we interpret them as uncertainties
  of flavor-universal anomalous parameters 
  $\rho_\text{NC}^\prime$ and $\kappa^\prime$.}.
At the LHeC, uncertainties are expected at the level of a few 
per mille, i.e.\ of similar size as those of the LEP+SLD 
combination. As expected, the scenario LHeC-60b yields the 
smallest uncertainties, while from the LHeC-50a scenario one 
should expect the largest ones.

The $\rho_\text{NC}^\prime$-$\kappa^\prime$ fit can be 
interpreted as a simultaneous determination of 
$\sin^2\theta_{\textrm{W},f}^\textrm{eff}$ and a universal 
modification of the normalization of NC weak couplings by 
\rhop. We find an uncertainty of 
$\Delta \sin^2\theta_{\textrm{W},f}^\textrm{eff} 
= \pm 0.00023$ ($\pm0.00071$) for LHeC-60b (LHeC-50a).

\begin{figure}[tb!]
  \centering
  \includegraphics[width=0.48\textwidth]{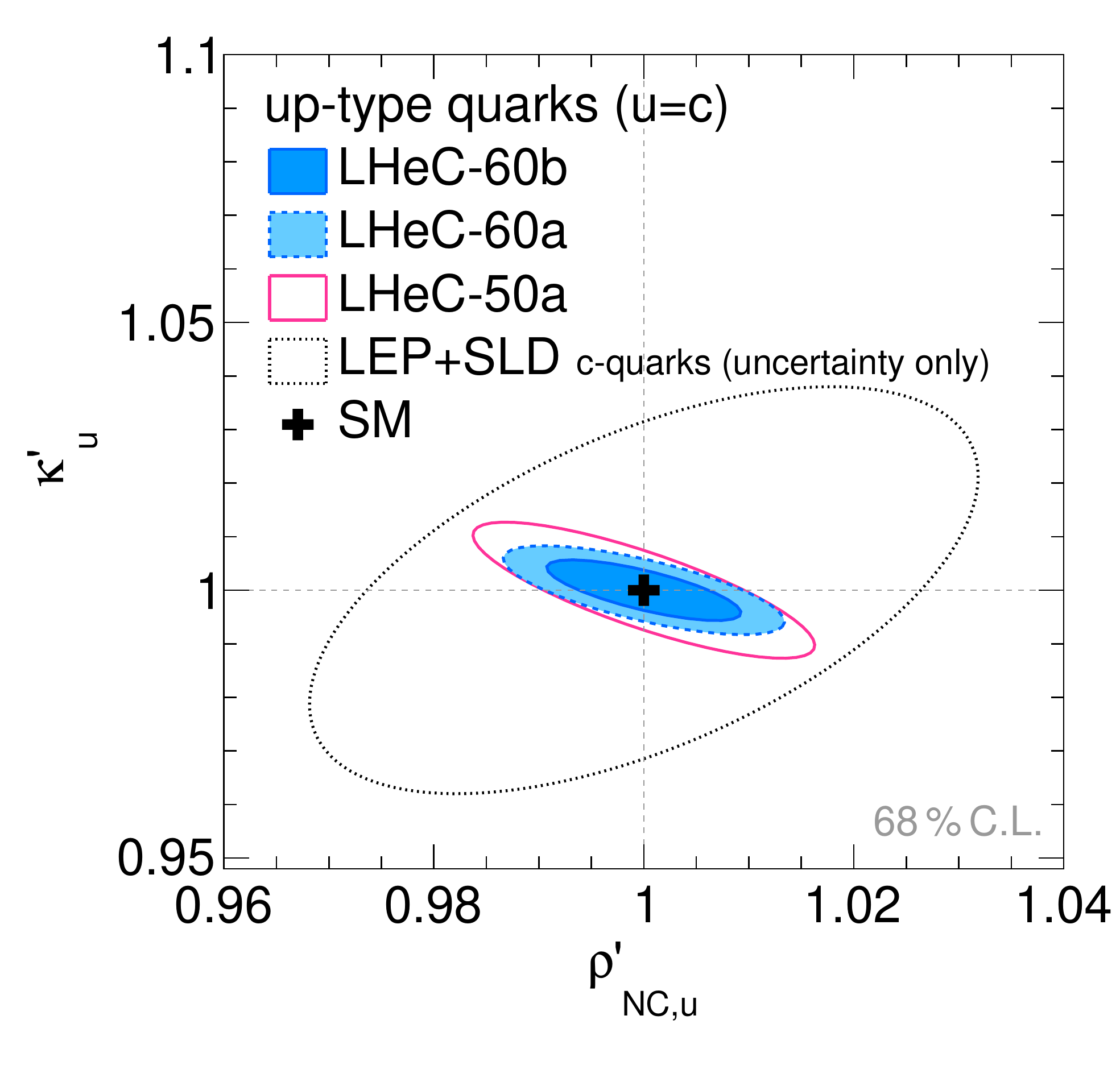}
  \hskip0.02\textwidth
  \includegraphics[width=0.48\textwidth]{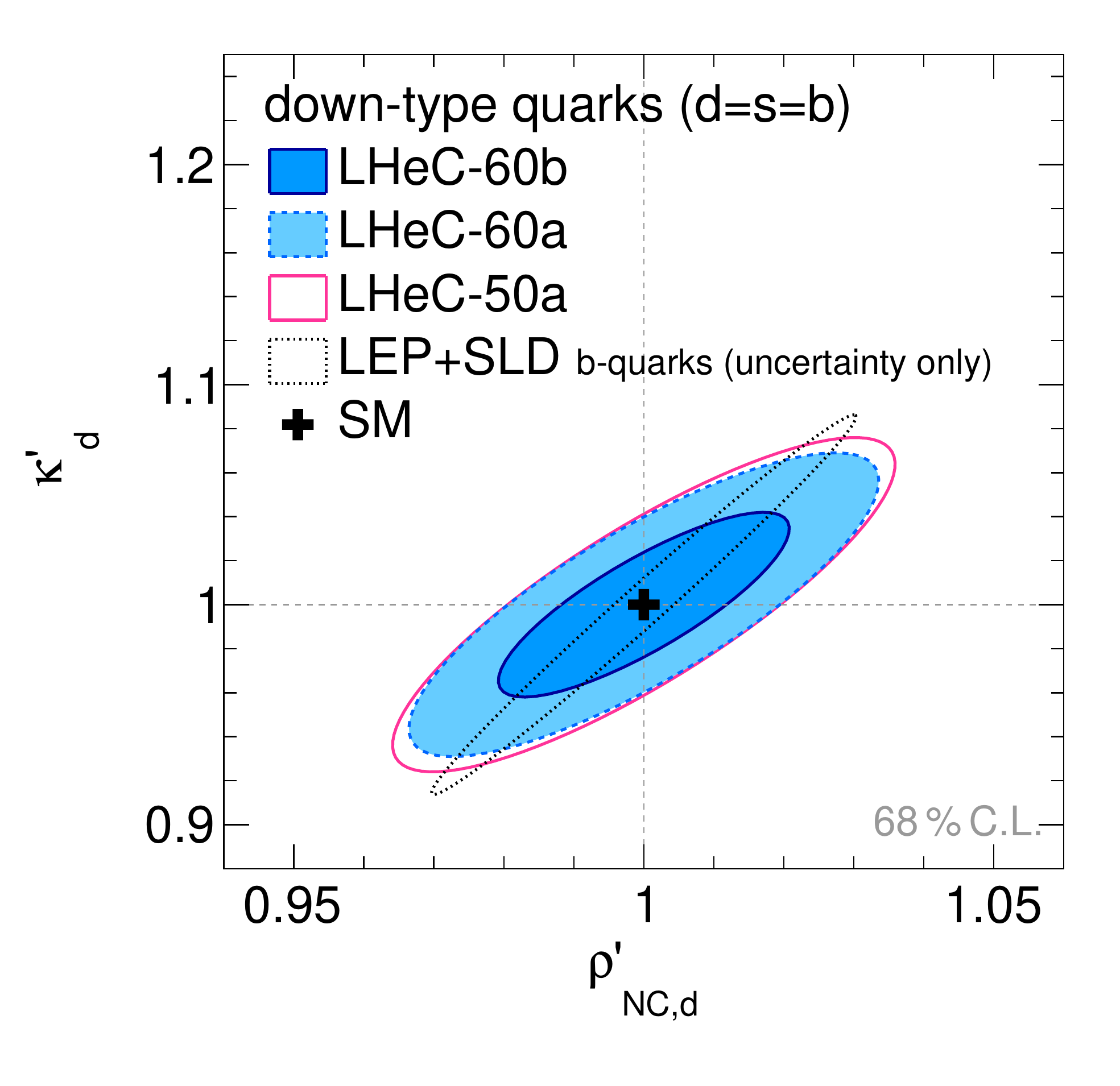}
  \caption{
    Expectations at the 68\,\% confidence level for the 
    simultaneous determination of anomalous up- and down-type 
    quark couplings, assuming electron couplings fixed at 
    their SM value. In the left panel, for up-type quarks, 
    the results are compared with uncertainties from LEP+SLD 
    for charm-quark anomalous couplings. The right panel for 
    down-type quarks shows a comparison of LHeC results with 
    LEP+SLD~\cite{ALEPH:2005ab} determinations of bottom-quark couplings. 
  }
\label{fig:rhokappa:ud}
\end{figure}

Next, we allow the anomalous form factors to be different 
for up- and down-type quarks, but assume the couplings of the 
electron as predicted by the SM. A precise knowledge of the 
up- and down-type couplings is particularly intersting since 
this may help to narrow down possible explanations of the 
flavor-structure of the SM. We perform a fit of the four 
anomalous parameters ($\rho_{\textrm{NC},u}^\prime$, 
$\kappa_{u}^\prime$, $\rho_{\textrm{NC},d}^\prime$, 
and $\kappa_{d}^\prime$). The resulting contours at 68\,\%~C.L.\ 
for a combination of two of the free parameters is shown in 
Fig.~\ref{fig:rhokappa:ud} (left panel for up-type, right panel 
for down-type quarks). The high-precision data from LEP+SLD did 
not allow for a full flavor-separated determination of quark 
couplings; however there are determinations of the couplings of 
the second- and third-generation quarks, charm and bottom, based 
on a data analysis using flavor tagging. It is interesting 
to compare the LHeC analysis, which is dominated by light-quark 
couplings, with these LEP+SLD results for heavy quarks. This is 
shown in Fig.~\ref{fig:rhokappa:ud} and we find that the 
uncertainties for up-type quarks are superior to those from 
LEP+SLD and comparable in the case of down-type quarks. 
The results for different LHeC scenarios are summarized in 
Tab.~\ref{tab:rhokappap}.

\begin{table}[bht!]
  \footnotesize
  \centering
  \begin{tabular}{lccccc}
    \toprule
    Fit parameters & Parameter & \multicolumn{4}{c}{Expected uncertainty}  \\
    \cmidrule(lr){3-6}
    &  & LHeC-60b& LHeC-60a& LHeC-50b & LHeC-50a  \\ 
    \midrule
    \rhop{,u}+\kapp{u}+\rhop{,d}+\kapp{d}+PDF
                            & \rhop{,u}  &  $\pm0.006$   &  $\pm0.009$   &  $\pm0.011$   &  $\pm0.014$  \\
                            & \kapp{u}   &  $\pm0.004$   &  $\pm0.005$   &  $\pm0.009$   &  $\pm0.011$  \\
                            & \rhop{,d}  &  $\pm0.014$   &  $\pm0.022$   &  $\pm0.024$   &  $\pm0.033$  \\
                            & \kapp{d}   &  $\pm0.028$   &  $\pm0.045$   &  $\pm0.050$   &  $\pm0.071$  \\
    \addlinespace
    \rhop{,u}+\kapp{u}+PDF  & \rhop{,u}  &  $\pm0.004$   &  $\pm0.006$   &  $\pm0.006$   &  $\pm0.009$  \\
                            & \kapp{u}   &  $\pm0.002$   &  $\pm0.003$   &  $\pm0.004$   &  $\pm0.005$  \\
    \addlinespace
    \rhop{,d}+\kapp{d}+PDF  & \rhop{,d}  &  $\pm0.008$   &  $\pm0.014$   &  $\pm0.013$   &  $\pm0.019$  \\ 
                            & \kapp{d}   &  $\pm0.014$   &  $\pm0.024$   &  $\pm0.023$   &  $\pm0.034$  \\ 
    \addlinespace
    \rhop{,f}+\kapp{f}+PDF  & \rhop{,f}  &  $\pm0.0015$   &  $\pm0.0025$   &  $\pm0.0033$   &  $\pm0.0043$  \\
                            & \kapp{f}   &  $\pm0.0010$   &  $\pm0.0015$   &  $\pm0.0025$   &  $\pm0.0031$  \\
    \addlinespace
    \rhop{,f}+PDF           & \rhop{,f}  &  $\pm0.0010$   &  $\pm0.0017$   &  $\pm0.0012$   &  $\pm0.0020$  \\
    \bottomrule
  \end{tabular}
  \caption{ 
    Overview of results for the \rhop{} and \kapp{} fits 
    in different LHeC scenarios. From top to bottom we 
    list results for 4-, 2- and 1-parameter+PDF fits. 
  }
  \label{tab:rhokappap}
\end{table}                                                                                                                                                                   


\begin{figure}[tb!]
  \centering
  \includegraphics[width=0.48\textwidth]{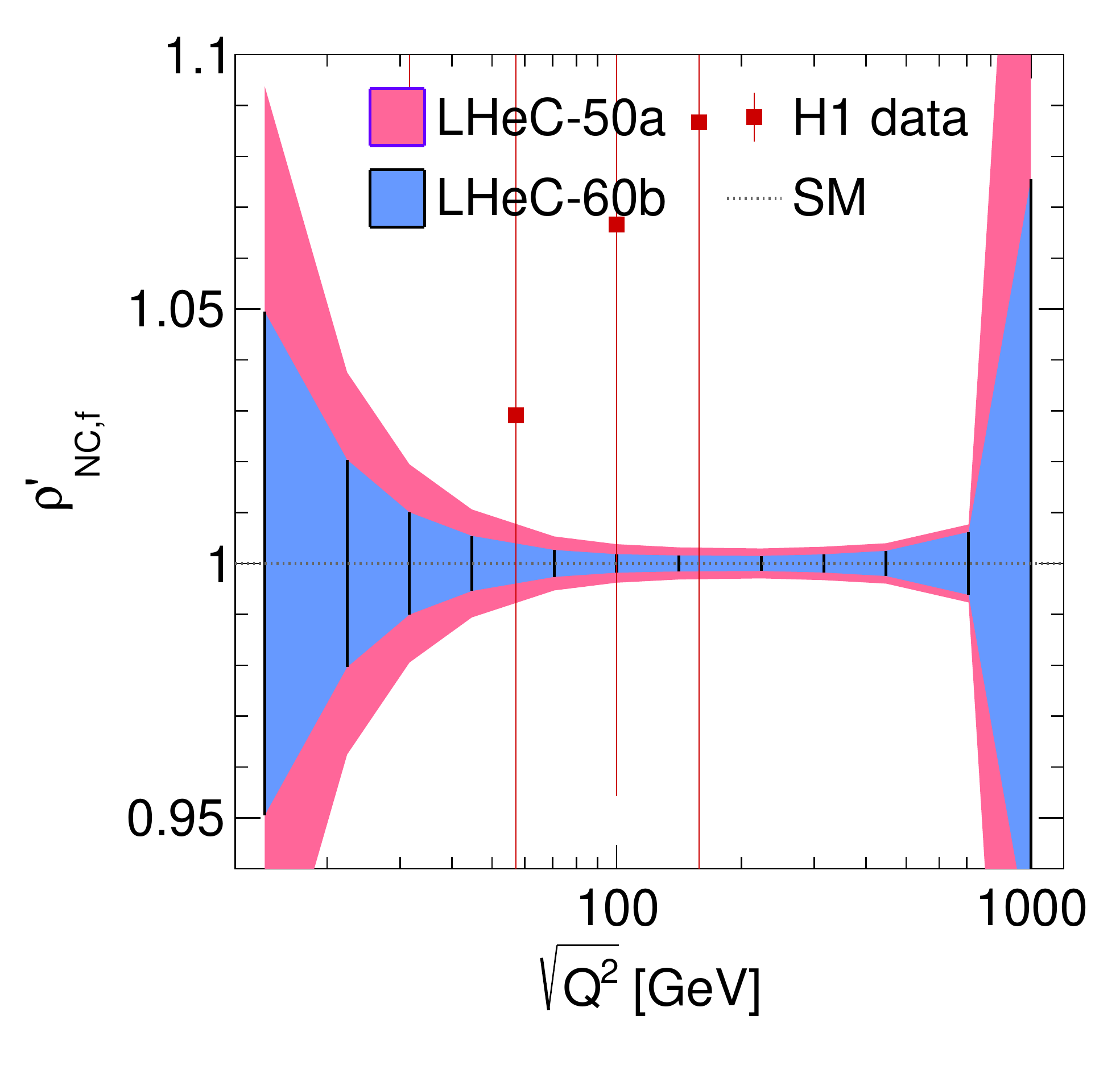}
  \hskip0.02\textwidth
  \includegraphics[width=0.48\textwidth]{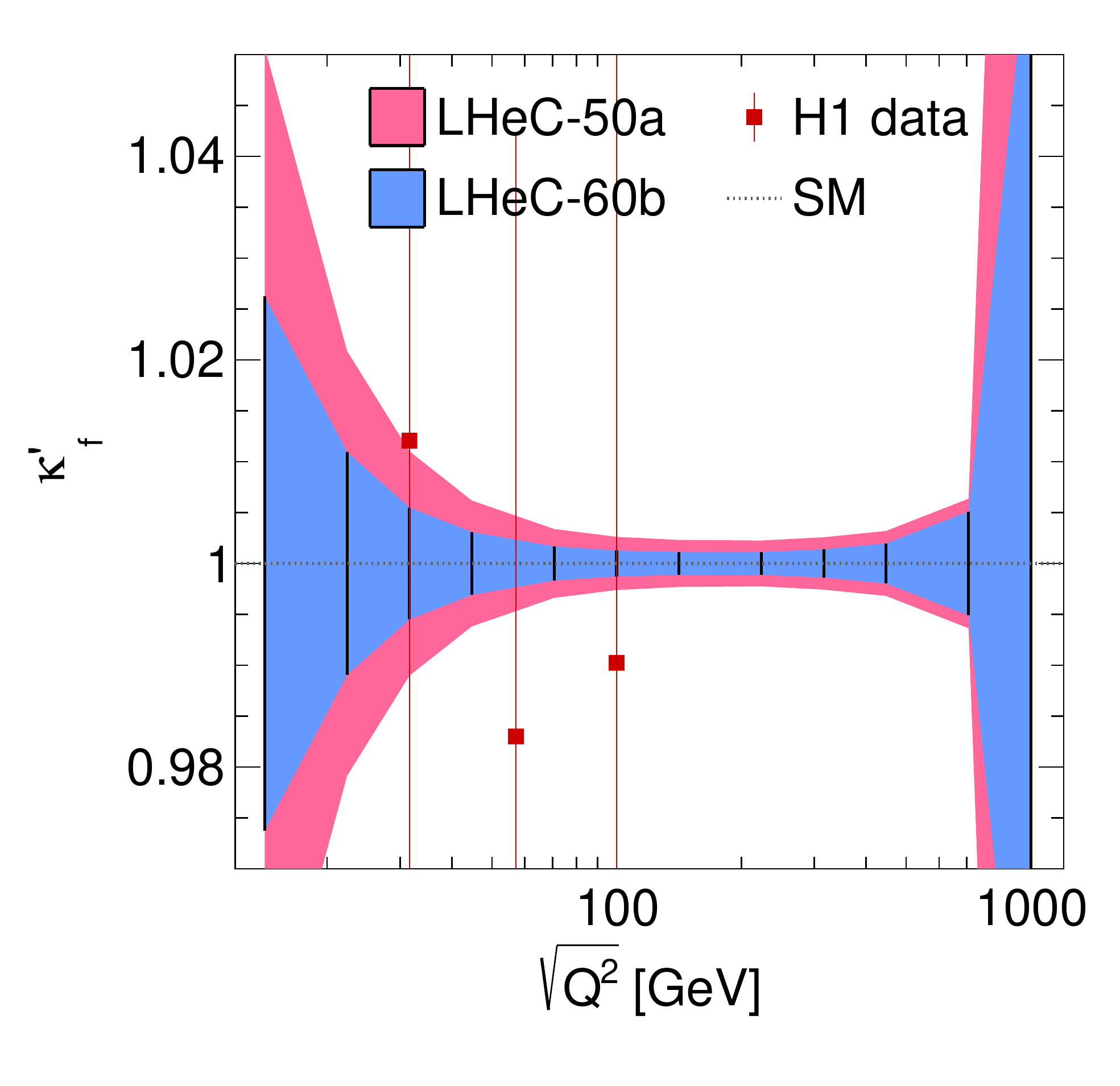}
  \caption{
    Scale dependence of the anomalous form factors
    $\rho^{\prime}_{\text{NC},f}(\mu^2)$ (left) and 
    $\kappa^{\prime}_f(\mu^2)$ (right) with $\mu^2 = - Q^2$ 
    for the scenarios  
    LHeC-50a and LHeC-60b. The highest precision is obtained 
    in the region of about $\Qsq\approx 20\,000\,\GeVsq$ for 
    scenario LHeC-60b.
    The expected uncertainties are compared to measured values by
    H1~\cite{Spiesberger:2018vki}.
    }
\label{fig:rhokappaQ2}
\end{figure}

The fact that DIS at the LHeC covers a huge range of \Qsq\ values 
allows us to perform a test of SM couplings which is not feasible 
at other experiments: one can determine the scale dependence of 
the anomalous form factors. Indeed, many models predict 
flavor-specific and \Qsq-dependent modifications. In order to 
study such a test, we perform fits of 
$\rho_\text{NC}^\prime$ and $\kappa^\prime$ to LHeC 
data split into twelve subsets with different \Qsq\ ranges. Our 
findings are shown in Fig.~\ref{fig:rhokappaQ2} for the scenarios 
LHeC-60a and LHeC-50a, where we include, for comparison, results 
obtained from H1 data~\cite{Spiesberger:2018vki}. At the LHeC 
we expect highest precision in the region of about $\Qsq \approx 
20\,000~\GeVsq$. In the worst case, for scenario LHeC-50a, we 
can expect uncertainties $\Delta \rhop\ = \pm 0.0029$ and $\Delta 
\kapp\ = \pm 0.0023$, while the best-case scenario LHeC-60b can 
provide a determination of the non-standard parameters with 
$\Delta \rhop\ = \pm 0.0015$ and $\Delta \kapp\ = \pm 0.0011$, 
i.e.\ in this case, the uncertainties obtained in the two extreme 
LHeC scenarios differ by a factor of 2.


\section{Electroweak effects in charged-current scattering}
\label{sec:ewcc}

The LHeC provides a unique opportunity to investigate 
charged-current scattering processes over many orders of 
magnitude in the momentum transfer $Q^2$ in a single experiment.
This is a consequence not only of the excellent detector 
performance like precise tracking, highly granular calorimetry 
and high-bandwidth triggers; particularly important is the fact 
that in CC DIS the event kinematics can be fully reconstructed 
from the measurement of the hadronic final state and the incoming 
electron beam four-momentum. 

Higher-order EW corrections to the CC DIS cross sections are 
collected in the effective couplings of the fermions to the $W$ 
boson as shown in Eqs.~(\ref{eq:w23el-NLO}, \ref{eq:w23po-NLO}). 
To allow for physics beyond the SM, we introduce new anomalous, 
primed parameters, $\rho^\prime_{\text{CC},\, eq}$ and  
$\rho^\prime_{\text{CC},\, e\bar{q}}$ 
\cite{Spiesberger:2018vki}, in a similar way as for the 
case of NC scattering. The modified CC structure functions 
then become 
\begin{eqnarray}
  W_2^- &=& 
  x \left( (\rho_{\text{CC}, eq}\rhopW{,eq})^2 U + 
  (\rho_{\text{CC},e\bar{q}}\rhopW{,e\bar{q}})^2 \overline{D} \right)
  \, , 
  \label{eq:rhocc1}
  \\
  xW_3^- &=& 
  x \left( (\rho_{\text{CC},eq}\rhopW{,eq})^2 U - 
  (\rho_{\text{CC},e\bar{q}}\rhopW{,e\bar{q}})^2 \overline{D} \right)
  \, ,
  \label{eq:rhocc2}
\\
  W_2^+ &=& 
  x \left( (\rho_{\text{CC},eq}\rhopW{,eq})^2 \overline{U}+ 
  (\rho_{\text{CC},e\bar{q}}\rhopW{,e\bar{q}})^2 D \right)
  \, ,
  \label{eq:rhocc3}
  \\
  xW_3^+ &=& 
  x \left( (\rho_{\text{CC},e\bar{q}}\rhopW{,e\bar{q}})^2 D - 
  (\rho_{\text{CC},eq}\rhopW{,eq})^2 \overline{U} \right)
  \, .
  \label{eq:rhocc4}
\end{eqnarray}

\begin{figure}[tb!]
  \centering
  \includegraphics[width=0.54\textwidth]{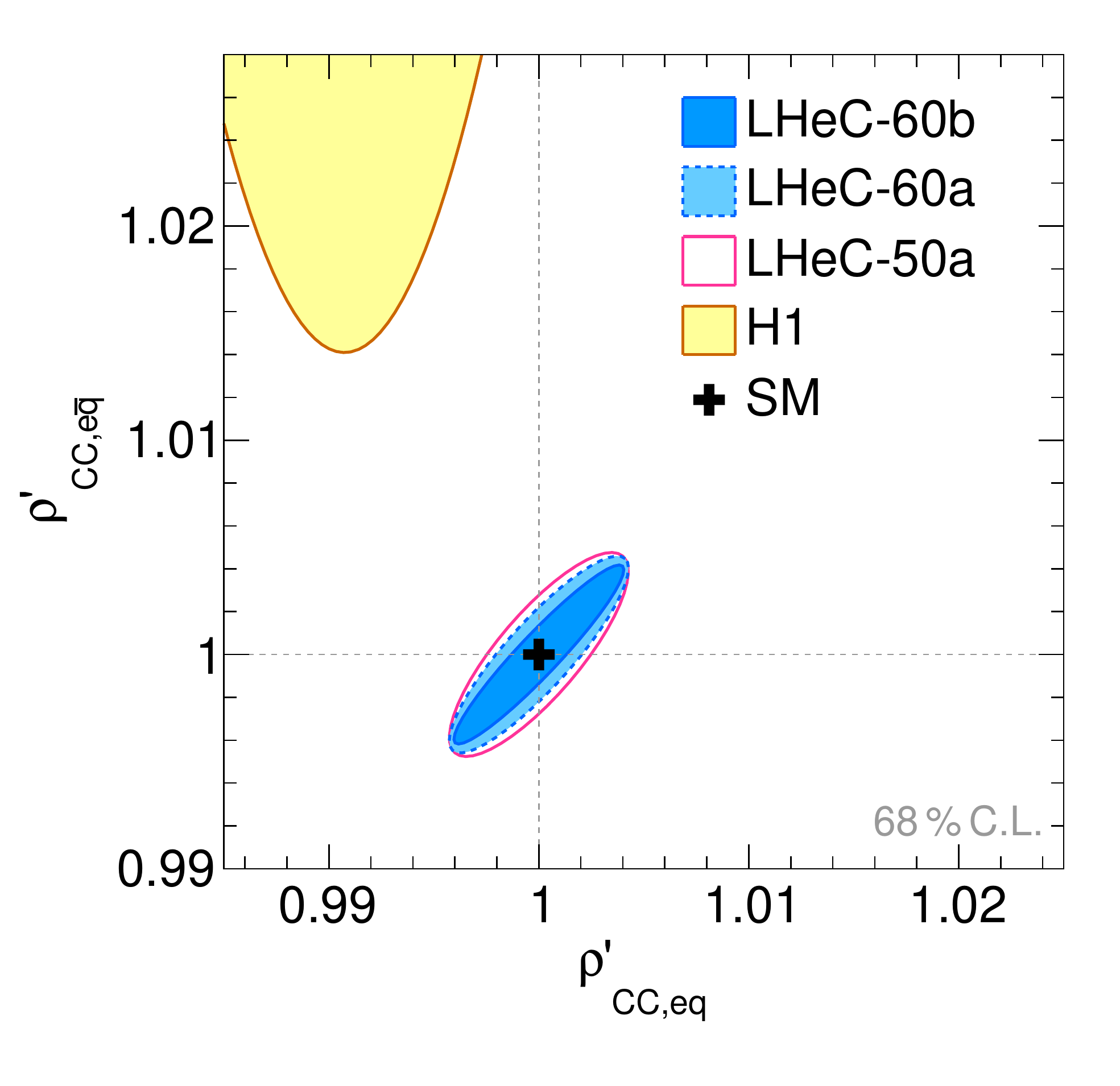} 
  \caption{
  Expected uncertainties of anomalous CC coupling parameters 
  \rhopW{,eq} and \rhopW{,e\bar{q}} for three different LHeC 
  scenarios compared with results from the H1
  measurement~\cite{Spiesberger:2018vki}. 
  }
\label{fig:rhoCC}
\end{figure}

\begin{table}[bt!]
  \footnotesize
  \centering
  \begin{tabular}{lccccc}
    \toprule
    Fit parameters & Parameter & \multicolumn{4}{c}{Expected uncertainty} \\
    \cmidrule(lr){3-6}
    & & LHeC-60b& LHeC-60a& LHeC-50b & LHeC-50a  \\
    \midrule
    \rhopW{,eq}+\rhopW{,e\bar{q}}+PDF  & \rhopW{,eq}         &   $\pm0.0027$  & $\pm0.0028$ & $\pm0.0027$ & $\pm0.0028$ \\
    \rhopW{,eq}+\rhopW{,e\bar{q}}+PDF  & \rhopW{,e\bar{q}}   &   $\pm0.0028$  & $\pm0.0030$ & $\pm0.0028$ & $\pm0.0031$ \\
    \addlinespace
    \rhopW{,eq}+PDF        & \rhopW{,eq}         &   $\pm0.0008$  & $\pm0.0013$ & $\pm0.0010$ & $\pm0.0015$ \\
    \addlinespace
    \rhopW{,e\bar{q}}+PDF  & \rhopW{,e\bar{q}}   &   $\pm0.0009$  & $\pm0.0014$ & $\pm0.0012$ & $\pm0.0018$ \\
    \addlinespace
    \rhopW{,f}+PDF  & \rhopW{,f}                 &   $\pm0.0017$ &   $\pm0.0019$ &   $\pm0.0016$ &   $\pm0.0018$ \\
    \addlinespace
    \midrule
    \addlinespace
    \rhopW{,f}+\kapp{f}+PDF  & \rhopW{,f} &  $\pm0.0017$   &  $\pm0.0019$   &  $\pm0.0016$   &  $\pm0.0018$  \\ 
                             & \kapp{f}   &  $\pm0.0006$   &  $\pm0.0011$   &  $\pm0.0009$   &  $\pm0.0015$  \\ 
    \bottomrule
  \end{tabular}
  \caption{
    Expected uncertainties of anomalous CC coupling parameters 
    \rhopW{,eq} and \rhopW{,e\bar{q}} from 2-parameter (upper 
    part) and 1-parameter fits (lower part). The last 
    two lines show the results from a fit combining the 
    CC parameter $\rhopW{}$ with the NC $\kappa^\prime$ 
    parameter (see Eq.~(\ref{eq:sw2eff-chapter10})). 
  }
\label{tab:rhoCC}
\end{table}
The prospects for a determination of these anomalous couplings 
with LHeC data are obtained by performing a fit of the two 
parameters $\rho^\prime_{\text{CC},eq}$ and 
$\rho^\prime_{\text{CC},e\bar{q}}$ together with the PDFs.
The expected uncertainties for the LHeC-50a and LHeC-60a 
scenarios are displayed in Fig.~\ref{fig:rhoCC} and collected 
in Tab.~\ref{tab:rhoCC}. We find that these parameters can be 
determined with a relative uncertainty of better than 0.3\,\%.
For the LHeC-60b scenario, even smaller uncertainties can be 
achieved and we find in 1-parameter+PDF fits relative uncertainties 
below one per mille. We can also consider a fit combining the 
CC parameters $\rho^\prime_{\text{CC},eq} = 
\rho^\prime_{\text{CC},e\bar{q}} =: \rho^\prime_{\text{CC},f}$ 
with the anomalous NC parameter $\kappa^\prime$ (see 
Eq.~(\ref{eq:sw2eff-chapter10})). Results for this case are 
also shown in Tab.~\ref{tab:rhoCC}.
Since the determination of the $\rho^\prime_{\text{CC}}$ parameters are strongly
correlated with the normalization uncertainty of the data, the study
benefits from the simultaneous analysis of NC and CC DIS data.
By doing so, not only the PDFs are constrained, but also systematic
uncertainties that are common to NC and CC DIS data, mainly the 
luminosity uncertainty, are
reduced by the NC DIS data, and therefore smaller uncertainties
are obtained in this analysis than in a fit with CC DIS data alone.

\begin{figure}[tb!]
  \centering
  \includegraphics[width=0.54\textwidth,trim=0 23 0 15,clip]{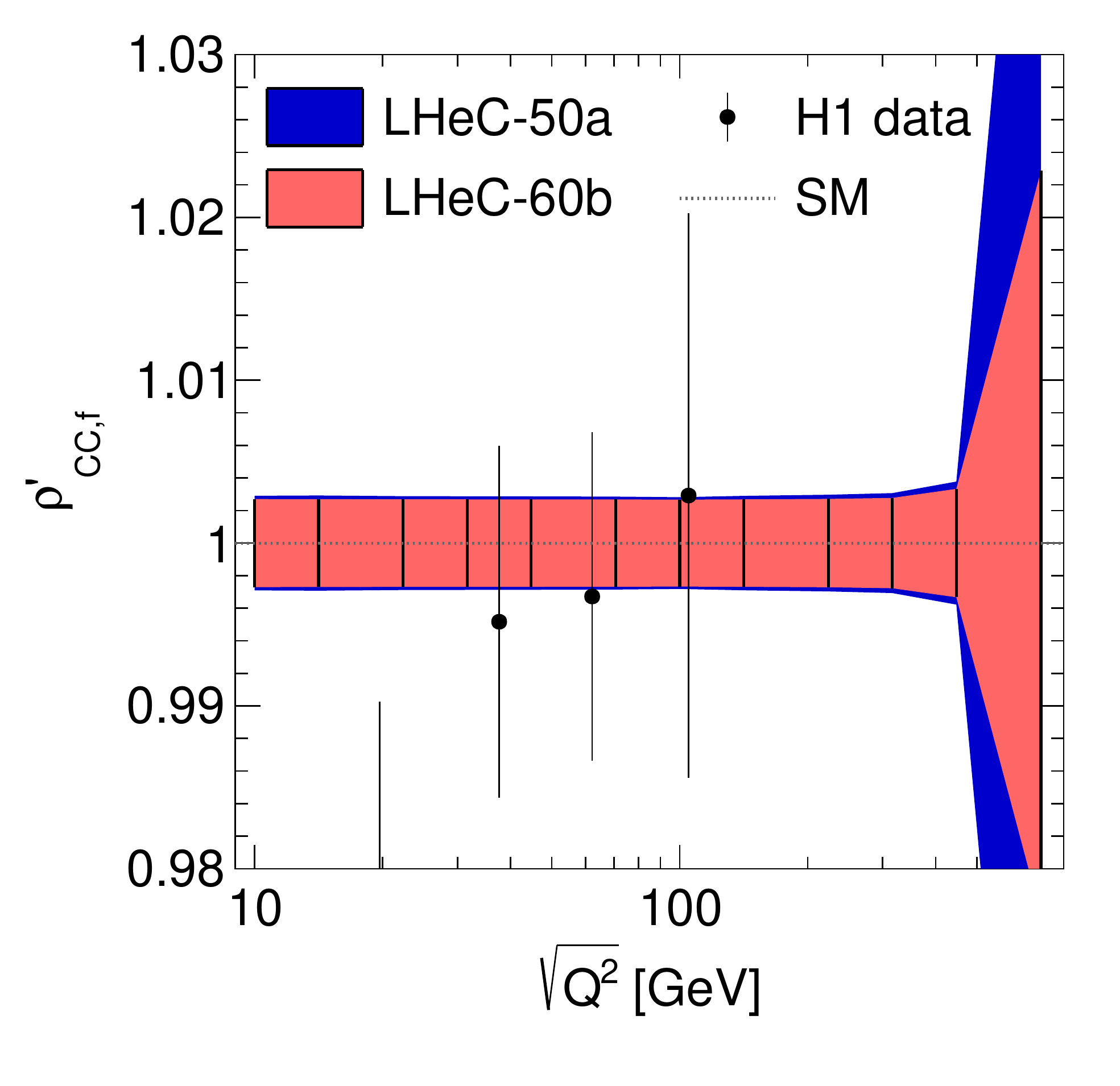}
  \caption{
  Scale dependent determination of the anomalous CC coupling 
  parameters, assuming $\rho^\prime_{\text{CC},eq} = 
  \rho^\prime_{\text{CC},e\bar{q}} = \rho^\prime_{\text{CC},f}$.
  For comparison, values from H1~\cite{Spiesberger:2018vki} are also displayed.
  }
\label{fig:rhoCCvsQ2}
\end{figure}
Event rates at the LHeC are expected to be large and will 
cover a large \Qsq range between $10^2$~GeV$^2$ and almost 
$1000^2$~GeV$^2$. It is therefore possible to determine 
the anomalous CC couplings in different \Qsq ranges. Our 
results are shown in Fig.~\ref{fig:rhoCCvsQ2}, assuming one 
flavor-independent parameter $\rho^\prime_{\text{CC},f} =  
\rho^\prime_{\text{CC},eq} = \rho^\prime_{\text{CC},e\bar{q}}$. 
The \Qsq range is split into twelve bins, and for each bin the 
coupling parameter $\rho^\prime_{\text{CC},f}$ was allowed to 
vary independently in a EW+PDF fit. We find uncertainties below 
0.3~\% in the \Qsq bins up to about $500^2$~GeV$^2$. They are 
dominated by the normalization uncertainties of the data. Higher 
center-of-mass energies, i.e.\ with $E_e=60~\GeV$ instead of 
50~\GeV, has therefore only a moderate impact on the size of the 
uncertainties in the central \Qsq region. However, a larger beam 
energy allows one to extend the measurement to higher \Qsq values.


\section{SM weak neutral-current couplings}
\label{sec:nccouplings}

The NC DIS cross sections are determined by products of the weak 
neutral-current coupling constants of the quarks. They enter 
through the $\gamma Z$ interference and $Z$ exchange terms in 
the generalized structure functions defined in Eqs.~(\ref{eq:last1}, 
\ref{eq:last2}). Here we focus on the inclusive measurement of 
DIS cross sections and do not discuss the possibility that 
individual quark flavors might be identified in the final state 
(e.g.\ for charm and bottom). Therefore a full flavor separation 
of quark couplings will not be possible. However, mainly two 
effects allow us to separate the up-type and the down-type quark 
contributions to the cross section: first, they carry different 
electric charge and contribute with different weights to the 
$\gamma Z$ interference terms; second, they affect, through 
$\tilde{F}_3^\pm$, the dependence on the polarization and the 
lepton charge. In fact, these effects due to the weak interaction 
are important predominantly at higher values of \Qsq, corresponding 
to large $x$, where the up- and down-valence quark PDFs dominate. 
A determination of vector and axial-vector couplings of up-type 
and down-type quarks can therefore be interpreted, with high 
precision, as a determination of the coupling constants of the 
up- and down-quarks, i.e.\ of light quarks only. Only for the 
down-type couplings, a contribution from strange quarks has
some relevant size. 

\begin{figure}[tb!]
  \centering
  \includegraphics[width=0.48\textwidth]{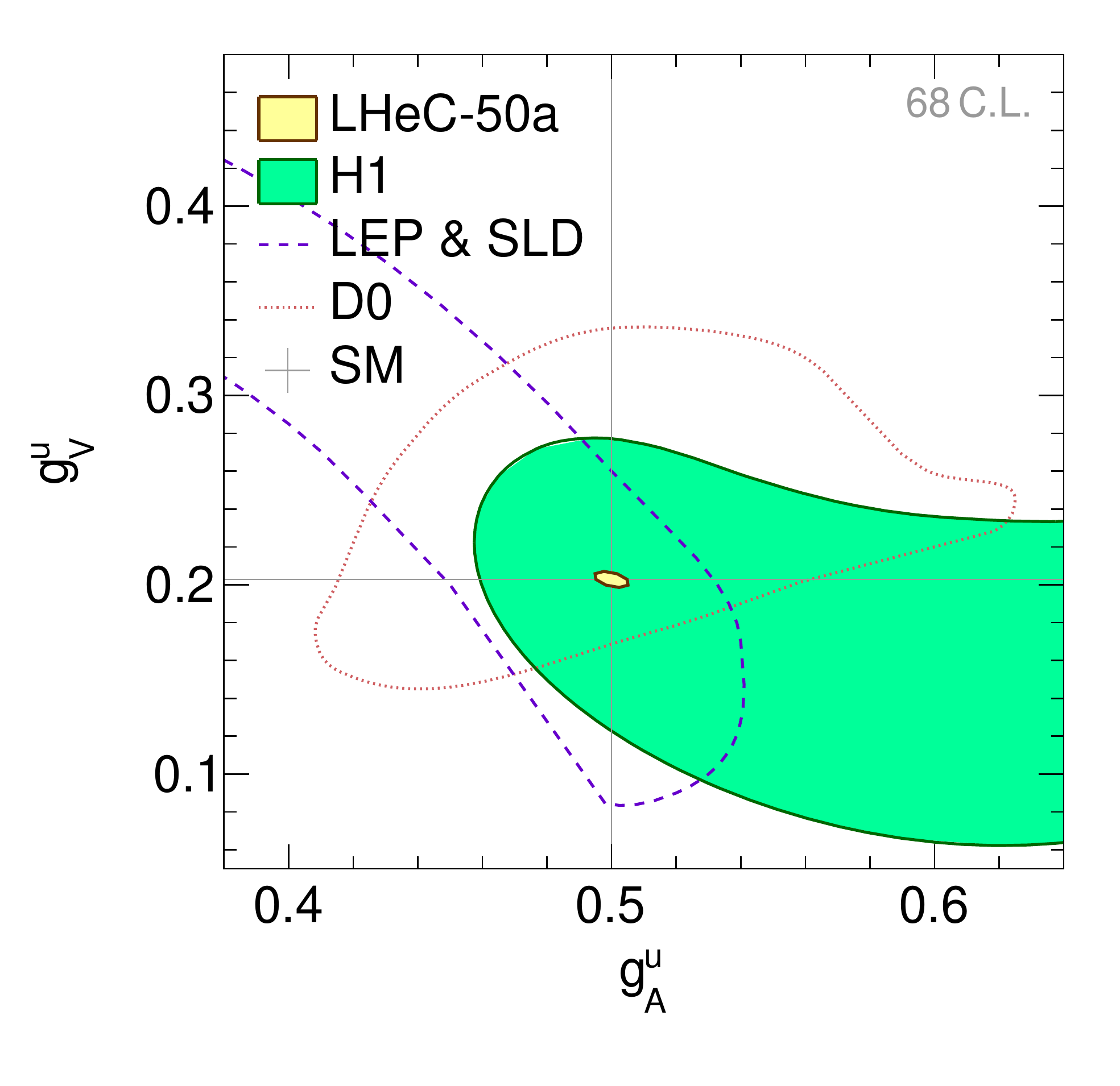}
  \hskip0.02\textwidth
  \includegraphics[width=0.48\textwidth]{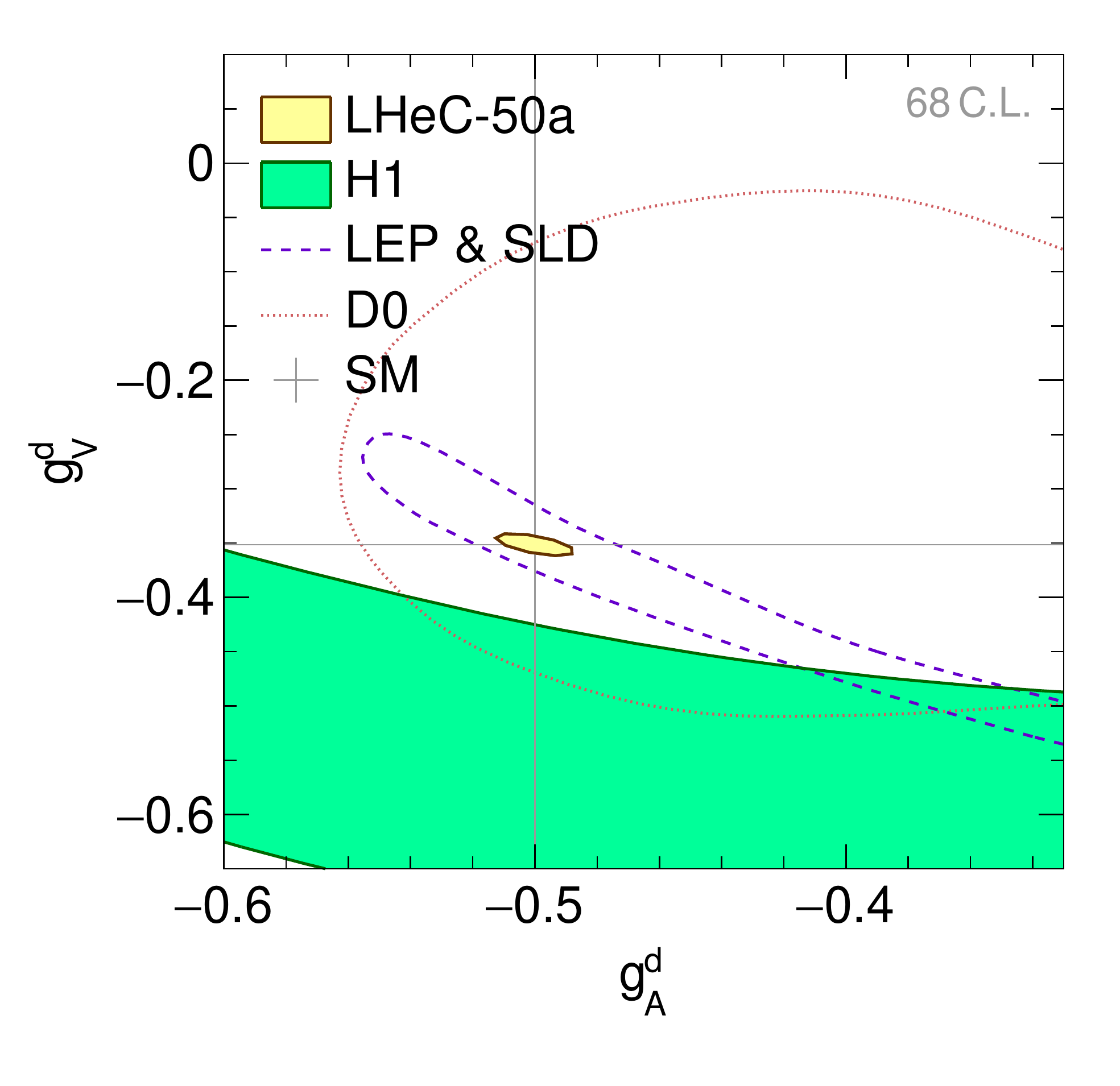}
  \caption{
    Weak-neutral-current vector and axial-vector couplings of 
    up-type quarks to the $Z$-boson (left), and those of the 
    down-type quarks (right) at 68\,\% confidence level for
    simulated LHeC data with $E_e=50\,\GeV$ (scenario LHeC-50a).
    The LHeC expectations are compared with results from the 
    combined LEP+SLD experiments~\cite{ALEPH:2005ab} and single 
    measurements by D0~\cite{Abazov:2011ws} 
    and H1~\cite{Spiesberger:2018vki}.
    The standard model expectations are at the crossing of the
    horizontal and vertical lines. 
  }
\label{fig:couplings}
\end{figure}

We perform an EW+PDF fit where the vector and axial-vector 
couplings of up-type ($u$ and $c$) and down-type quarks ($d$, 
$s$ and $b$), denoted as $g_V^{u}$, $g_V^{d}$ and $g_A^{u}$, 
$g_A^{d}$, are free parameters in one single fit. Other 
parameters, in particular the lepton couplings, are 
fixed\,\footnote{A fit to determine the electron couplings 
  will not be competitive with corresponding determinations 
  from $Z$-pole observables (see last two lines of 
  Tab.~\ref{tab:couplings}). 
}. The fit parameters 
are identified with the coupling constants defined in the Born 
approximation, and \Qsq\ dependent higher-order corrections are 
calculated in the SM formalism in the 1-loop approximation.
We have verified that the results of this analysis is consistent 
with those of the anomalous form factors described above in  
Sec.~\ref{sec:rhokappa} (cf.\ Fig.~\ref{fig:rhokappa:ud}).
The resulting uncertainties of the fits are summarized in 
Tab.~\ref{tab:couplings}
for different LHeC scenarios. Fig.~\ref{fig:couplings} shows 
the results for the LHeC-50a scenario, compared with the current 
most precise measurements.
All other LHeC scenarios result in even smaller uncertainties.

\begin{table}[t!]
\footnotesize
  \centering
  \begin{tabular}{cr@{\hskip4pt}lcccc}
    \toprule
    Coupling  &  \multicolumn{2}{c}{PDG} & \multicolumn{4}{c}
       {Expected uncertainties} \\
       \cmidrule(lr){4-7}
     parameter &  &  & LHeC-60b & LHeC-60a & LHeC-50b & LHeC-50a \\
    \midrule
    $\au$  & $0.519$   & $^{+0.028}_{-0.033}$  & $\pm0.0015$ & $\pm0.0022$ & $\pm0.0027$ & $\pm0.0035$ \\
    $\ad$  & $-0.527$  & $^{+0.040}_{-0.028}$  & $\pm0.0034$ & $\pm0.0055$ & $\pm0.0059$ & $\pm0.0083$ \\
    $\vu$  & $0.266$   & $\pm0.034$            & $\pm0.0010$ & $\pm0.0015$ & $\pm0.0020$ & $\pm0.0028$ \\
    $\vd$  & $-0.38$   & $^{+0.04}_{-0.05}$    & $\pm0.0027$ & $\pm0.0046$ & $\pm0.0059$ & $\pm0.0067$ \\
    \midrule
    $\au$  &  &  & $\pm0.0009$ & $\pm0.0015$ & $\pm0.0015$ & $\pm0.0022$   \\
    $\vu$  &  &  & $\pm0.0005$ & $\pm0.0008$ & $\pm0.0009$ & $\pm0.0013$   \\
    \midrule
    $\ad$  & & & $\pm0.0020$ & $\pm0.0034$ & $\pm0.0032$ & $\pm0.0047$   \\
    $\vd$  & & & $\pm0.0013$ & $\pm0.0024$ & $\pm0.0019$ & $\pm0.0031$   \\
    \midrule
    $\gae$  & $-0.50111$ & $\pm 0.00035$ & $\pm0.0009$ & $\pm0.0014$ & $\pm0.0018$ & $\pm0.0025$   \\
    $\ve$   & $0.03817$  & $\pm 0.00047$ & $\pm0.0014$ & $\pm0.0022$ & $\pm0.0031$ & $\pm0.0041$   \\
    \bottomrule
\hline
  \end{tabular}
  \caption{
    Present values and uncertainties of the light-quark 
    ($\au$, $\ad$, $\vu$, $\vd$) and electron ($\gae$, $\ve$) 
    weak neutral-current couplings from the PDG~\cite{PDG2020} 
    and the uncertainties expected for inclusive DIS measurements 
    in different LHeC scenarios, obtained in a simultaneous EW+PDF
    fit. The upper section shows results of a fit where all 4 
    couplings are free fit parameters, the lower three sections 
    are results from two-parameter+PDF fits.  
    }
\label{tab:couplings}
\end{table}

Current determinations of light-quark couplings from $e^+e^-$, 
$ep$ or $p\bar{p}$ collisions all appear with a similar precision. 
Future measurements at the LHeC, however, will greatly improve 
the measurement of these EW parameters. The scenario with $E_e = 
60~\GeV$ and the optimistic assumptions for systematic uncertainties 
is particularly promising, see Tab.~\ref{tab:couplings}. In this 
table we also show results from fits where only two couplings are 
free fit parameters while the other couplings are fixed at their 
SM value. We find that the uncertainties of light-quark couplings 
can be improved by more than an order of magnitude through LHeC 
data.

At the LHeC, the vector and axial-vector couplings can be 
disentangled without any sign-ambiguity, since the DIS cross 
sections receive important contributions from the interference 
of photon and $Z$-boson exchange diagrams. This is in contrast 
with $Z$-pole observables where only squares of the couplings 
are accessible. The determination of quark NC couplings will 
provide a unique possibility for testing the EW SM theory. 
Such a measurement cannot be performed with a comparably high 
precision in other experiments.

\clearpage

\section{Summary and Conclusion}
\label{sec:conclusions}

The proposed LHeC experiment at CERN's HL-LHC will provide a 
unique opportunity for high precision electroweak physics in 
neutral- and charged-current interactions in a yet completely 
unexplored kinematic regime of spacelike momentum transfer. 

In this article we have  simulated inclusive NC and CC 
deep-inelastic scattering cross section data at electron-proton 
center-of-mass energies of 1.2 and 1.3\,TeV, and assessed their 
sensitivity to a number of parameters of the electroweak theory 
and the sensitivity to possible generic extensions beyond the SM.
Our theoretical calculations include next-to-next-to-leading 
order QCD corrections to the DGLAP evolution of PDFs and their 
relation to the structure functions and the full set of one-loop 
electroweak corrections to electron--(anti-)quark $t$-channel 
scattering in the on-shell renormalization scheme. 
Our simulated pseudo-data comprise a full set of experimental systematic
uncertainties and have been used also elsewhere to study the 
prospects for a determination of parton distribution functions. 
The latter are implicitly also included in our analysis framework,
while we extend the phenomenological analysis towards electroweak 
parameters.

%
\begin{table}[b!]
  \small
  \centering
  \begin{tabular}{lcccccc}
    \\
    \toprule
     & & & \multicolumn{4}{c}{Expected uncertainty}  \\
    \cmidrule(lr){4-7}
    Fit parameters & Parameter & Unit & LHeC-60b& LHeC-60a& LHeC-50b & LHeC-50a  \\
    \midrule
    \mw+PDF      &  \mw & MeV  &  $\pm6$ & $\pm10$ & $\pm8$ & $\pm12$ \\
    \addlinespace
    \mw+\mz+PDF  &  \mw & MeV  &  $\pm17$ & $\pm28$ & $\pm24$ & $\pm37$ \\
                 &  \mz & MeV  &  $\pm19$ & $\pm31$ & $\pm26$ & $\pm40$ \\
    \addlinespace
    \sw+PDF      &  \sw & $10^{-5}$   &  $\pm15$ & $\pm25$ & $\pm22$ & $\pm34$ \\
    \addlinespace
    \mt+PDF      &  \mt & GeV  &  $\pm1.1$ & $\pm1.8$ & $\pm1.4$ & $\pm2.2$ \\
    \addlinespace
    \mW+\mt+PDF  &  \mW & MeV  &  $\pm18$ & $\pm29$ & $\pm24$ & $\pm38$\\
                 &  \mt & GeV  &  $\pm3.3$ & $\pm5.4$ & $\pm4.4$ & $\pm6.9$\\
    \addlinespace
    $\ln\mH$+PDF      & $\ln\tfrac{\mH}{\GeV}$ &   &  $\pm0.11$ & $\pm0.17$ & $\pm0.13$ & $\pm2.0$ \\
    $\ln\mH$+PDF      &    \mH & GeV  &  $^{+14}_{-13}$ &  $^{+24}_{-20}$  &  $^{+17}_{-15}$ &  $^{+28}_{-23}$ \\
    \bottomrule
  \end{tabular}
  \caption{
    Prospects for the determination of Standard Model parameters 
    from simulated inclusive NC and CC DIS data at the LHeC.
    Scenarios for electron beam energies of $E_e=50$ and 
    60\,\GeV, and with two assumptions for experimental 
    uncertainties, denoted scenarios `a' and `b', are studied.
  }
  \label{tab:SM-summary}
\end{table}

The sensitivity of inclusive NC/CC DIS data at the LHeC to 
important independent parameters of the electroweak Standard Model 
are summarized in Tab.~\ref{tab:SM-summary}. At the LHeC, the high 
experimental precision for SM parameters is due to the fact that 
a large kinematic range of space-like momentum transfer can be 
used to obtain a large amount of DIS data points that all 
contribute to the sensitivity. For our prospects, we perform 
combined fits of EW parameters and PDFs to simulated NC and CC 
DIS data. 
The large number of cross section data in different $Q^2$ bins 
and including both NC and CC scattering, as well as scattering 
with positrons and with polarized electrons, is equivalent to 
a large number ratios of bin cross sections. In particular, the 
ratios of NC cross sections at large and small $Q^2$ carry 
information about electroweak parameters. Therefore, correlated 
(normalization) uncertainties cancel to a large extent and play 
a minor role, while uncorrelated uncertainties are reduced by the 
implicit averaging over a set of several hundreds of independent 
measurements. This explains why we observe that the uncertainties 
for most of the electroweak parameters are found at the per mille 
level while the uncertainties for individual cross section data 
points are in the order of percent.
For instance the weak mixing angle can be determined with a high 
experimental uncertainty of down to $\pm1.5\cdot10^{-4}$ which 
corresponds to 0.65 per mille, as shown in Tab.~\ref{tab:SM-summary}.
The large number of data points at different \Qsq\ will also 
allow a determination of the scale dependence of the weak mixing 
angle in the spacelike regime of about $25<\sqrt{\Qsq}<700\,\GeV$. 
Experimental uncertainties below $\pm20\cdot10^{-4}$ will be 
possible in the range $60\lesssim\sqrt{\Qsq}\lesssim450\,\GeV$. 
This analysis will be equivalent to a determination of the 
\emph{running} of the weak mixing angle, an opportunity which 
is unique to the LHeC. Only small and well-known correction 
factors from theory will be needed to turn the result of such 
measurements into a determination of the effective weak mixing 
angle, or the weak mixing angle in the $\overline{\rm MS}$ 
scheme, which then can be compared with other measurements.

In the on-shell scheme, the weak mixing angle is defined by 
the ratio of the weak boson masses, \mw\ and \mz. The 
measurement of inclusive DIS cross sections can therefore 
be interpreted as an indirect determination of the $W$-boson 
mass. We find from our combined \mw+PDF fit an experimental 
uncertainty down to $\pm6\,\MeV$. This is compatible with a 
rough estimate, based on simple error propagation, of an expected 
relative uncertainty at the level of $10^{-4}$. This high precision 
will improve all present measurements and provide a highly valuable 
validation of future improved direct \mw\ measurements with 
$\mathcal{O}(\MeV)$ accuracy.

The high precision of the Born-level parameters of the EW theory 
also suggests a determination of the dominant parameters of the 
higher-order EW corrections, most notably the top-quark mass \mt. 
The value of \mt\ can be determined with an uncertainty down to 
$\pm 1.1\,\GeV$.
This provides a precise indirect determination 
which is complementary to direct measurements, since it is theoretically 
clean and free from ambiguities due to unknown QCD corrections.
The dependence of higher-order corrections on the Higgs-boson 
mass, \mH, is logarithmic only, i.e.\ sub-dominant. Therefore, 
its uncertainty is fairly large and compares in size with the 
indirect determinations from the LEP+SLD data.
Non-Standard Model contributions to one-loop weak boson self 
energy corrections, usually described by the so-called 
\emph{oblique} parameters $S$, $T$ and $U$, can also be determined 
independently when NC and CC DIS data are considered together, and
uncertainties of a few percent are expected.

A simultaneous determination of the weak boson masses, \mw\ and 
\mz, or a simultaneous determination of \mw\ together with \mt, 
yield moderate uncertainties. These, however, compare well with  
the uncertainties that can be achieved nowadays in global EW fits. 
We have studied in a simplified formalism the potential impact of
LHeC NC/CC DIS data to such a global EW fit and found only small 
improvements; the correlation between \mW\ and \mt\ or \mw\ and 
\mz\ are not very different in DIS than in other observables. 

The $\textrm{SU}(2) \times \textrm{U}(1)$ gauge symmetry predicts 
the weak NC couplings of fermions, \vf\ and \af, in a unique way. 
Modifications of the SM can be studied in a rather 
model-independent manner by considering these coupling constants 
as free parameters, 
or alternatively, by introducing multiplicative anomalous
factors to the $\rho_{\text{NC},f}$ and $\kappa_f$ form factors,
which can be considered to be flavor and \Qsq\ dependent. 
While the NC coupling parameters have been measured at the $Z$-pole 
with high precision, in particular for leptons and $b$ quarks, 
the couplings of the light quarks, $u$ and $d$, are experimentally 
measured only with a rather poor precision.
The LHeC, in contrast,
has a high sensitivity to the NC couplings of light quarks, and
DIS data will provide the unique opportunity to 
determine the light-quark couplings with per mille accuracy, 
either for the vector and axial-vector coupling constants (\vu, 
\au, \vd\ and \ad), or for the anomalous form factors, 
$\rhop{f}$ and $\kapp{f}$.
In addition, their \Qsq\ dependence 
can be determined with percent precision in the range 
$60\lesssim\sqrt{\Qsq}\lesssim600\,\GeV$. 

Many precise data for parameters of the NC sector of the EW 
theory can be found in the literature, based on measurements 
extending to highest energies. In contrast, high-precision 
measurements of CC interactions are often restricted to low 
energies. Here, the LHeC will offer once more unique opportunities. 
Anomalous electron--quark and electron--anti-quark form factors 
can be determined with an accuracy in the order of 
$\mathcal{O}(0.1\,\%)$, and their potential \Qsq\ dependence 
can be measured with per mille uncertainties in a \Qsq\ range 
up to about $400^2\,\GeVsq$.

%
We have compared LHeC inclusive DIS pseudo data at two different 
center-of-mass energies, $\sqrt{s}=1.2$ and 1.3\,\TeV, corresponding 
to an electron beam energy of 50 or 60\,\GeV, respectively. Our  
studies confirm the presumption that higher center-of-mass 
energies would be quite valuable for the precise determination 
of EW parameters. Of course, with higher $\sqrt{s}$, also the 
range of \Qsq\ dependent studies can be extended to some extent. 
Further improvements will be obtained by higher resolution of 
the detector, which would allow us to obtain measurements for 
a larger set of independent data points. Altogether, the expected 
uncertainties from the four LHeC scenarios studied in this article 
differ by about a factor of 2 to 3. Such an improvement is 
indeed interesting in view of the fact that many parameters are 
measured with comparable uncertainties in $e^+e^-$ or hadron-hadron 
interactions.

In this study, we employed calculations in the on-shell
renormalization scheme including the full set of one-loop EW 
corrections. For our main purpose, i.e.\ to investigate prospects 
for the uncertainties of EW parameters, this provides a valid 
framework. However, once real LHeC data are available, a more 
careful study of higher-order corrections will be necessary.
On the one hand, one has to match the high experimental precision 
of the data with correspondingly accurate theoretical predictions.
On the other hand, a conclusive test of the SM can only be achieved 
by comparing as many as possible different measurements, i.e.\ 
also observables in processes other than DIS. Therefore, a 
consistent framework for the calculation of higher-order 
corrections to all observables in question will be needed. 
For example, different definitions of the effective weak mixing 
angle have to be matched to each other. We believe that our study 
provides a motivation for the investigation and calculation of 
higher-order corrections to DIS observables beyond the 1-loop 
level, and in different renormalization schemes. 

%
In conclusion, the high center-of-mass energy, the large kinematic 
range, and the large integrated luminosity at the LHeC will allow 
for the first time to perform precision electroweak measurement 
at high momentum transfer in NC and CC DIS.
Rich aspects of the electroweak theory can be probed with highest 
precision without being limited by uncertainties from parton 
distribution functions. Instead, electroweak parameters and 
PDFs can be determined in simultaneous fits taking into account 
their mutual correlations. In many cases, the results are 
complementary to direct measurements in $e^+e^-$ or hadron-hadron 
collisions, such as the indirect determination of the $W$-boson 
mass or the top-quark mass. A unique measurement can be expected 
for the scale-dependence of the weak mixing angle, as well as 
of non-SM coupling parameters. A particularly important and 
outstanding precision is expected for the determination of weak 
neutral-current couplings of the light quarks.

\vspace{1cm}
\section*{Acknowledgements}
We very gratefully acknowledge the numerous interactions with members
of the LHeC Study Group of which we had been a part also for the here presented study. 
We would like to thank S.~Schmitt, A.~Sch\"oning and Z.~Zhang for many related
discussions during the earlier stages of this work.
We thank J.~Erler and R.~Kogler for valuable discussions on the
interpretation of $Z$-pole data, and S.~Kluth for continuous support
and valuable comments to the manuscript.

\clearpage

\appendix
\section{A simplified global EW fit including LHeC data}
\label{app:globalfit}

The validity of the Standard Model is frequently studied in 
so-called `global EW 
fits'~\cite{deBlas:2016ojx,Haller:2018nnx,PDG2020}. A large 
number of precision measurements are simultaneously fitted 
and compared within a consistent theoretical framework. On 
the one hand, such a general approach allows one to perform 
a comprehensive consistency test of the SM; on the other hand 
one can expect to obtain the most precise determination of 
SM parameters. In this appendix, we are interested in estimating 
the potential impact of LHeC inclusive NC/CC DIS data on the 
parameter determination within such a global fit. We describe 
in the following a simplified approach which is suited for our 
purpose. 

It is common to consider a set of data as input for a global 
EW fit comprising $\alpha$ and \gf, the mass parameters \mz, 
\mw, \mH, and \mt, partial decay widths of the gauge bosons, 
$\Gamma_W$ and $\Gamma_Z$, the effective weak mixing angle
$\sweffl$, partial decay widths and 
asymmetry observables in $e^+e^-$ collisions~\cite{ALEPH:2005ab} 
$R_\ell^0$, $R_b^0$, $R_c^0$, $A_\ell$, $A_c$, $A_b$, 
$A_\textrm{FB}^{0,\ell}$, $A_\textrm{FB}^{0,c}$, 
$A_\textrm{FB}^{0,b}$, QCD contributions to the QED vacuum 
polarization $\Delta\alpha_\textrm{had}^5$, and sometimes also 
measurements of $\alpha_\textrm{s}$, $m_b$, and $m_c$. With the 
assumption that $\alpha$ and \gf\ are known with ultimate 
precision, the fit parameters are chosen as \mZ, \mH, \mt, 
$\Delta\alpha_\textrm{had}^5$ and $\alpha_\textrm{s}$. 
Sometimes also $m_b$ and $m_c$ are taken as parameters to be 
determined by the fit. 

Since we are interested only in estimating uncertainties and 
how these are affected by LHeC data, rather than in the actual 
fit values, some simplifications can be made in the following. 
One can observe that some of the fit parameters are determined with 
a high precision directly from measurements. This is the case 
for \mz, \mH, $m_b$, $m_c$, $\Delta\alpha_\textrm{had}^5$ and 
$\alpha_\textrm{s}$. If these direct data were not taken into 
account in the global fit, these parameters could be determined 
only with a much larger uncertainty. We do not expect that this 
will change if LHeC data are taken into account. Many other 
observables, like the $Z$-pole cross section ratios and 
asymmetries have much larger experimental uncertainties than 
predicted by the global fit. However, in the case of \mW\ and 
\mt{}, the uncertainties from direct measurements is of a 
similar size as those from a global fit. These parameters are 
therefore of special interest since for them one might expect 
a significant impact from LHeC data. 

Therefore, we perform a simplified global EW fit, where we 
consider only the determination of the SM parameters \mZ, \mW, 
\mH\ and \mt. All other parameters do not provide additional 
constraints or are known with good precision. The calculations 
are performed in the on-shell scheme. The fit parameters are 
chosen as \mW, \mt\ and \mH. When using input data for $\alpha$, 
\gf\ and the SM masses from PDG16~\cite{Olive:2016xmw} we obtain 
uncertainties of $\Delta\mw=\pm5.6\,\MeV$ ($\pm6.2\,\MeV$) and 
$\Delta\mt=\pm0.87\,\GeV$ ($\pm2.3\,\GeV$) when the data for 
\mW\ and \mt\ are included (excluded) in the fit. This result 
is very similar to the one of the general global EW 
fit~\cite{Haller:2018nnx,deBlas:2016ojx,Erler:2019hds} and 
confirms that the simplifications of our approach are justified 
for our purpose. 

\begin{figure}[tb!]
  \centering
  \includegraphics[width=0.48\textwidth,trim=0 20 0 10,clip]{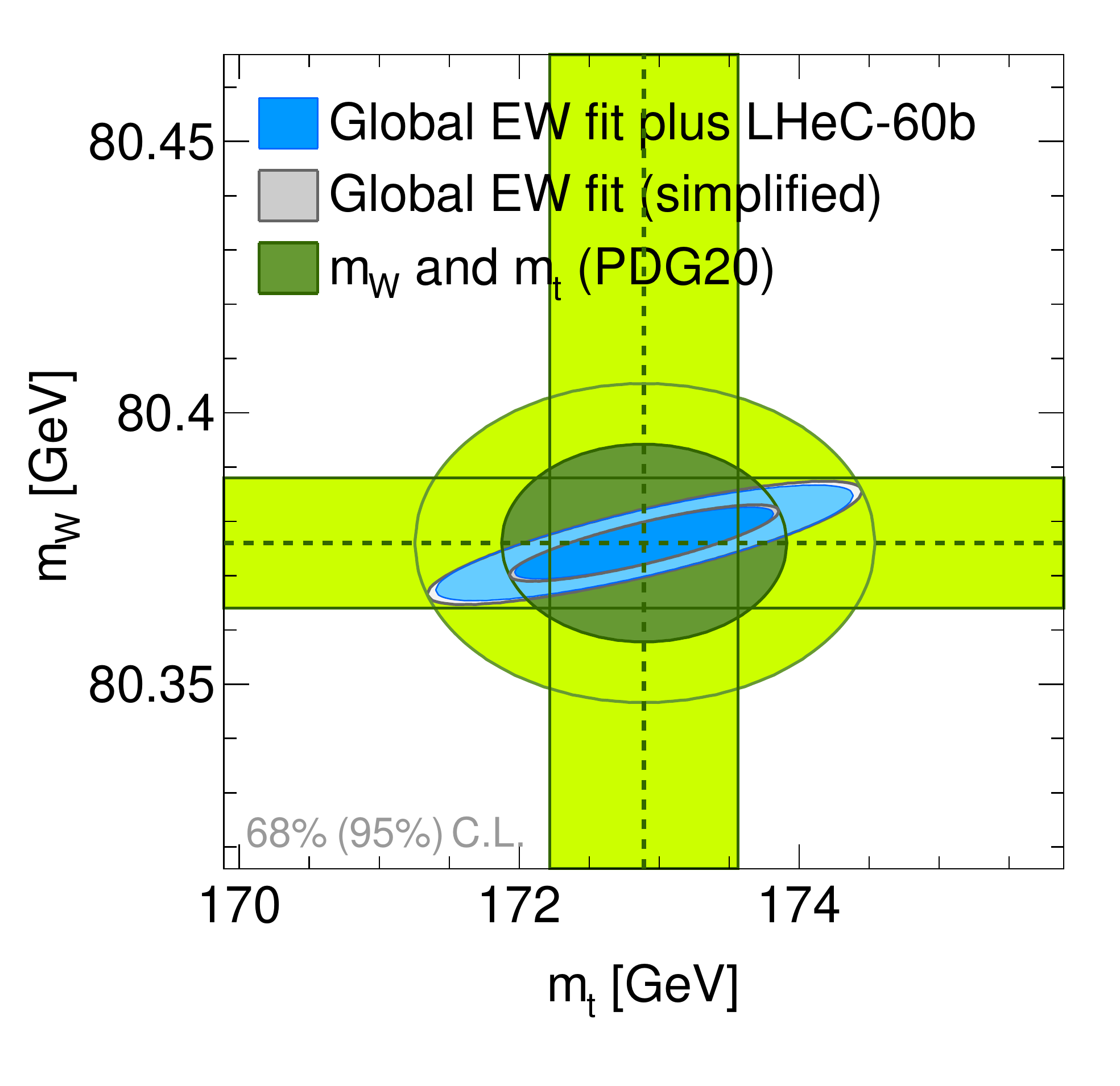} 
  \hskip0.02\textwidth
  \includegraphics[width=0.48\textwidth,trim=0 20 0 10,clip]{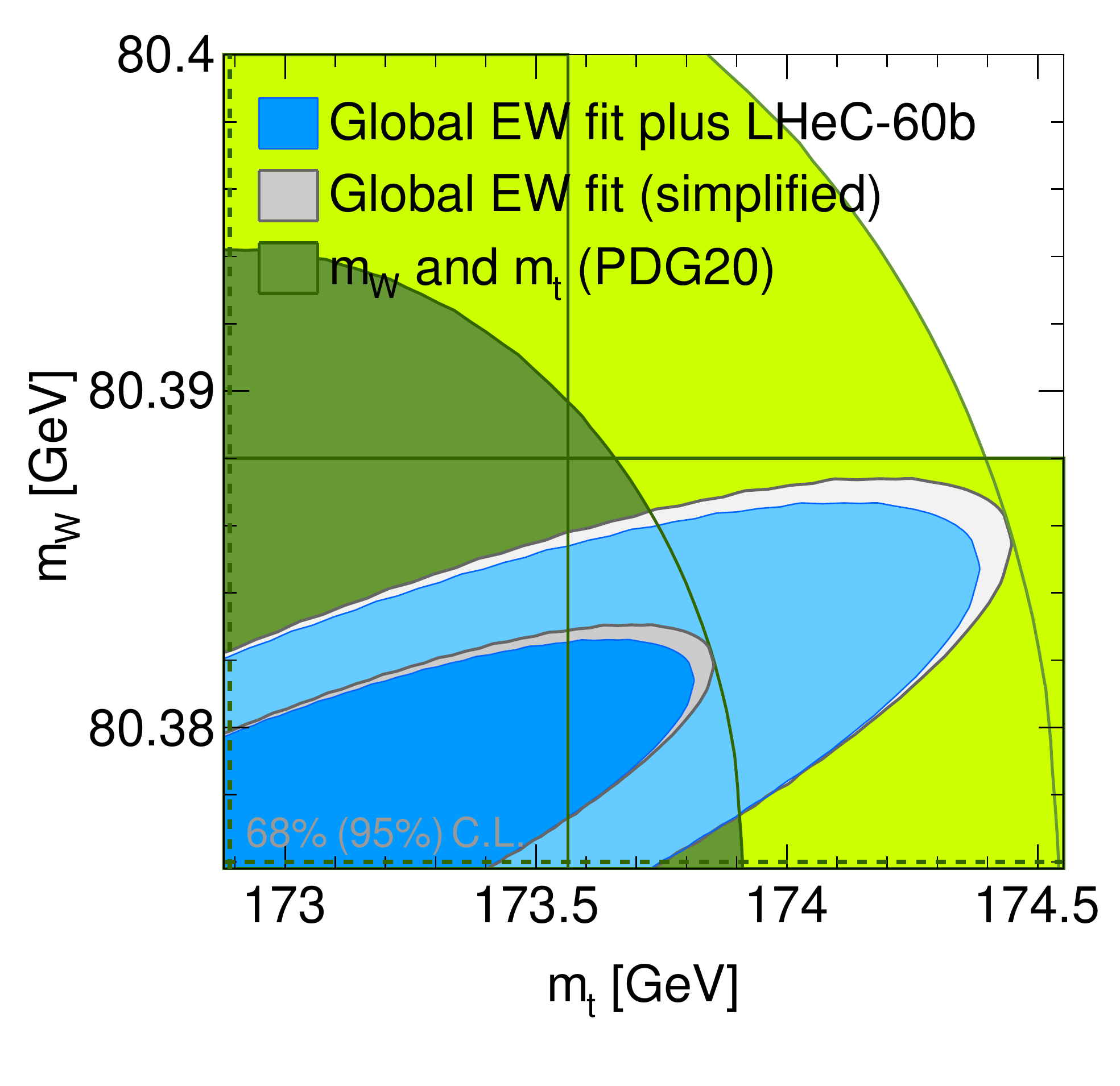} 
  \caption{
    Left: Expected uncertainties from a simplified global EW fit
    including LHeC-60b inclusive NC/CC DIS data. The results are 
    compared with expected uncertainties from the simplified 
    global EW fit without LHeC data and to the direct measurements 
    of \mW\ and \mt.
    Right: A magnified view of the left figure.
  }
\label{fig:globfit}
\end{figure}
\begin{table}[bht!]
  \footnotesize
  \centering
  \begin{tabular}{lcccc}
  \\
    \toprule
    Input data & Fit parameters & $\Delta\mw$ [MeV] & $\Delta\mt$ [GeV] & Correlation $\rho_{\mW\mt}$  \\
    \midrule
    \mw, \mt, \mZ, \mH\ (PDG16)     & \mw, \mt  &   $\pm5.6$ & $\pm0.87$  &  $+0.8$\\
    \addlinespace
    \mw, \mt\ (PDG20)              & \mw, \mt  &   $\pm12$ & $\pm0.67$   &  (0)\\
    \mw, \mt, \mZ, \mH\ (PDG20)     & \mw, \mt  &   $\pm4.6$ & $\pm0.63$  &  $+0.83$ \\
    \mw, \mt, \mZ, \mH, NC/CC DIS  & \mw, \mt, PDF  &   $\pm4.5$ & $\pm0.63$  & $+0.83$ \\
    \bottomrule
  \end{tabular}
  \caption{
    Results for expected uncertainties of \mw\ and \mt\ from a
    simplified global fit, with and without NC/CC DIS data using the
    LHeC-60b scenario.
  }
  \label{tab:global}
\end{table}
We can therefore proceed and include LHeC-60b inclusive NC/CC 
DIS data into the fit. Moreover, we include the PDF parameters 
as well, and we use recent values for \mW, \mt, and \mH\ from 
PDG20~\cite{PDG2020}. When including LHeC-60b NC/CC DIS data, 
we obtain a very moderate improvement for \mw: 
$\Delta\mw=\pm4.6\,\MeV$ changes to $\Delta\mw = \pm 4.5\,\MeV$.
The 68\,\% C.L. contours in the \mW-\mt--plane are displayed in
Fig.~\ref{fig:globfit}.
Numerical values of the validation fits, and the study including
LHeC-60b data are collected in Tab.~\ref{tab:global}.
No significant improvement of $\Delta\mt$ 
is seen. This was to be expected, since \mt\ contributes to NC/CC 
DIS only through higher-order corrections. Obviously, already 
other than the LHeC NC/CC DIS data are sufficient to provide good 
constraints on \mW\ and \mt. LHeC DIS data probe essentially the 
same relation of \mW\ and \mt\ and they cannot compete with direct 
measurements of these mass parameters.

Therefore, the strength of the LHeC EW analysis is its 
complementary with measurements in $e^+e^-$ or $pp$ collisions, 
in particular the fact that the dominant uncertainty in the 
$W$-boson mass determination is due to its correlation with the 
PDF determination. The impact of the latter may be considerably 
reduced if measurements at the LHeC will become available 
\cite{Azzi:2019yne}. 
Commonly, the global fit is repeated with direct measurements of 
the parameters of interest excluded, i.e.\ without \mw\ and \mt.
This was studied in Sec.~\ref{sec:mt}.

\clearpage
\begin{flushleft}
\bibliography{lhec_ew}
\end{flushleft}

\end{document}